\documentclass[10pt,journal]{IEEEtran} %
\usepackage{amsmath,amsfonts}
\usepackage{algorithm}
\usepackage{algpseudocode}
\usepackage{array}
\usepackage[caption=false,textfont=rm,justification=centering]{subfig}
\usepackage{textcomp}
\usepackage{stfloats}
\usepackage{url}
\usepackage{verbatim}
\usepackage{graphicx}
\usepackage{cite}
\usepackage[nolist, nohyperlinks]{acronym}
\usepackage{booktabs} 
\usepackage{pifont}

\usepackage[hidelinks]{hyperref}
\usepackage[nameinlink]{cleveref}

\crefname{equation}{}{}
\crefname{figure}{\figurename}{figures}  
\Crefname{figure}{\figurename}{Figures}
\crefname{table}{Table}{tables}     
\Crefname{table}{Table}{Tables} 
\crefname{section}{Sec.}{sections}
\Crefname{section}{Section}{Sections}
\crefname{algorithm}{Algorithm}{algorithms}
\Crefname{algorithm}{Algorithm}{Algorithms}

\usepackage{circuitikz}
\usepackage{tikz}
\usetikzlibrary{decorations.pathreplacing,calligraphy, arrows.meta, shapes.misc, shapes.geometric, positioning,patterns.meta,spy}
\tikzset{cross/.style={cross out, draw=black, minimum size=2*(#1-\pgflinewidth), inner sep=0pt, outer sep=0pt},}
\usepackage{pgf}
\usepackage{pgfplots}
\usetikzlibrary{calc}
\usepackage{contour}
\usepackage{xparse}
\colorlet{bboxcolor}{white} %

\newcommand\tickWidth{0.5pt}
\newcommand\majorTick{0.75mm}
\newcommand\minorTick{0.5mm}
\newcommand\tickSize{7pt}
\newcommand\tickSkip{7pt}
\newcommand\tickFont{\fontsize{\tickSize}{\tickSkip}\selectfont}
\newcommand\labelFont{\footnotesize}

\newcommand\legendRectEdge{1mm}

\newcommand\splitSize{3pt}
\newcommand\multiSize{5pt}
\newcommand\shadowOff{0.375mm}
\newcommand\dnnWidth{0.85cm}
\newcommand\dnnHeight{0.5cm}
\newcommand\pfWidth{\dnnWidth}
\newcommand\pfHeight{\dnnHeight}
\newcommand\nodeDistance{2.5mm}
\newcommand\doubleInputOff{1.5mm}
\newcommand\specWidth{7mm}
\newcommand\specHeight{5mm}
\newcommand\delayOff{5mm}
\newcommand\condShiftX{-2.5mm}
\newcommand\condShiftY{2.5mm}
\newcommand\condArrowOff{-2.5mm}

\newcommand\boxStartX{-1.3cm}
\newcommand\boxStartY{-0.25cm} %
\newcommand\boxEndY{2.0cm} %

\newcommand\ssfOutputShift{1mm}
\newcommand\flowBigFont{\normalsize}
\newcommand\flowMathFont{\small}

\newcommand\sigSymbol{*}
\newcommand\sigSymbolOff{1.5mm}

\newcommand\sigSymbolTwo{**}
\newcommand\sigSymbolOffTwo{3mm}
\newif\ifshowtrackingdnn

\newcommand\ambixWidth{0.5pt}

\tikzstyle{rotBox} = [
  rectangle,
  draw=black,
  minimum width=0.6cm,
  minimum height=0.6cm,
  inner sep=0pt,
  outer sep=0pt,
  align=center,
  text width=0.6cm,
  text height=0.6cm,
  text depth=0cm
]

\NewDocumentCommand{\dnnSymb}{m O{gray} O{lightgray} O{0.85}}{%
  \scalebox{#4}{%
    \raisebox{0.2mm}{%
    \begin{tikzpicture}
      \def\r{0.25mm} %
      \def\xleft{-0.45}
      \def\xmidlone{-0.3}
      \def\xmidltwo{-0.15}
      \def\xmidrtwo{0.15}
      \def\xmidrone{0.3}
      \def\xright{0.45}
      \def\yspace{1.25mm}
      \def\numNmid{5}
      \def\numNin{3}

      \foreach \i in {1,...,\numNin} {
        \node[circle,fill=#2,inner sep=0pt,minimum size=2*\r]
          (ll\i) at (\xleft,{\yspace*(\i-\numNin+ (\numNmid - \numNin) / 2)}) {};
      }

      \foreach \i in {1,...,\numNmid} {
        \node[circle,fill=#2,inner sep=0pt,minimum size=2*\r]
          (mlone\i) at (\xmidlone,{\yspace*(\i-\numNin)}) {};
      }

      \foreach \i in {1,...,\numNmid} {
        \node[circle,fill=#2,inner sep=0pt,minimum size=2*\r]
          (mltwo\i) at (\xmidltwo,{\yspace*(\i-\numNin)}) {};
      }

      \foreach \i in {1,...,\numNmid} {
        \node[circle,fill=#2,inner sep=0pt,minimum size=2*\r]
          (mrtwo\i) at (\xmidrtwo,{\yspace*(\i-\numNin)}) {};
      }

      \foreach \i in {1,...,\numNmid} {
        \node[circle,fill=#2,inner sep=0pt,minimum size=2*\r]
          (mrone\i) at (\xmidrone,{\yspace*(\i-\numNin)}) {};
      }

      \foreach \i in {1,...,\numNin} {
        \node[circle,fill=#2,inner sep=0pt,minimum size=2*\r]
          (r\i) at (\xright,{\yspace*(\i-\numNin+ (\numNmid - \numNin) / 2)}) {};
      }

      \foreach \li in {1,...,\numNin}{
        \foreach \mi in {1,...,\numNmid}{
          \draw[#3,line width=0.25pt] (ll\li) -- (mlone\mi);}
      }
      \foreach \li in {1,...,\numNmid}{
        \foreach \mi in {1,...,\numNmid}{
          \draw[#3,line width=0.25pt] (mlone\li) -- (mltwo\mi);}
      }
      \foreach \li in {1,...,\numNmid}{
        \foreach \mi in {1,...,\numNmid}{
          \draw[#3,line width=0.25pt] (mltwo\li) -- (mrtwo\mi);}
      }
      \foreach \li in {1,...,\numNmid}{
        \foreach \mi in {1,...,\numNmid}{
          \draw[#3,line width=0.25pt] (mrtwo\li) -- (mrone\mi);}
      }
      \foreach \mi in {1,...,\numNmid}{
        \foreach \ri in {1,...,\numNin}{
          \draw[#3,line width=0.25pt] (mrone\mi) -- (r\ri);}
      }

      \node[anchor=center] at (0, 0) {\small\contour{white}{#1}};
    \end{tikzpicture}}%
  }%
}

\tikzstyle{dnnBox} = [
  rectangle,
  draw=black,
  minimum width=\dnnWidth,
  minimum height=\dnnHeight,
  inner sep=0pt,
  outer sep=0pt,
  align=center,
  text width=\dnnWidth,
  text height=\dnnHeight,
  text depth=0cm
]

\tikzstyle{delay} = [
  rectangle,
  draw=black,
  minimum width=0.3cm,
  minimum height=0.3cm,
  inner sep=0pt,
  outer sep=0pt,
  align=center,
  text width=0.3cm,
  text height=0.3cm,
  text depth=0cm,
  fill=white
]
\newcommand{\delaySymb}{
  \makebox[1cm][l]{\hspace{-1mm}\raisebox{0.25mm}{
  \begin{tikzpicture}
  \node {\tiny$z^{\hspace{-0.5mm}\scalebox{0.5}[1.0]{$-$}\hspace{-0.25mm}\scalebox{0.8}{$1$}}$};
  \end{tikzpicture}
  }
  }
}
\tikzstyle{split} = [circle, fill=black, minimum size=\splitSize, inner sep=0pt]
\tikzstyle{unfoldedSplit} = [circle, fill=black, minimum size=\unfoldedSplitSize, inner sep=0pt]
\tikzstyle{add} = [circle, fill=white, minimum size=\addSize, inner sep=0pt, draw=black]
\tikzset{
  multi/.style={
    circle,
    draw,           %
    fill=white,     %
    minimum size=\multiSize, inner sep=0pt,
    path picture={
      \draw[draw, line width=0.5pt]
        (path picture bounding box.south west) -- (path picture bounding box.north east)
        (path picture bounding box.north west) -- (path picture bounding box.south east);
    }
  }
}
\tikzstyle{arrow} = [line width=\ambixWidth,
    -{Computer Modern Rightarrow[line cap=round,width=6pt,length=2.5pt,line width=1pt]},
    draw=black]

\newcommand\myPM{$\pm$}

\newcommand{\myExp}[1]{
    10$^{\scalebox{0.6}[1.0]{$-$}#1}$
}

\newcommand{\tinySuper}[1]{\scalebox{0.6}[0.6]{$#1$}}

\newcommand{\picLegend}[1]{\raisebox{-0.275ex}{\includegraphics[height=0.75em]{#1}}}  %
\newcommand{\picLegendLarge}[1]{\raisebox{-0.35ex}{\includegraphics[height=0.85em, width=2em]{#1}}} 
\newcommand{\picTable}[1]{\raisebox{-0.275ex}{\includegraphics[height=0.85em]{#1}}}  %
\newcommand{\gridIcon}[1][1.1]{%
  \tikz[baseline=0.2ex, scale=#1, xscale=1.25]{
    \fill[tab_orange, opacity=0.5] (0mm,1mm) rectangle (1mm,2mm);
    \fill[tab_blue,   opacity=0.5] (1mm,1mm) rectangle (3mm,2mm);
    \fill[tab_orange, opacity=0.5] (3mm,1mm) rectangle (4mm,2mm);

    \fill[tab_blue,   opacity=0.5] (0mm,0mm) rectangle (1mm,1mm);
    \fill[tab_orange, opacity=0.5] (1mm,0mm) rectangle (3mm,1mm);
    \fill[tab_blue,   opacity=0.5] (3mm,0mm) rectangle (4mm,1mm);
  }%
}

\tikzset{
  boundaryhatch/.style={
    pattern={
      Lines[
        angle=-45,
        distance=2pt,
        line width=0.4pt,
      ]
    },
    pattern color=red
  }
}

\newcommand\unfoldedFlowMathFont{\footnotesize}
\newcommand\dnnBoxWidth{0.6 * \dnnWidth} %
\newcommand\dnnBoxHeight{0.6 * \dnnHeight}
\newcommand\unboundedBoxStartY{-1.25cm} %
\newcommand\unboundedBoxEndY{1.8cm} %
\newcommand\numRep{2}
\newcommand\unfoldedNoisyOff{-0.8cm}  %
\newcommand\unfoldedBigFont{\scriptsize}
\newcommand\unfoldedSSFshift{5mm} %
\newcommand\unfoldedSSFlabelShift{1.25mm}
\colorlet{bboxcolorUnfolded}{white} %
\newcommand\unfoldedSplitSize{2.5pt}

\newcommand\interferingSpeaker{I}

\DeclareRobustCommand{\particlePictogram}[1][yellow!60!orange]{%
  \tikz[baseline=-0.6ex]{%
    \begin{scope}
      \clip (0,-3.5pt) rectangle (4mm,3.5pt);

      \foreach \amplitude / \phase in {
      1pt/0pt,1.5pt/2pt,2pt/8pt
      }{%

        \pgfmathsetlengthmacro{\negativephase}{-\phase}%

        \begin{scope}[xshift=\negativephase]
          \draw[
            color=#1,
            line width=0.5pt,
            decorate,
            decoration={
              snake,
              amplitude=\amplitude,
              segment length=10pt
            }
          ]
            (0,0) -- (4mm+10pt,0);
        \end{scope}
      }%
    \end{scope}
  }%
}

\DeclareRobustCommand{\meanpictogram}{%
  \tikz[baseline=-0.5ex,x=1em,y=1ex]{
    \draw[black, dash pattern=on 2pt off 1pt, line width=0.8pt] (0,0) -- (1,0);
  }%
}

\DeclareRobustCommand{\confidencepictogram}{%
  \tikz[baseline=-0.5ex,x=1em,y=1ex]{
    \fill[gray!35] (0,-0.6) rectangle (1,0.6);
    \draw[black, line width=0.4pt] (0,0.6) -- (1,0.6);
    \draw[black, line width=0.4pt] (0,-0.6) -- (1,-0.6);
  }%
}

\usepackage{multirow}

\usepackage{xcolor}
\usepackage[normalem]{ulem} %

\definecolor{diffColor}{RGB}{90,150,210}
\definecolor{diffOldColor}{RGB}{200,0,0}

\newif\ifdiff  %
\newif\ifnewdiff  %

\difffalse

 \newdifffalse

\usepackage{soul}
\soulregister\cite7
\soulregister\cref7
\soulregister\Cref7
\soulregister\citen7
\soulregister\ac7
\soulregister\acp7
\soulregister\acfp7
\soulregister\acl7
\soulregister\acs7
\soulregister\textit7

\newcommand{\diffComm}[2]{%
  \ifdiff
    {\setstcolor{diffOldColor}\textcolor{diffOldColor}{\st{#1}}}%
    \ {\textcolor{diffColor}{#2}}%
  \else
    #2%
  \fi
}

\DeclareRobustCommand{\newdiffComm}[2]{%
  \ifnewdiff
    {\setstcolor{diffOldColor}\textcolor{diffOldColor}{\st{#1}}}%
    \ {\textcolor{diffColor}{#2}}%
  \else
    #2%
  \fi
}
\newcommand{\diffState}[2]{%
  \ifdiff
    \State {\textcolor{diffOldColor}{\st{#1}}}%
    \Statex \hspace{\algorithmicindent}\hspace{\algorithmicindent}\hspace{\algorithmicindent}{\textcolor{diffColor}{#2}}%
  \else
    \State #2%
  \fi
}
\newenvironment{diffEnv}
  {%
    \ifdiff
      \color{diffColor}%
    \else
      \color{black}%
    \fi
  }
  {}

\usepackage{xstring} %

\acrodefplural{nns}[NNs]{neural networks}
\acrodefplural{irm}[IRMs]{ideal ratio masks}
\acrodefplural{ibm}[IBMs]{ideal binary masks}
\acrodefplural{doa}[DoAs]{directions of arrival}
\acrodefplural{rirs}[RIRs]{room impulse responses}

\begin{acronym}
\acro{stft}[STFT]{short-time Fourier transform}
\acro{rir}[RIR]{room impulse response}
\acro{istft}[iSTFT]{inverse short-time Fourier transform}
\acro{doa}[DoA]{direction of arrival}
\acro{irm}[IRM]{ideal ratio mask}
\acro{mvdr}[MVDR]{minimum-variance distortionless response}
\acro{nn}[NN]{neural network}
\acro{dnn}[DNN]{deep neural network}
\acro{gcc_phat}[GCC-PHAT]{generalized cross-correlation phase transform}
\acro{ssl}[SSL]{sound source localization}
\acro{srp_phat}[SRP-PHAT]{steered response power with phase transform}
\acro{mse}[MSE]{mean squared error}
\acro{mae}[MAE]{mean angular error}
\acro{sdr}[SDR]{signal-to-distortion ratio}
\acro{sisdr}[SI-SDR]{scale-invariant signal-to-distortion ratio}
\acro{pesq}[PESQ]{perceptual evaluation of speech quality}
\acro{wer}[WER]{word error rate}
\acro{pit}[PIT]{permutation-invariant training}
\acro{lbt}[LBT]{location-based training}
\acro{asr}[ASR]{automatic speech recognition}
\acro{bce}[BCE]{binary cross-entropy}
\acro{jnf}[JNF]{joint non-linear filter}
\acro{nn}[NN]{neural network}
\acro{dnn}[DNN]{deep neural network}
\acro{mccruse}[MC-CRUSE]{multi-channel convolutional recurrent U-net architecture for speech enhancement}
\acro{estoi}[ESTOI]{extended short-time objective intelligibility}
\acro{snr}[SNR]{signal-to-noise ratio}
\acro{sir}[SIR]{signal-to-interference ratio}
\acro{iid}[i.i.d.]{independent and identically distributed}
\acro{wrt}[w.r.t.]{with respect to}
\acro{ssf}[SSF]{spatially selective filter}
\acro{cv}[CV]{constant velocity}
\acro{rw}[RW]{random walk}
\acro{ar}[AR]{autoregressive}
\acro{tse}[TSE]{target speaker extraction}
\acro{tst}[TST]{target speaker tracking}
\acro{tsl}[TSL]{target speaker localization}
\acro{dnsmos}[DNSMOS]{deep noise suppression mean opinion score}
\acro{map}[MAP]{maximum a posteriori}
\acro{kf}[KF]{Kalman filter}
\acro{pf}[PF]{particle filter}
\acro{das}[DaS]{delay-and-sum}
\acro{ae}[AE]{angular error}
\acro{acc}[ACC]{accuracy}
\acro{miso}[MISO]{multiple-input and single-output}
\acro{mimo}[MIMO]{multiple-input and multiple-output}
\acro{mmse}[MMSE]{minimum mean squared error}
\acro{ls}[LS]{least-squares}
\acro{ema}[EMA]{exponential moving average}
\acro{mac}[MAC]{multiply-accumulate operation}
\acro{ipd}[IPD]{inter-channel phase difference}
\acro{seld}[SELD]{sound event localization and detection}
\acro{rds}[RDS]{recurrent deep stacking}
\acro{vad}[VAD]{voice activity detection}
\acro{rtf}[RTF]{real-time factor}
\end{acronym}

\newcommand\projectPage{
    \scriptsize \texttt{https://sp-uhh.github.io/autoregressive-spatial-filters/}
}
\newcommand\codePage{
    \scriptsize \texttt{https://github.com/sp-uhh/autoregressive-spatial-filters}
}

\newcommand\perturb{\nu}
\newcommand\idxParticles{n}
\newcommand\numParticles{N}
\newcommand\particleSuperscript[1]{#1}
\newcommand\condSymb{\, | \,}
\newcommand{\delayIdx}[2][1]{%
    \scriptstyle #2\hspace{0.25mm}\raisebox{0.4ex}{\rule{0.3em}{0.06ex}}#1%
}

\DeclareMathAlphabet\mathbfcal{OMS}{cmsy}{b}{n}

\definecolor{tab_blue}{RGB}{31, 119, 180}
\definecolor{tab_green}{RGB}{44, 160, 44}
\definecolor{tab_orange}{RGB}{255, 127, 14}
\definecolor{tab_purple}{RGB}{148, 103, 189}
\definecolor{tab_brown}{RGB}{120, 42, 38} %
\definecolor{tab_pink}{RGB}{227, 119, 194}
\definecolor{tab_cyan}{RGB}{23, 190, 207}
\definecolor{orig_green}{RGB}{44, 160, 44}
\definecolor{orig_purple}{RGB}{148, 103, 189}
\definecolor{custom_green}{RGB}{149.5, 207.5, 149.5}
\definecolor{custom_purple}{RGB}{201.5, 179, 222}
\definecolor{dark_gray}{RGB}{128, 128, 128}

\begin{document}

\title{Autoregressive Guidance of Deep Spatially Selective Filters using Bayesian Tracking for Efficient Extraction of Moving Speakers}

\author{Jakob Kienegger,~\IEEEmembership{Student Member,~IEEE,} and Timo Gerkmann,~\IEEEmembership{Senior Member,~IEEE}
\thanks{\phantom{Manuscript received xx ???? xxxx; revised xx ??? 202x and xx ???? 202x;} \\ \phantom{accepted xx ???? 202x. Date of publication xx ??? 202x; date of current} \\ \phantom{version xx ???? 202x.}}
\thanks{This work was supported by the Deutsche Forschungsgemeinschaft (DFG, German Research Foundation) under Grant 508337379. Computational resources were provided by the Regional Computer Center (RRZ) of the University of Hamburg and the Erlangen National High Performance Computing Center (NHR@FAU) under Project f104ac. NHR is funded by the Federal Government and the State of Bavaria. Hardware at NHR@FAU and RRZ received partial DFG funding under Grants 440719683 and 498394658.}
\thanks{The authors are with the Signal Processing Group, Department of Informatics, University of Hamburg, 22527 Hamburg, Germany (e-mail: jakob.kienegger@uni-hamburg.de; timo.gerkmann@uni-hamburg.de).
}%
\thanks{\phantom{Digital Object Identifier xxx}}
}
\markboth{Journal of \LaTeX\ Class Files,~Vol.~XX, No.~X, Month~XXXX}%
{Shell \MakeLowercase{\textit{et al.}}: A Sample Article Using IEEEtran.cls for IEEE Journals}

\maketitle

\begin{abstract}
Deep spatially selective filters achieve high-quality enhancement with real-time capable architectures for stationary speakers of known directions.
To retain this level of performance in dynamic scenarios \diffComm{when}{where} only the speakers' initial directions are given, accurate, yet computationally lightweight tracking algorithms become necessary.
Assuming a frame-wise causal processing style, temporal feedback allows for leveraging the enhanced speech signal to improve tracking performance. In this work, we investigate strategies to incorporate the enhanced signal into lightweight tracking algorithms and autoregressively guide deep spatial filters.
Our proposed Bayesian tracking algorithms are compatible with arbitrary deep spatial filters.
To increase the realism of simulated trajectories during development and evaluation, we \diffComm{propose and publish a novel dataset}{develop a synthetic data generation framework} based on the social force motion model.
Results validate that the autoregressive incorporation significantly improves the accuracy of our Bayesian trackers, resulting in superior enhancement with none or only negligibly increased computational overhead.
Real-world recordings complement these findings and demonstrate the generalizability of our methods to unseen\diffComm{, challening}{} acoustic conditions.
\end{abstract}

\begin{IEEEkeywords}
Multichannel speaker extraction, direction of arrival (DoA) estimation, moving speakers, Bayesian tracking.
\end{IEEEkeywords}

\section{Introduction}\label{sec:introduction}
\IEEEPARstart{S}{peech} enhancement aims to improve the quality and intelligibility of a recorded speech signal by removing noise and reverberation.
In a scenario with multiple speakers, such as the \textit{cocktail party problem}~\cite{cherry53cocktail_party}, \newdiffComm{additional,}{} overlapping speech signals of \newdiffComm{other }{}competing speakers \newdiffComm{ represent a}{are} particularly challenging\newdiffComm{ noise type}{}, due to their similar and non-stationary statistical properties.
\newdiffComm{If these interferences are of similar level as the desired target speaker}{Consequently}, an ambiguity arises who to enhance and who to suppress.
Target speaker extraction~(TSE)\acused{tse} solves this problem by utilizing additional information, referred to as cues, to distinguish the desired from competing speakers.
Conditioned on \newdiffComm{one or multiple}{} cues, recent advances in \ac{nn}-driven methods demonstrate exceptional speech enhancement performance under ever more challenging conditions, see\newdiffComm{}{, e.g.,} \cite{zmolikova23tse_overview}\newdiffComm{ for an overview}{}.

\newdiffComm{When recordings from a microphone array are available}{With microphone array recordings}, the \newdiffComm{target speaker's}{target's} position provides an effective cue for speech enhancement.
\newdiffComm{Leveraging this information}{With this cue}, a \ac{ssf} can be steered toward the desired location to extract the corresponding speech signal.
In practice, this cue is commonly restricted to the target’s azimuth orientation relative to the microphone array, referred to as the \ac{doa}~\cite{tesch24ssf_journal, bohlender24sep_journal}.
For stationary and directionally distinct target speakers, deep non-linear \acp{ssf} can achieve high spatial selectivity~\cite{tesch23deep_nonliner_filter_multichannel}, resulting in strong interference suppression.
Consequently, when provided with accurate \ac{doa} information, recently proposed \acp{ssf} demonstrate state-of-the-art enhancement performance while retaining computationally lightweight \ac{nn} architectures \cite{tesch24ssf_journal, bohlender24sep_journal,jing25e2e_ssf,pandey12directional_speech_extraction,gu24rezero,choi25mimo_ssf_tse,padilla25hearing_aids_ssf}.

Highly constrained recording setups, such as a seated conference meeting with a centrally placed microphone array~\cite{chen20libricss}, may legitimate the assumption of stationary and directionally distinct speaker locations.
However, more general settings like the dinner party scenario considered in \cite{barker18chime5}, clearly violate these assumptions.
The resulting time-varying \acp{snr} due to changing speaker-to-array distances and directionally ambiguous constellations, e.g., crossing speakers, \newdiffComm{significantly}{substantially} increase the difficulty of the enhancement task.
While deep \acp{ssf} are capable of resolving such ambiguities by utilizing temporal context to learn the target's temporal-spectral characteristics~\cite{kienegger25wg_ssf}, the need for precise directional guidance bears an additional challenge.
Since continuous knowledge of the target speaker’s \ac{doa} throughout the recording, referred to as \textit{strong} guidance, is in general unavailable, \textit{weakly} guided \ac{tse} relies only on the initial direction and incorporates a tracking algorithm to automate the steering of the \ac{ssf}~\cite{kienegger25wg_ssf,kienegger25sg_ssf,kienegger25deep_joint_ar_ssf}.
However, accurate tracking of a moving target speaker under difficult acoustic conditions typically requires resource-intensive \acp{nn}~\cite{diaz21srp_phat, bohlender21ssl_temporal_context, yang2022srp_dnn, wang2023fnssl_full_band_narrow_band_tracking,xiao25tf_mamba_ssl}, increasing the computational burden of the \ac{tse} pipeline.

\diffComm{While most speech enhancement systems operate offline, increasing demand in telecommunications, assistive technologies, and consumer electronics drives research toward real-time solutions~\cite{tan18cruse, defossez20demucs, braun21cruse}.
Typically implemented as frame-wise causal versions of offline methods, these approaches suffer from a fundamental disadvantage due to being restricted to current and past data during processing~\cite{luo19conv_tasnet, chao24SEmamba}.
However, recent works show that their sequential nature can also benefit the enhancement performance.
An \ac{ar} \ac{nn} architecture with the processed signal as feedback facilitates improved exploitation of the temporal correlations of speech to preserve waveform continuity~\cite{andreev23iterative_autoregression, pan24paris_autoregressive_separation,shen25arise}. 
Pseudo-\ac{ar} training strategies allow these methods to maintain parallelizability while generalizing to frame-wise 
inference.}{Driven by applications in telecommunications, assistive technologies, and consumer electronics, many speech enhancement systems are implemented as frame-wise causal methods restricted to current and past data~\cite{tan18cruse}.
Besides being a necessity for low latency enhancement, causal methods also allow to elegantly include \ac{ar} feedback of the processed signal to better exploit temporal correlations of speech and preserve waveform continuity \cite{andreev23iterative_autoregression, pan24paris_autoregressive_separation,shen25arise}, with pseudo-AR training strategies maintaining parallelizability while generalizing to frame-wise inference.}

Instead of leveraging autoregression within the speech enhancement architecture, we have previously proposed to incorporate the processed speech signal for tracking, resulting in an \ac{ar}-guided, or \textit{self-steering}, \ac{ssf}~\cite{kienegger25sg_ssf,kienegger25deep_joint_ar_ssf}.
While we focused on a neural tracker in~\cite{kienegger25deep_joint_ar_ssf}, our work in~\cite{kienegger25sg_ssf} demonstrated how the estimates of \diffComm{a slightly}{an} adapted \ac{ssf} architecture can effectively compensate for the limited modeling capabilities of a lightweight, statistics-based algorithm.
\diffComm{}{In particular, we modified a \ac{ssf} to estimate the target speech at all microphones instead of at a single microphone, and then used it to replace the unprocessed recording for tracking.}
\newdiffComm{}{While dedicated neural systems may still have advantages when speaker identity has to be maintained through long silence, 
our method offers efficient, interpretable model-based spatial guidance.
}

In this work, \diffComm{we further develop our approach from \cite{kienegger25sg_ssf} by exploring different strategies to incorporate the \ac{ssf} into Bayesian tracking frameworks.}{we generalize the concept and develop new strategies to incorporate the enhanced speech signal without modifying the \ac{ssf} architecture.}
\diffComm{Specifically, we propose modified filtering formulations for widely used}{For all approaches, we derive novel tracking formulations based on}  Kalman \cite{traa13wrapped_kalman_filter} and particle filter \cite{ward03basic_particle_filter} algorithms.
To improve realism of simulated speaker trajectories\diffComm{ during development and evaluation, we publish a novel synthetic dataset}{, we develop a synthetic data generation methodology} based on the social force motion model~\cite{helbing95social_force_model}.
Results demonstrate that the \ac{ar} incorporation of the processed speech signal consistently increases tracking accuracy,
yielding significantly improved enhancement\newdiffComm{ performance}{}.
A detailed analysis demonstrates the generalization capabilities of our methods to real-world recordings in unseen\newdiffComm{, challenging}{} \diffComm{}{acoustic }conditions.

The remainder of this paper is organized as follows.
\Cref{sec:problem_def} formulates the problem and notation, along with an introduction of steerable \acp{ssf} and Bayesian estimators for tracking in \cref{sec:ssf}.
Our proposed Bayesian tracking formulations for \ac{ar} guidance are presented in \cref{sec:self_tse}.
\Cref{sec:dataset} introduces our \diffComm{novel synthetic dataset}{data simulation framework}, followed by an overview of the experimental setup in \cref{sec:experimental_setup}.
Performance and generalization capabilities are discussed in \cref{sec:evaluation}.

\section{Problem Definition}\label{sec:problem_def}
We consider a noisy and reverberant \newdiffComm{recording scenario}{scene} captured by a planar \newdiffComm{omni-directional microphone array with M channels}{array of $M$ omni-directional microphones}. 
The multichannel observation at the $m$-th microphone $y^m$ is modeled as the sum of anechoic target speech signals $s^m$ and noise $v^m$, where $v^m$ comprises interfering speech, environmental and measurement noise, and the reverberant components of the target speech.
In the \ac{stft} domain, which we denote by capital letters, the multichannel observation can be written as
\begin{equation}\label{eq:signal_model}
    \mathbf{Y}_{tk} = \mathbf{S}_{tk} + \mathbf{V}_{tk} \, \in \mathbb{C}^M,
\end{equation}
with $t$ and $k$ indexing frame and frequency bins respectively and vectorization (indicated in boldface) conducted over the $M$ microphone channels.
In this work, we aim to reconstruct the anechoic target speech at a predefined reference microphone, denoted by $S_{tk}$. 
Under far-field conditions, amplitude differences across microphones are negligible, and the remaining inter-channel time delays w.r.t. the reference microphone can be modeled using a steering vector $\mathbf{d}_{tk}$ [\citen{benesty24microphone_arrays}, Sec.~3.1], giving
\begin{equation}\label{eq:farfield_signal_model}
    \mathbf{Y}_{tk} = \mathbf{d}_{tk} S_{tk} + \mathbf{V}_{tk} \ .
\end{equation} 
For a planar array with microphones at a similar height as the target speaker, the steering vector can be approximated as depending only on the target's azimuth direction $\theta_t$, i.e., 
\begin{equation}\label{eq:steering_vector_approximation}
    \mathbf{d}_{tk} \approx \mathbf{d}_{k}(\theta_t) \, .
\end{equation}
Consequently, we refer to $\theta_t$ as the \acl{doa} (\acs{doa}) throughout this work, implicitly excluding elevation.

\newcommand\flowVspace{-10pt}
\newcommand\flowHspace{-5pt}
\begin{figure*}[t!]
\vspace*{-10pt}
\centering
\subfloat[Concatenation of \acs{tst} and \acs{ssf}.\label{fig:weak_ssf}]{
  \hspace*{\flowHspace}\begin{minipage}{0.30\linewidth}
    \centering

\newcommand\boxEndX{4.2cm}
\begin{tikzpicture}[node distance=1.5*\nodeDistance and 1*\nodeDistance, background rectangle/.style={fill=none}] 
\path[use as bounding box, draw, dashed, bboxcolor] (\boxStartX,\boxStartY) rectangle (\boxEndX,\boxEndY);

\pgfmathsetlengthmacro{\condShift}{
    \pfHeight + 0.5 * \nodeDistance
}
\pgfmathsetlengthmacro{\ssfOff}{
    \dnnWidth - 1.25 * \nodeDistance
}
\pgfmathsetlengthmacro{\ssfShift}{
    \pfWidth - \dnnWidth
}

\node [] (noisy_input) {};
\foreach \x in {0, ..., 2}{
 \pgfmathtruncatemacro{\colorPerc}{round(\x*30 + 15)}
\pgfmathsetlengthmacro{\sOff}{(\x - 1) * \shadowOff}
\node [xshift=-\sOff - \specWidth / 2, yshift=-\sOff] at (noisy_input.center) {
  \pgfimage[width=\specWidth, height=\specHeight]{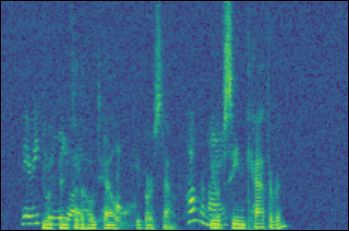}
};
}
\node [anchor=east, xshift=-\specWidth] at (noisy_input) {\flowMathFont$\mathbf{Y}_t$};

\node [right=of noisy_input] (split_input) {};
\node at (split_input) (split_input_shift) {};
\node[above right=of split_input_shift, minimum width=\pfWidth, minimum height=\pfHeight, align=center,
      inner sep=0pt,
      outer sep=0pt] {
\pgfimage[width=\pfWidth, height=\pfHeight]
{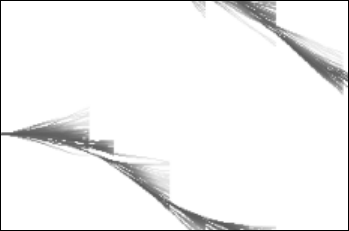}
};
\node[dnnBox, above right=of split_input_shift, minimum width=\pfWidth, minimum height=\pfHeight, align=center,
      inner sep=0pt,
      outer sep=0pt] (tst) {};
\node at (tst.center) {\flowBigFont\contour{white}{TST}};
\node[right=of tst] (tst_knick) {};

\node [above=of noisy_input, xshift=\condShiftX, yshift=\condShiftY] (cond) {
    \input{images/flowchart/weak_init_doa.tikz}
};

\node [dnnBox] at (tst_knick |- split_input) (ssf) {\dnnSymb{}};
\node [right=of ssf] (ar_split) {};
\node[xshift=\ssfOutputShift] at (ar_split) (ssf_output) {};
\node [xshift=\specWidth, anchor=west] at (ssf_output) {\flowMathFont$\hat{S}_t$};
\node [xshift=\specWidth / 2 + 2 * \shadowOff, draw=tab_orange, line width=1pt, inner sep=0.5pt] at (ssf_output.west) {
  \pgfimage[width=\specWidth, height=\specHeight]{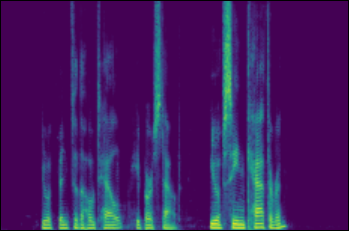}
};
\node at (ssf.center) {\flowBigFont\contour{white}{SSF}};

\foreach \x in {0, ..., 2}{
\pgfmathsetlengthmacro{\sOff}{(\x - 1) * \shadowOff}
\ifthenelse{\x = 2}{
    \def\shadowColor{black}
}{
    \pgfmathtruncatemacro{\colorPerc}{round(\x*10 + 20)}
    \def\shadowColor{black!\colorPerc!white}
}

\node [yshift=-\sOff, xshift=-\sOff] at (split_input.center) (split_input_\x) {};
\node [yshift=-\sOff, xshift=-\sOff] at (split_input.center) (split_input_\x) {};
\node [split, fill=\shadowColor] at (split_input_\x) (split_input_shift_\x) {};
\draw [arrow, draw=\shadowColor] (split_input_shift_\x) |- ($(tst.west) + (0,-\sOff)$);
\draw [arrow,color=\shadowColor] (split_input_\x.center)
        -- ($(ssf.west) + (0,-\sOff)$);
\draw [color=\shadowColor] ($(noisy_input.east) + (-\sOff,-\sOff)$)
        -- (split_input_\x.center);
\ifnum\x=2
\node [split, fill=\shadowColor] at (split_input_\x.center) {};
\fi
\draw[color=\shadowColor] (split_input_\x.center) -- (split_input_\x.center);
}

\draw [arrow] (tst) -| ($(ssf.north) + (0,0)$);
\draw[arrow] ($(cond.east) + (\condArrowOff, 0)$) -| (tst);
\draw[arrow] (ssf) -- (ssf_output);

\node[xshift=4*\shadowOff + 0.5 * \specWidth, draw=tab_orange, line width=1pt, inner sep=0.5pt,       
  outer sep=2pt] at ($(ssf.north |- tst.east) + (0,0)$) (weak_doa) {
    \pgfimage[width=\specWidth, height=\specHeight]{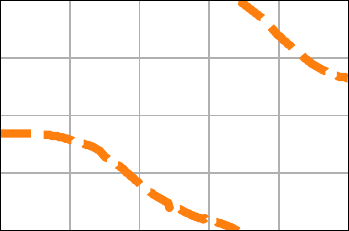} 
};
\node[anchor=west, xshift=0] at (tst.north east) {\flowMathFont$\hat{\theta}_t$
};
\node[anchor=south] at (cond -| tst) {\flowMathFont$\theta_0$
};


\end{tikzpicture}
    \vspace\flowVspace
  \end{minipage}
  }
 \hfill
\subfloat[Modify TST for AR use of processed speech.\label{fig:weak_ssf_miso}]{
  \hspace*{\flowHspace}\begin{minipage}{0.34\linewidth}
    \centering

\newcommand\boxEndX{5.1cm}
\begin{tikzpicture}[node distance=1.5*\nodeDistance and 1*\nodeDistance, background rectangle/.style={fill=none}] 
\path[use as bounding box, draw, dashed, bboxcolor] (\boxStartX,\boxStartY) rectangle (\boxEndX,\boxEndY);

\pgfmathsetlengthmacro{\condShift}{
    \pfHeight + 0.5 * \nodeDistance
}
\pgfmathsetlengthmacro{\ssfOff}{
    \dnnWidth - 1.25 * \nodeDistance
}
\pgfmathsetlengthmacro{\ssfShift}{
    \pfWidth - \dnnWidth
}

\node [] (noisy_input) {};
\foreach \x in {0, ..., 2}{
 \pgfmathtruncatemacro{\colorPerc}{round(\x*30 + 15)}
\pgfmathsetlengthmacro{\sOff}{(\x - 1) * \shadowOff}
\node [xshift=-\sOff - \specWidth / 2, yshift=-\sOff] at (noisy_input.center) {
  \pgfimage[width=\specWidth, height=\specHeight]{images/flowchart/stft_noisy.pdf}
};
}
\node [anchor=east, xshift=-\specWidth] at (noisy_input) {\flowMathFont$\mathbf{Y}_t$};

\node [right=of noisy_input] (split_input) {};
\node[yshift=\delayOff] at (split_input) (split_input_shift) {};
\node[right=of split_input_shift, yshift=\delayOff, minimum width=\pfWidth, minimum height=\pfHeight, align=center,
      inner sep=0pt,
      outer sep=0pt] {
\pgfimage[width=\pfWidth, height=\pfHeight]
{images/flowchart/particle_posterior.pdf}
};
\node[dnnBox, right=of split_input_shift, minimum width=\pfWidth, minimum height=\pfHeight, align=center, yshift=\delayOff,
      inner sep=0pt,
      outer sep=0pt] (tst) {};
\node at (tst.center) {\flowBigFont\contour{white}{TST}};
\node[right=of tst] (tst_knick) {};

\node [above=of noisy_input, xshift=\condShiftX, yshift=\condShiftY] (cond) {
    \input{images/flowchart/weak_init_doa.tikz}
};

\node [dnnBox] at (tst_knick |- split_input) (ssf) {\dnnSymb{}};
\node [split, right=of ssf, xshift=5mm] (ar_split) {};
\node[right=of ar_split, xshift=\ssfOutputShift] (ssf_output) {};
\node [xshift=\specWidth, anchor=west] at (ssf_output) {\flowMathFont$\hat{S}_t$};
\node [xshift=\specWidth / 2 + 2 * \shadowOff, draw=tab_orange, line width=1pt, inner sep=0.5pt] at (ssf_output.west) {
  \pgfimage[width=\specWidth, height=\specHeight]{images/flowchart/stft_enhanced.pdf}
};
\node at (ssf.center) {\flowBigFont\contour{white}{SSF}};

\node at (ar_split |- cond) (ar_knick1) {};
\node at ($(tst |- ar_knick1) + (\doubleInputOff, 0)$) (ar_knick2) {};
\draw[arrow] (ar_split) -- (ar_knick1.center) -- (ar_knick2.center) -- ($(tst.north) + (\doubleInputOff, 0)$);
\node[delay, yshift=\delayOff] at (ar_split) (ar_delay) {\delaySymb};

\foreach \x in {0, ..., 2}{
\pgfmathsetlengthmacro{\sOff}{(\x - 1) * \shadowOff}
\ifthenelse{\x = 2}{
    \def\shadowColor{black}
}{
    \pgfmathtruncatemacro{\colorPerc}{round(\x*10 + 20)}
    \def\shadowColor{black!\colorPerc!white}
}

\node [yshift=-\sOff, xshift=-\sOff] at (split_input.center) (split_input_\x) {};
\node [yshift=-\sOff, xshift=-\sOff] at (split_input.center) (split_input_\x) {};
\node [split, fill=\shadowColor] at (split_input_\x) (split_input_shift_\x) {};
\draw [arrow, draw=\shadowColor] (split_input_shift_\x) |- ($(tst.west) + (0,-\sOff)$);
\draw [arrow,color=\shadowColor] (split_input_\x.center)
        -- ($(ssf.west) + (0,-\sOff)$);
\draw [color=\shadowColor] ($(noisy_input.east) + (-\sOff,-\sOff)$)
        -- (split_input_\x.center);
\ifnum\x=2
\node [split, fill=\shadowColor] at (split_input_\x.center) {};
\fi
\draw[color=\shadowColor] (split_input_\x.center) -- (split_input_\x.center);
\node[delay, draw=\shadowColor] at ($(split_input_shift) + (-\sOff,-\sOff)$) {\delaySymb};
}

\draw [arrow] (tst) -| ($(ssf.north) + (0,0)$);
\draw[arrow] ($(cond.east) + (\condArrowOff, 0)$) -| ($(tst.north) + (-\doubleInputOff, 0)$);
\draw[arrow] (ssf) -- (ssf_output);

\node[xshift=4*\shadowOff + 0.5 * \specWidth, draw=tab_orange, line width=1pt, inner sep=0.5pt,       
  outer sep=2pt] at ($(ssf.north |- tst.east) + (0,-2.75mm)$) (weak_doa) {
    \pgfimage[width=\specWidth, height=\specHeight]{images/flowchart/particle_tracking.pdf} 
};
\node[anchor=west, xshift=0] at (tst.north east) {\flowMathFont$\hat{\theta}_t$
};
\node[anchor=south] at ($(cond -| tst) + (-\doubleInputOff-1mm, 0)$){\flowMathFont$\theta_0$};
\node[anchor=south] at ($(cond -| tst) + (\doubleInputOff + 2.5mm, 0)$){\flowMathFont$\hat{S}_{\delayIdx{t}}$};

\node [anchor=east, xshift=0.5mm, yshift=-1mm] at (split_input_shift |- tst) {\flowMathFont$\mathbf{Y}_{\hspace{-0.25mm}{\delayIdx{t}}}$};









\end{tikzpicture}
    \vspace\flowVspace
  \end{minipage}
  }
 \hfill
\subfloat[Multichannel (MIMO) SSF for MIMO-AR.\label{fig:weak_ssf_mimo}]{
  \hspace*{\flowHspace}\begin{minipage}{0.31\linewidth}
    \centering

\newcommand\boxEndX{4.2cm}
\begin{tikzpicture}[node distance=1.5*\nodeDistance and 1*\nodeDistance, background rectangle/.style={fill=none}] 
\path[use as bounding box, draw, dashed, bboxcolor] (\boxStartX,\boxStartY) rectangle (\boxEndX,\boxEndY);

\pgfmathsetlengthmacro{\condShift}{
    \pfHeight + 0.5 * \nodeDistance
}
\pgfmathsetlengthmacro{\ssfOff}{
    \dnnWidth - 1.25 * \nodeDistance
}
\pgfmathsetlengthmacro{\ssfShift}{
    \pfWidth - \dnnWidth
}

\node [] (noisy_input) {};
\foreach \x in {0, ..., 2}{
 \pgfmathtruncatemacro{\colorPerc}{round(\x*30 + 15)}
\pgfmathsetlengthmacro{\sOff}{(\x - 1) * \shadowOff}
\node [xshift=-\sOff - \specWidth / 2, yshift=-\sOff] at (noisy_input.center) {
  \pgfimage[width=\specWidth, height=\specHeight]{images/flowchart/stft_noisy.pdf}
};
}
\node [anchor=east, xshift=-\specWidth] at (noisy_input) {\flowMathFont$\mathbf{Y}_t$};

\node [dnnBox, right=of noisy_input, xshift=\ssfOutputShift] (ssf) {\dnnSymb{}};
\node at (ssf.center) {\flowBigFont\contour{white}{SSF}};

\node[right=of ssf.center, yshift=2*\delayOff, xshift=1mm, minimum width=\pfWidth, minimum height=\pfHeight, align=center,
      inner sep=0pt,
      outer sep=0pt] {
\pgfimage[width=\pfWidth, height=\pfHeight]
{images/flowchart/particle_posterior.pdf}
};
\node[dnnBox, right=of ssf.center, yshift=2*\delayOff, xshift=1mm, minimum width=\pfWidth, minimum height=\pfHeight, align=center,
      inner sep=0pt,
      outer sep=0pt] (tst) {};
\node at (tst.center) {\flowBigFont\contour{white}{TST}};
\node[right=of tst] (tst_knick) {};
\node (split_input) at (tst_knick |- ssf) {};
\node[yshift=\delayOff] at (split_input) (split_input_shift) {};

\node [above=of noisy_input, xshift=\condShiftX, yshift=\condShiftY] (cond) {
    \input{images/flowchart/weak_init_doa.tikz}
};

\node[xshift=\ssfOutputShift, yshift=-\shadowOff, right=of split_input] (ssf_output) {};
\node [xshift=\specWidth, anchor=west] at (ssf_output) {\flowMathFont$\hat{S}_t$};
\node [xshift=\specWidth / 2 + 2 * \shadowOff, draw=tab_orange, line width=1pt, inner sep=0.5pt] at (ssf_output.west) {
  \pgfimage[width=\specWidth, height=\specHeight]{images/flowchart/stft_enhanced.pdf}
};

\foreach \x in {0, ..., 2}{
\pgfmathsetlengthmacro{\sOff}{(\x - 1) * \shadowOff}
\ifthenelse{\x = 2}{
    \def\shadowColor{black}
}{
    \pgfmathtruncatemacro{\colorPerc}{round(\x*10 + 20)}
    \def\shadowColor{black!\colorPerc!white}
}
\node [yshift=-\sOff, xshift=-\sOff] at (split_input.center) (split_input_\x) {};
\node at (split_input_\x) (split_input_shift_\x) {};
\draw [arrow, draw=\shadowColor] (split_input_shift_\x.center) |- ($(tst.east) + (0,-\sOff)$);
\draw [draw=\shadowColor] ($(ssf.east) + (0,-\sOff)$) -- (split_input_shift_\x.center);
\draw [arrow, color=\shadowColor] ($(noisy_input.east) + (-\sOff,-\sOff)$)
        -- ($(ssf.west) + (0,-\sOff)$);
\ifnum\x=2
\node [split, fill=\shadowColor] at (split_input_\x.center) {};
\fi
\draw[color=\shadowColor] (split_input_\x.center) -- (split_input_\x.center);
\node[delay, draw=\shadowColor] at ($(split_input_shift) + (-\sOff,-\sOff)$) {\delaySymb};
}

\draw [arrow] (tst) -| ($(ssf.north) + (0,0)$);
\draw[arrow] ($(cond.east) + (\condArrowOff, 0)$) -| (tst);
\draw[arrow] ($(ssf.east) + (0, -\shadowOff)$) -- (ssf_output);

\node[xshift=-4*\shadowOff - 0.5 * \specWidth, draw=tab_orange, line width=1pt, inner sep=0.5pt,       
  outer sep=2pt] at ($(ssf.north |- tst.west) + (0,-2.75mm)$) (weak_doa) {
    \pgfimage[width=\specWidth, height=\specHeight]{images/flowchart/particle_tracking.pdf} 
};
\node [xshift=1mm, anchor=south] at (tst -| split_input) {\flowMathFont$\hat{\mathbf{S}}_{\delayIdx{t}}$};
\node[anchor=east, xshift=0] at (tst.north west) {\flowMathFont$\hat{\theta}_t$
};
\node[anchor=south] at (cond -| tst) {\flowMathFont$\theta_0$
};



\end{tikzpicture}
    \vspace\flowVspace
  \end{minipage}
  }
\caption{Weakly guided speaker extraction using \acl{tst} (\acs{tst}) to estimate the target’s direction $\theta_t$ from starting direction $\theta_0$ and guide a \acl{ssf} (\acs{ssf}) for enhancement. We propose an autoregressive (AR) integration of the processed speech for improved guidance in (b) and (c).}
\label{fig:flowchart}
\end{figure*}
\section{Steering Spatial Filters}\label{sec:ssf}
\subsection{Strongly Guided Target Speaker Extraction}\label{sec:strong_tse}
\newdiffComm{Spatially selective filters (\acp{ssf}) exploit positional information to extract a sound source originating from a designated direction. 
In this work, we follow the common convention of using only the target speaker’s azimuth \ac{doa} $\theta_t$
 for guidance \cite{tesch24ssf_journal, bohlender24sep_journal}. 
 When the \ac{doa} is known throughout the entire recording, the \ac{ssf} can be directly employed for \acfp{tse} by continuously steering it toward the target speaker,}{
 \Acfp{ssf} extract a sound source from a designated direction, here guided solely
by the target speaker's azimuth \ac{doa} $\theta_t$
\cite{tesch24ssf_journal, bohlender24sep_journal}. If $\theta_t$ is known
throughout the recording, the \ac{ssf} can be steered continuously toward
the target for \ac{tse},} a scenario we refer to as \textit{strong} guidance.
 
In a frame-wise causal \ac{stft}-domain processing pipeline, the \ac{ssf} has access to the current and all previous broadband multichannel observations $\mathbf{Y}_{1:t}$ together with the \ac{doa}s $\theta_{1:t}$ for computing the speech estimate $\hat{S}_{tk}$.
However, for online inference, re-evaluating all prior input values for each new frame~$t$ becomes computationally intractable. 
Instead, temporal context can be embedded into a hidden state $\mathbf{z}_{\delayIdx{t}}$,
yielding a sequential processing style, which, conditioned on $\mathbf{z}_{\delayIdx{t}}$, solely depends on current multichannel observation $\mathbf{Y}_t$ and \ac{doa} $\theta_{t}$,
\begin{equation}\label{eq:miso_ssf_sequential}
    \hat{S}_{tk} = \mathcal{F}_k\!\left( \mathbf{Y}_{t}, \theta_{t} \condSymb \mathbf{z}_{\delayIdx{t}}\right) \, ,
\end{equation}
with $\mathcal{F}_k$ denoting the \ac{ssf}. Updating the hidden state $\mathbf{z}_{t}$ frame-by-frame yields a computationally efficient formulation suitable for real-time speech enhancement 
\cite{tan18cruse}.

\subsection{Bayesian Tracking for Weakly Guided Speaker Extraction}\label{sec:weak_tse}
The dependency of strongly guided \ac{tse} on continuous ground-truth directional cues greatly limits practical applicability. 
\textit{Weakly} guided \ac{tse} \cite{kienegger25wg_ssf} relaxes this constraint and solely relies on the target speaker’s initial \ac{doa} $\theta_0$.
\newdiffComm{To continue
using a SSF for enhancement,}{Starting with $\theta_0$,} a \ac{tst} algorithm must be incorporated to replace the continuous oracle guidance with \ac{doa} estimates $\hat{\theta}_t$.
When tracking solely relies on the noisy observations $\mathbf{Y}_{1:t}$, the \ac{tst} and \ac{ssf} algorithms can be directly concatenated, as shown in \cref{fig:weak_ssf}.
In this work, we focus on recursive Bayesian filters for \ac{tst}, which model the posterior $p(\theta_t \condSymb \mathbf{Y}_{1:t}, \theta_0)$, referred to as \textit{filtering distribution}~\cite{sarkka13bayseian_filering}.
The \ac{doa} is inferred via a central tendency measure of the filtering distribution, e.g., the mean, yielding the \ac{mmse} estimate
\begin{equation}\label{eq:posterior_estimate}
    \hat{\theta}_t = \mathbb{E}\{ \theta_t \condSymb \mathbf{Y}_{1:t}, \theta_0 \} \, .
\end{equation}
Recursive Bayesian filters rely on a generative state-space model which specifies how the state $\theta_t$ evolves over time and generates the observations $\mathbf{Y}_{1:t}$.
Assuming Markov properties [\citen{sarkka13bayseian_filering}, Sec.~4.1], the \emph{state-transition} is fully specified by $p(\theta_t \condSymb \theta_{\delayIdx{t}})$, 
and 
the observation $\mathbf{Y}_t$ is conditionally independent of all past states and observations given the current state $\theta_t$.
This allows to recursively update the filtering distribution 
\begin{equation}\label{eq:kalman_recursion}
p(\theta_t \condSymb \mathbf{Y}_{1:t}, \theta_0)
\propto p(\mathbf{Y}_{t} \condSymb \theta_t) \, p(\theta_t \condSymb \mathbf{Y}_{1:\delayIdx{t}}, \theta_0) \, ,
\end{equation}
via \emph{likelihood} $p(\mathbf{Y}_{t} \condSymb \theta_t)$, and the \emph{predictive distribution (prior)} written as a function of the state transition $p(\theta_t \condSymb \theta_{\delayIdx{t}})$
\begin{equation}\label{eq:kalman_predict}
\hspace{-1.8mm}p(\theta_t \condSymb \mathbf{Y}_{1:\delayIdx{t}}, \theta_0) = \hspace{-1mm}\int_{\theta_{\delayIdx{t}}} \hspace{-4mm}
p(\theta_t \condSymb \theta_{\delayIdx{t}})
\, p(\theta_{\delayIdx{t}} \condSymb \mathbf{Y}_{1:{\delayIdx{t}}}, \theta_0)
\, \mathrm{d}\theta_{\delayIdx{t}} \, .\hspace{-1mm}
\end{equation}

\textbf{\textit{Kalman Filter}} 
The \ac{kf} [\citen{sarkka13bayseian_filering}, Sec.~4.3], \cite{kalman60original_formulation} is a recursive Bayesian filter defined for a linear-Gaussian state-space model.
Given this condition, both the filtering and predictive distributions in \cref{eq:kalman_recursion,eq:kalman_predict} remain Gaussian, yielding a tractable recursion while providing the optimal \ac{mmse} estimate via \cref{eq:posterior_estimate}.
In tracking applications, the state-transition model is often extended by first or higher-order derivatives to enforce smooth trajectories~\cite{rong03survey_target_tracking}.
In this work, we adopt a white-noise acceleration model~\cite{zhong12pf_avs_2d_ssl,dong20pf_doa_coprime}, which assumes linear dynamics for \ac{doa} $\theta_t$ and azimuth velocity $\dot{\theta}_t$
\begin{equation}\label{eq:cv_motion_model}
    \begin{bmatrix}
    \theta_{t} \\
    \dot{\theta}_{t}
    \end{bmatrix}\!=\!\begin{bmatrix}
        1 & \Delta T \\
        0 & 1
    \end{bmatrix}\!\begin{bmatrix}
    \theta_{\delayIdx{t}} \\
    \dot{\theta}_{\delayIdx{t}}
    \end{bmatrix}\!+\!\begin{bmatrix}
        \Delta T^2/ 2 \\
        \Delta T
    \end{bmatrix} \perturb_t \, , \ \perturb_t \sim \mathcal{N}\!\left(0, \sigma_\perturb^2\right) \, .
\end{equation}
With one-dimensional process noise $\nu_t$, the joint state transition $p\big(\theta_t, 
\dot{\theta}_t \condSymb \theta_{\delayIdx{t}}, \dot{\theta}_{\delayIdx{t}} \big)$ is Gaussian but degenerate (rank-deficient).
Eliminating $\nu_t$ yields
$\theta_t\!=\!\theta_{\delayIdx{t}}\hspace{0.1mm}+\,\Delta T / 2\,  \big(\dot{\theta}_{t}\hspace{0.3mm}+\hspace{0.3mm}\dot{\theta}_{\delayIdx{t}}\big)$, thus, a deterministic link between $\theta_t$ and $\dot{\theta}_t$, reducing the two-dimensional model to only one effective degree of freedom.

While in commonly used signal models, e.g.  \cref{eq:farfield_signal_model}, the \ac{doa} $\theta_t$ enters non-linearly via the steering vector $\mathbf{d}_{tk}$, the \ac{kf} requires a linear relationship between \ac{stft} coefficients $\mathbf{Y}_{t}$ and $\theta_t$. 
To fulfill this property, Traa et al. \cite{traa13wrapped_kalman_filter},~[\citen{traa13multichannel_separation_tracking}, Sec.~4.1.2] utilize the \ac{doa} estimate $\Phi_{t}(\mathbf{Y}_{t})$ and implicitly assume it is a sufficient statistic of $\mathbf{Y}_t$ regarding $\theta_t$.
In particular, $\Phi_{t}$ is the aggregation of narrow-band \ac{doa} estimates $\phi_{tk}(\mathbf{Y}_{tk})$, which minimize the linear-phase \ac{ls} error between corresponding direct-path and noisy \acp{ipd} across all microphone pairs, see, e.g., \cite{thiergart16ls_doa_estimator}.
Given equal inter-channel spacings, such as in a uniform circular array of three microphones, the sufficient statistic $\Phi_{t}$ 
can be expressed as
\begin{equation}\label{eq:doa_estimator}
    \Phi_t = \arg\!\left( \sum_{k=1}^{K/2} g_k \,e^{\mathrm{j}\phi_{tk}} \right)\,, \  g_k = \begin{cases}
        1, & k \leq K_\mathrm{A} \\
        0, & \mathrm{else}
    \end{cases} \, ,
\end{equation}
 with $ p(\mathbf{Y}_t \condSymb \theta_t) \propto p(\Phi_t \condSymb \theta_t)$ regarding $\theta_t$. The weights $g_k$ exclude frequency bins above $K_\mathrm{A}$ suffering from spatial aliasing \cite{thiergart16ls_doa_estimator}. 
\diffComm{}{For non-uniform inter-channel spacings, 
spatial aliasing must be considered separately for each microphone pair at each frequency bin.
This is achieved by replacing the ordinary-\ac{ls} narrowband \ac{doa} estimates $\phi_{tk}$ with weighted-\ac{ls} counterparts, which exclude aliasing-affected microphone pairs.}
 However, \diffComm{due to the inherent circularity, this}{even when assuming a linear-angular relationship between statistic $\Phi_{t}$ and \ac{doa} $\theta_t$,
 the inherent circularity} results in a wrapped Gaussian state-space
[\citen{mardia00directional_statistics}, Sec.~3.5.7]
\begin{equation}\label{eq:wrappedkf_measurement_model}
     \Phi_t\condSymb\theta_t \sim \mathcal{WN}\!\left(\theta_t, \sigma_\Phi^2 \right) \, .
\end{equation}
To maintain tractability, Traa et al. use mode‑matching to project the wrapped Gaussian back to an ordinary Gaussian.

\textbf{\textit{Particle Filter}}
Instead of relying on linear-Gaussian assumptions, the \ac{pf} \cite{gordon93pf_bootstrap_filter, arulampalam02pf_tutorial} approximates the filtering distribution using a weighted set of samples, known as particles. 
Specifically, the \ac{pf} models $p(\theta_{1:t} \condSymb \mathbf{Y}_{1:t}, \theta_0)$, i.e., the joint filtering distribution of \ac{doa} sequence $\theta_{1:t}$, factorizing as
\begin{equation}\label{eq:pf_joint_posterior}
    p(\theta_{1:t} \condSymb \mathbf{Y}_{1:t}, \theta_0) \propto p(\mathbf{Y}_{1:t} \condSymb \theta_{0:t}) \, p(\theta_{1:t} \condSymb \theta_0) \, .
\end{equation}
The \textit{bootstrap filter} [\citen{sarkka13bayseian_filering}, Alg.~7.5],~\cite{gordon93pf_bootstrap_filter} is a variant of the \ac{pf} representing the joint filtering distribution using Monte Carlo samples $\theta_{1:t}^\particleSuperscript{\idxParticles}$ [\citen{sarkka13bayseian_filering}, Sec.~2.5] from the joint prior $p(\theta_{1:t} \condSymb \theta_0)$ 
\begin{equation}\label{eq:pf_joint_posterior_approx}
    p(\theta_{1:t} \condSymb \mathbf{Y}_{1:t}, \theta_0) \approx \sum_{\idxParticles=1}^\numParticles w_{t}^\particleSuperscript{\idxParticles}  \delta\!\left(\theta_{1:t}-\theta_{1:t}^\particleSuperscript{\idxParticles}\right)\, ,
\end{equation}
with $\delta(\cdot)$ denoting the Dirac delta function and normalized weights $w_{t}^\particleSuperscript{\idxParticles}$.
The filtering distribution $p(\theta_{t} \condSymb \mathbf{Y}_{1:t}, \theta_0)$ is trivially obtained 
via marginalization.
Under Markov assumptions, the joint prior in \cref{eq:pf_joint_posterior} factorizes, enabling sequential sampling of $\theta_{t}^\particleSuperscript{\idxParticles}$ from $p(\theta_{t} \condSymb \theta_{\delayIdx{t}}^\particleSuperscript{\idxParticles})$ and recursive weight updates via
\begin{equation}\label{eq:pf_recursive_weights}
w_t^\particleSuperscript{\idxParticles} \propto \prod_{t'=1}^t p(\mathbf{Y}_{t'} \condSymb \theta_{t'}^\particleSuperscript{\idxParticles}) \propto p(\mathbf{Y}_{t} \condSymb \theta_{t}^\particleSuperscript{\idxParticles}) \, w_{\delayIdx{t}}^\particleSuperscript{\idxParticles} \, .
\end{equation}
This multiplicative recursion can lead to weight degeneracy, resulting in near-zero weights for the majority of particles. 
Using the effective number of particles $N^{\scriptscriptstyle{(\mathrm{eff})}}_t$ as an indicator
\begin{equation}\label{eq:pf_eff_sample_size}
    N^{\scriptscriptstyle{(\mathrm{eff})}}_t=1 \bigg{/} \sum_{n=1}^N \left(w_t^\particleSuperscript{n}\right)^2 \, ,
\end{equation}
adaptive resampling schemes [\citen{sarkka13bayseian_filering}, Sec.~7.4],~\cite{arulampalam02pf_tutorial} counter weight degeneracy by resampling particles according to weights  $w_t^\particleSuperscript{n}$ if $N^{\scriptscriptstyle{(\mathrm{eff})}}_t$ falls below a threshold.
Since $\theta_t$ is embedded in the multidimensional Gaussian state-space in \cref{eq:cv_motion_model}, simulating $p(\theta_t \condSymb \theta_{{\delayIdx{t}}})$ is in general achieved via Rao-Blackwellization [\citen{sarkka13bayseian_filering}, Sec.~7.5]. 
However, in this case the degeneracy of the state-space allows for sampling particles $\theta_t^\particleSuperscript{\idxParticles}$ directly from $p(\theta_t \condSymb \theta_{{\delayIdx{t}}}^\particleSuperscript{\idxParticles}, \dot{\theta}_{{\delayIdx{t}}}^\particleSuperscript{\idxParticles})$
using the recursive relationship
\begin{equation}\label{eq:velocity_recursion}
    \dot{\theta}^\particleSuperscript{\idxParticles}_{t} \condSymb \theta^\particleSuperscript{\idxParticles}_{1:t},\theta_0 = 2 / \Delta T \,  %
    (\theta^\particleSuperscript{\idxParticles}_{t} - \theta^\particleSuperscript{\idxParticles}_{\delayIdx{t}}) - \dot{\theta}^\particleSuperscript{\idxParticles}_{\delayIdx{t}} \, ,
\end{equation}
where we set the initial angular velocity $\dot{\theta}_0$ to zero.
Given that the \ac{pf} does not require Gaussianity, we may now use a directional distribution to model the circular nature of the \ac{doa} estimation problem.
\diffComm{Under far-field conditions, amplitude differences between microphones are negligible and the \acp{ipd} of the normalized \ac{stft} coefficients $\mathbfcal{Y}_{tk} = \mathbf{Y}_{tk} / \lVert \mathbf{Y}_{tk} \rVert$ contain all spatial information.
Thus, we assume sufficiency for estimating \ac{doa} $\theta_t$ and model $\mathbfcal{Y}_{tk}$ via the complex Watson distribution [\citen{mardia00directional_statistics}, Sec.~14.7],\cite{drude15complex_watson_doa_estimation,wang18localization_watson_DNN}, which is invariant 
to global phase shifts, centered at}{Since inter-channel magnitude and phase differences between the noisy  \ac{stft} coefficients bear all directional information, we assume sufficiency of the normalized statistics $\mathbfcal{Y}_{tk} = \mathbf{Y}_{tk} / \lVert \mathbf{Y}_{tk} \rVert$ for estimating \ac{doa} $\theta_t$ and model $\mathbfcal{Y}_{tk}$ via the complex Watson distribution [\citen{mardia00directional_statistics}, Sec.~14.7],\cite{drude15complex_watson_doa_estimation,wang18localization_watson_DNN}, which is invariant 
to global phase shifts.
For a compact array under far-field conditions, amplitude differences between microphones become negligible and we model $\mathbfcal{Y}_{tk}$ centered at plane-wave} steering vector $\mathbf{d}_{tk}/\sqrt{M}$ with concentration $\kappa$.
For independent \ac{stft} bins, this gives
\begin{equation}\label{eq:watson_likelihood}
    \mathbfcal{Y}_{tk} \condSymb \theta_t \sim \mathcal{CW}(\mathbf{d}_{tk}/\sqrt{M}, \kappa)
\end{equation}
with $p(\mathbf{Y}_t \condSymb \theta_t) \propto p(\mathbfcal{Y}_{t} \condSymb \theta_t)$ regarding $\theta_t$, which, together with the state-transition model in \cref{eq:cv_motion_model}, defines the \ac{pf}'s recursive 
approximation of the filtering distribution $p(\theta_{t} \condSymb \mathbf{Y}_{1:t}, \theta_0)$. 
\diffComm{}{For array configurations with non-negligible magnitude differences, the Watson likelihood in \cref{eq:watson_likelihood} can be adapted by replacing the plane-wave steering vector with a normalized inter-microphone transfer function, such as a head-related transfer function for wearable hearing devices.}%
\newdiffComm{}{ However, deep \acp{ssf} may still be array-dependent, while recent work improves robustness to microphone count~\cite{luo20tac_transform_average_concatenate} and placement~\cite{mannanova24var_array_meta_learning}.
}

\section{Autoregressively Guided Spatial Filters}\label{sec:self_tse}
\newdiffComm{By including an independent upstream \acl{tst} (\acs{tst}) algorithm, the concatenative speaker extraction (\acs{tse}) pipeline shown in \cref{fig:weak_ssf} may appear like the natural approach}{Concatenating an independent upstream tracking \acl{tst} (\acs{tst}) algorithm (\cref{fig:weak_ssf}) may appear as the natural way} to automate the steering of a \acl{ssf} (\acs{ssf}).
However, in a frame-wise causal and sequential processing framework, the previously enhanced speech signal $\hat{S}_{\delayIdx{t}}$ is available at frame $t$ and can be incorporated to improve extraction.
In the resulting \ac{ar} pipeline, $\hat{S}_{\delayIdx{t}}$ can either serve as \diffComm{}{an} auxiliary guide for enhancement \cite{andreev23iterative_autoregression, pan24paris_autoregressive_separation, shen25arise}, to improve tracking performance \cite{kienegger25sg_ssf}, or both~\cite{kienegger25deep_joint_ar_ssf}. 
\diffComm{
Extending our conference paper~\cite{kienegger25sg_ssf},
in this work, we aim 
to increase the tracking accuracy of Kalman and particle filters, while retaining minimal computational overhead. 
The processed speech signal is either additionally incorporated into the Bayesian filtering formulations as shown in \cref{fig:weak_ssf_miso} or directly used to replace the noisy observation $\mathbf{Y}_t$, see \cref{fig:weak_ssf_mimo}.}{In this work, we aim to increase the accuracy of Bayesian tracking algorithms, while retaining minimal computational overhead. 
Building on our conference paper~\cite{kienegger25sg_ssf}, we modify the \ac{ssf} to estimate the target's multichannel direct-path speech signal $\hat{\mathbf{S}}_{t}$ instead of $\hat{S}_{t}$ at a single reference microphone.
By retaining the spatial cues of the target speaker, $\hat{\mathbf{S}}_{t}$ can replace the unprocessed recording $\mathbf{Y}_t$ as tracking input, see \cref{fig:weak_ssf_mimo}.
While we previously utilized a particle filter~\cite{kienegger25sg_ssf}, in this work, we generalize our \ac{ar} framework to arbitrary Bayesian trackers and include a Kalman filter to further reduce computational cost.
Additionally, we develop novel Bayesian formulations for Kalman and particle filters which are capable of jointly utilizing processed and unprocessed speech signals, as shown in \cref{fig:weak_ssf_miso}.
By avoiding architectural modifications of the \ac{ssf}, this further increases the applicability of our \ac{ar} frameworks.}

\subsection{Extended Bayesian Filtering Formulations}\label{sec:adapt_tracking}
To include the enhanced speech from the \ac{ssf} in the presented Bayesian tracking algorithms, we extend the generative framework by introducing the clean speech signals $S_{1:{\delayIdx{t}}}$ as latent observations that complement $\mathbf{Y}_{1:t}$ for estimating the \ac{doa} $\theta_t$.
However, without a matched speech signal at frame $t$, noisy \ac{stft} coefficients $\mathbf{Y}_t$ are less informative for tracking, since the target's spectral characteristics cannot be exploited.
We therefore omit $\mathbf{Y}_{t}$ and solely rely on the predictive distribution $p(\theta_t \condSymb \mathbf{Y}_{1:\delayIdx{t}}, S_{1:\delayIdx{t}}, \theta_0)$ for \ac{doa} estimation, yielding 
\begin{equation}\label{eq:miso_posterior_prediction}
    \hat{\theta}_t = \mathbb{E}\{ \theta_{t} \condSymb \mathbf{Y}_{1:{\delayIdx{t}}}, S_{1:{\delayIdx{t}}}, \theta_0 \} \, .
\end{equation}
During inference, we use the enhanced \ac{stft} segments $\hat{S}_{1:{\delayIdx{t}}}$ as plug-in approximation for samples of clean speech $S_{1:{\delayIdx{t}}}$.
While this neglects processing degradation from the \ac{ssf}, we demonstrate that a high tracking accuracy can be achieved.

\textbf{\textit{Kalman Filter}} 
Assuming independence between the single channel clean speech signal $S_t$ and \ac{doa} $\theta_t$,
the predictive distribution can be further factorized, resulting in the recursion
\begin{equation}\label{eq:kalman_predictive_recursion}
\begin{aligned}
\hspace{-1mm}p(\theta_t \condSymb \mathbf{Y}_{1:\delayIdx{t}}, S_{1:\delayIdx{t}}, \theta_0)
\propto& \int_{\theta_{\delayIdx{t}}} \hspace{-3.5mm} p(\theta_t \condSymb \theta_{\delayIdx{t}}) \times \\
 p(\mathbf{Y}_{\delayIdx{t}} \condSymb S_{\delayIdx{t}}, \theta_{\delayIdx{t}})
\, p(&\theta_{\delayIdx{t}} \condSymb \mathbf{Y}_{1:{\delayIdx[2]{t}}}, S_{1:\delayIdx[2]{t}}, \theta_0)
\, \mathrm{d}\theta_{\delayIdx{t}} \, .
\end{aligned}
\end{equation}
To enforce linear-Gaussianity of the likelihood $p(\mathbf{Y}_{t} \condSymb S_{t}, \theta_{t})$ for tractability, we follow Traa et al. and employ the \ac{doa} estimator $\Phi_t$ in \cref{eq:doa_estimator}
as a sufficient statistic for $\theta_t$.
However, instead of uniformly aggregating the narrow-band \ac{doa} estimates $\phi_{tk}$ as done in \cref{eq:doa_estimator}, we propose incorporating $S_t$ to emphasize frequency bins dominated by the target speaker. This gives
\begin{equation}\label{eq:miso_doa_estimator}
    \hspace{-1.5mm}\Phi_t = \arg\!\left( \sum_{k=1}^{K/2} g_{tk} \,e^{\mathrm{j}\phi_{tk}} \right)\,, \  g_{tk} = \begin{cases}
        |S_{tk}|^2, & k \leq K_\mathrm{A} \\
        0, & \mathrm{else}
    \end{cases} \, ,
\end{equation}
with $p(\mathbf{Y}_{t} \condSymb S_{t}, \theta_{t})
\propto p(\Phi_t \condSymb S_{t}, \theta_{t})$ regarding $\theta_t$.
Nevertheless, as with the \ac{kf} from Traa et al. in \cref{sec:weak_tse}, the linear-phase \ac{doa} estimates $\phi_{tk}$ conceptually limit the efficiency of incorporating the enhanced speech, 
since only the bandwidth below the spatial aliasing frequency bin $K_\mathrm{A}$ \diffComm{can be utilized}{is unambiguous}.
Our subsequent \ac{pf} algorithm does not share this limitation.

\textbf{\textit{Particle Filter}} 
To obtain a bootstrap \ac{pf} formulation for computing the predictive mean in \cref{eq:miso_posterior_prediction}, we approximate the joint predictive distribution $p(\theta_{1:t} \condSymb \mathbf{Y}_{1:\delayIdx{t}}, S_{1:\delayIdx{t}}, \theta_0)$ via Monte Carlo sampling. 
Assuming that speech $S_t$ is independent of \ac{doa} $\theta_t$,  the joint predictive distribution factorizes as
\begin{equation}\label{eq:pf_joint_predictive}
\begin{aligned}
p(\theta_{1:t} \condSymb \mathbf{Y}_{1:\delayIdx{t}}, \, &S_{1:\delayIdx{t}}, \theta_0)
\propto \\
&p(\mathbf{Y}_{1:\delayIdx{t}} \condSymb S_{1:\delayIdx{t}}, \theta_{0:t})\,
p(\theta_{1:t} \condSymb \theta_0) \, .
\end{aligned}
\end{equation}
Under Markov properties, the Monte Carlo approximation for the filtering distribution after marginalization yields
\begin{equation}\label{eq:pf_miso_posterior_approx}
    p(\theta_{t} \condSymb \mathbf{Y}_{1:\delayIdx{t}}, S_{1:\delayIdx{t}}, \theta_0) \approx \sum_{\idxParticles=1}^\numParticles w_{\delayIdx{t}}^\particleSuperscript{\idxParticles}  \delta\!\left(\theta_{t}-\theta_{t}^\particleSuperscript{\idxParticles}\right)\, ,
\end{equation}
with recursively sampled particles 
$\theta_{t}^\particleSuperscript{\idxParticles}$ (\cref{sec:weak_tse}) and weights
\begin{equation}\label{eq:pf_miso_recursive_weights}
w_t^\particleSuperscript{\idxParticles} \propto p(\mathbf{Y}_{t} \condSymb S_{t}, \theta_{t}^\particleSuperscript{\idxParticles}) \, w_{\delayIdx{t}}^\particleSuperscript{\idxParticles}  \, .
\end{equation}
Since the \ac{pf} is not constrained to a linear relationship between observation $\mathbf{Y}_t$ and \ac{doa} $\theta_t$, we use the generative model in \cref{eq:farfield_signal_model}, which encodes $\theta_t$ via steering vector $\mathbf{d}_{tk}$.
\diffComm{Assuming noise \ac{stft} coefficients $\mathbf{V}_{tk}$ are uncorrelated across frequency \cite{hendriks09optimal_multichannel_MMSE} and follow a zero mean, proper complex Gaussian distribution [\citen{schreier10ssp_complex_valued_data}, Sec.~2.3.1] with covariance $\mathbf{R}_{tk}$, results in the likelihood}{In particular, we approximate the composite noise \ac{stft} coefficients $\mathbf{V}_{tk}$, which, among others, include reverberant target and interfering speech, with a zero mean, proper complex Gaussian distribution with covariance matrix $\mathbf{R}_{tk}$ [\citen{schreier10ssp_complex_valued_data}, Sec.~2.3.1], resulting in}
\begin{equation}\label{eq:pf_miso_likelihood}
\mathbf{Y}_{tk} \condSymb S_{tk}, \theta_{t} \sim \mathcal{CN}(\mathbf{d}_{tk}S_{tk}, \mathbf{R}_{tk}) \, .
\end{equation}
\diffComm{}{While the Gaussian distribution cannot capture multimodality or higher-order structure of the aggregated signals, it provides a practical \textit{trade-off} between model fidelity and parameterizability.
We further assume that the frequency-bin components of 
$\mathbf{V}_{tk}$ are mutually uncorrelated~\cite{hendriks09optimal_multichannel_MMSE},
while allowing for arbitrary spatial correlations between microphone channels in $\mathbf{R}_{tk}$ per frequency-bin.}
During inference, we \diffComm{use the noise estimate $\hat{\mathbf{V}}_{tk}$ defined as}{estimate $\mathbf{V}_{tk}$ via} 
\begin{equation}
    \hat{\mathbf{V}}_{tk} = \mathbf{Y}_{tk} - \mathbf{d}_{k}(\hat{\theta}_t) \hat{S}_{tk} \, 
\end{equation}
to recursively \diffComm{estimate the time-varying noise covariance matrix $\mathbf{R}_{t}$}{update a time-varying noise covariance matrix estimate $\hat{\mathbf{R}}_{t}$} using the \ac{ema} 
\begin{equation}\label{eq:pf_miso_ema_covariance}
\hat{\mathbf{R}}_{t} = (1 - \alpha^{\tinySuper{(\mathrm{EMA})}})\hat{\mathbf{V}}_t \hat{\mathbf{V}}^\mathrm{H}_t + \alpha^{\tinySuper{(\mathrm{EMA})}} \hat{\mathbf{R}}_{\delayIdx{t}}\, .
\end{equation}
\Cref{alg:bootstrap_filter} summarizes the proposed incorporation of speech estimates into the \ac{pf} using the generic bootstrap filter from Lehmann et al. [\citen{lehmann06pf_resampling_Neff}, Alg.~1] as \diffComm{}{a} foundational framework.

\begin{algorithm}[t!]
	\begin{algorithmic}[1]
		\caption{Proposed bootstrap \acl{pf} (\acs{pf}) for autoregressive (AR) incorporation with SSF estimates (MISO-AR).}\label{alg:bootstrap_filter}
        \Require \ac{doa} $\theta_0$, noise covariance $\hat{\mathbf{R}}_{0}$ and threshold $\tau^{\scriptscriptstyle(\mathrm{eff})}$
		\State Initialize $\{\theta_{0}^\particleSuperscript{\idxParticles}, \dot{\theta}_{0}^\particleSuperscript{\idxParticles}, w_{0}^\particleSuperscript{\idxParticles}\} \gets \{\theta_0, 0, 1/N\}$
		\For {$t=1, 2, 3, \dots$}
        \If {$t > 1$} 
        \State Obtain measurements $\mathbf{Y}_{\delayIdx{t}}$, $\hat{S}_{\delayIdx{t}}$
        \State Update noise covariance matrix $\hat{\mathbf{R}}_{\delayIdx{t}}$ using \cref{eq:pf_miso_ema_covariance}
        \State Update weights $\tilde{w}_{\delayIdx{t}}$ using \cref{eq:pf_miso_recursive_weights}
		\State Normalize weights $w_{\delayIdx{t}}^\particleSuperscript{\idxParticles} \gets \tilde{w}_{\delayIdx{t}}^\particleSuperscript{\idxParticles} / \sum_{n'} \tilde{w}_{\delayIdx{t}}^\particleSuperscript{{n'}}$
        \State Compute eff. sample size $N_{\delayIdx{t}}^{\scriptscriptstyle(\mathrm{eff})}$ using \cref{eq:pf_eff_sample_size}
		\If {$N_{\delayIdx{t}}^{\scriptscriptstyle(\mathrm{eff})} < \tau^{\scriptscriptstyle(\mathrm{eff})}N$} 
         \diffState{Resample $\idxParticles' \sim \texttt{Categorical}(w_{\delayIdx{t}}^\particleSuperscript{\idxParticles})$}{Sample $\idxParticles'$ according to $\mathrm{Pr}(n'=n) = w_{\delayIdx{t}}^\particleSuperscript{\idxParticles}$}
        \State Set $\{\theta_{\delayIdx{t}}^\particleSuperscript{\idxParticles}, \dot{\theta}_{\delayIdx{t}}^\particleSuperscript{\idxParticles}, w_{\delayIdx{t}}^\particleSuperscript{\idxParticles}\} \gets \{\theta_{\delayIdx{t}}^{\particleSuperscript{\idxParticles'}}, \dot{\theta}_{\delayIdx{t}}^{\particleSuperscript{\idxParticles'}}, 1/N\}$
		\EndIf 
		\EndIf
        \State Sample \ac{doa} particles $\theta_t^\particleSuperscript{\idxParticles}$ from $p(\theta_t  \condSymb\theta_{\delayIdx{t}}^\particleSuperscript{\idxParticles}, \dot{\theta}_{\delayIdx{t}}^\particleSuperscript{\idxParticles})$
        \State Update velocities $\dot{\theta}_t^\particleSuperscript{\idxParticles}$ using \cref{eq:velocity_recursion}
        \State Estimate \ac{doa} $\hat{\theta}_t \gets \arg\!\left( \sum_n w_{\delayIdx{t}}^\particleSuperscript{\idxParticles} e^{\mathrm{j} \theta_{t}^\particleSuperscript{\idxParticles}} \right)$
		\EndFor			
	\end{algorithmic}

\end{algorithm}

\subsection{Multiple-Input and Multiple-Output (MIMO) Spatial Filters}\label{sec:adapt_enhancement}
Instead of modifying the Bayesian filtering formulations to incorporate the enhanced speech signal as additional observation, it can also be used as a replacement for the noisy measurement $\mathbf{Y}_t$.
However, the original \ac{ssf} formulation in \cref{eq:miso_ssf_sequential}, which is \ac{miso} considering the channel dimension, yields a single-channel speech estimate $\hat{S}_t$ without spatial information and is therefore uninformative for tracking on its own.
Motivated by recent works reporting accurate localization in stationary scenarios \cite{li23gcc-speaker, chen24locselect, battula25MIMO_localization}, we propose to extend the \ac{miso} \ac{ssf} in \cref{eq:miso_ssf_sequential} to a \ac{mimo} formulation by estimating the target's full direct-path propagated speech signal 
$\hat{\mathbf{S}}_{tk}$~\cite{kienegger25sg_ssf}
\begin{equation}\label{eq:mimo_ssf_sequential}
    \hat{\mathbf{S}}_{tk} = \mathbfcal{F}_k\!\left( \mathbf{Y}_{t}, \theta_{t} \condSymb \mathbf{z}_{\delayIdx{t}}\right) \, .
\end{equation}
Given that the final layer of a deep \ac{ssf} is typically linear \cite{tesch24ssf_journal, bohlender24sep_journal}, the \ac{mimo} extension \newdiffComm{effects}{affects} the model complexity only marginally for reasonably sized arrays.
However, enforcing spatial cue preservation in the speech estimates introduces additional challenges for enhancement.
With the signal model in \cref{eq:signal_model} implying that direct-path speech $\mathbf{S}_t$ captures all information in $\mathbf{Y}_t$ for \ac{doa} $\theta_t$, we can drop the additional conditioning
\begin{equation}\label{eq:mimo_sufficient_statistics}
p(\theta_t \condSymb \mathbf{Y}_{1:\delayIdx{t}}, \mathbf{S}_{1:\delayIdx{t}}, \theta_0)
= p(\theta_t \condSymb \mathbf{S}_{1:\delayIdx{t}}, \theta_0) \, ,
\end{equation}
and use the predictive prior $p(\theta_t \condSymb \mathbf{S}_{1:\delayIdx{t}}, \theta_0)$ for estimation 
\begin{equation}\label{eq:mimo_posterior_mean}
\hat{\theta}_t = \mathbb{E}\{\theta_t \condSymb \mathbf{S}_{1:\delayIdx{t}}, \theta_0\} \, .
\end{equation}
By retaining the same generative model, $\mathbf{Y}_t$ can be directly substituted by $\mathbf{S}_t$ in the filtering formulations in \cref{sec:weak_tse}.
Similar to \cref{sec:adapt_tracking}, we use the enhanced signals $\hat{\mathbf{S}}_{1:{\delayIdx{t}}}$ as plug-in approximation for  $\mathbf{S}_{1:{\delayIdx{t}}}$ during inference.

\Cref{fig:weak_ssf_mimo} presents the resulting \ac{ar} \ac{tse} pipeline, which we denote as \textit{MIMO-AR}, opposed to \textit{MISO-AR} from \cref{sec:adapt_tracking} and \cref{fig:weak_ssf_miso}.
\diffComm{}{\Cref{fig:unfolded_flowcharts} complements the system overview in \cref{fig:flowchart} by illustrating the temporal processing order between tracking~(\ac{tst}) and enhancement~(\ac{ssf}).
Since the \ac{ssf} does not inherently preserve spatial information, our proposed \ac{miso}-AR Bayesian filters (\cref{fig:weak_ssf_miso}, \cref{fig:weak_ssf_miso_unfolded}), combine both enhanced speech and noisy recording for improved tracking.
On the other hand, by modifying the \ac{ssf} to retain the target speaker's spatial cues, the MIMO-AR configuration (\cref{fig:weak_ssf_mimo}, \cref{fig:weak_ssf_mimo_unfolded}) enables replacing unprocessed with processed speech signals in the original tracking formulations from \cref{sec:weak_tse}.}

\newcommand\flowVspaceUnfolded{-17.5pt}
\newcommand\flowHspaceUnfolded{-3.5pt}
\begin{figure}[t!]
\vspace*{-20pt}
\centering
\subfloat[Concat.\label{fig:weak_ssf_unfolded}]{
  \hspace*{-0pt}\begin{minipage}{0.29\linewidth}%
    \centering

\newcommand\unboundedBoxStartX{7.5mm}
\newcommand\unboundedBoxEndX{3.5cm}
\newcommand\unfoldedOff{1.2cm}
\begin{tikzpicture}[node distance=0.5*\nodeDistance and 1*\nodeDistance, background rectangle/.style={fill=none}] 
\path[use as bounding box, draw, dashed, bboxcolorUnfolded] (\unboundedBoxStartX,\unboundedBoxStartY) rectangle (\unboundedBoxEndX,\unboundedBoxEndY);

\pgfmathsetlengthmacro{\condShift}{
    \pfHeight + 0.5 * \nodeDistance
}
\pgfmathsetlengthmacro{\ssfOff}{
    \dnnWidth - 1.25 * \nodeDistance
}
\pgfmathsetlengthmacro{\ssfShift}{
    \pfWidth - \dnnWidth
}


\foreach \nRep in {1, ..., \numRep}{
\pgfmathsetlengthmacro{\xtstshift}{\nRep * \unfoldedOff}
 \node[draw,
    rectangle, xshift=\xtstshift, minimum width=\dnnBoxWidth, minimum height=\dnnBoxHeight, align=center,
      inner sep=0pt,
      outer sep=0pt] (tst_\nRep) {};
\node at (tst_\nRep.center) {\unfoldedBigFont TST};

\pgfmathsetlengthmacro{\xtstshift}{\nRep * \unfoldedOff}
 \node[draw,
    rectangle, above=of tst_\nRep, xshift=\unfoldedSSFshift, minimum width=\dnnBoxWidth, minimum height=\dnnBoxHeight, align=center,
      inner sep=0pt,
      outer sep=0pt] (ssf_\nRep) {};
\node at (ssf_\nRep.center) {\unfoldedBigFont SSF};

\node [unfoldedSplit, above=of ssf_\nRep, color=white] (split_ssf_\nRep) {};
\ifnum\nRep=1
    \node [below=of ssf_\nRep, yshift=\unfoldedNoisyOff] (noisy_input_\nRep) {\unfoldedFlowMathFont$\mathbf{Y}_{\delayIdx{t}}$};
\else
    \node [below=of ssf_\nRep, yshift=\unfoldedNoisyOff] (noisy_input_\nRep) {\unfoldedFlowMathFont$\mathbf{Y}_t$};
\fi 

\ifnum\nRep=1
    \node [left=of ssf_\nRep, xshift=\unfoldedSSFlabelShift]{\unfoldedFlowMathFont$\hat{\theta}_{\delayIdx{t}}$};
\else
    \node [left=of ssf_\nRep, xshift=\unfoldedSSFlabelShift] {\unfoldedFlowMathFont$\hat{\theta}_t\,$};
\fi 

\ifnum\nRep=1
    \node [above=of split_ssf_\nRep] (enhanced_\nRep) {\unfoldedFlowMathFont$\hat{S}_{\delayIdx{t}}$};
\else
    \node [above=of split_ssf_\nRep] (enhanced_\nRep) {\unfoldedFlowMathFont$\hat{S}_t$};
\fi 

\draw [arrow] (tst_\nRep.north) |- (ssf_\nRep.west);
\draw [arrow] (ssf_\nRep.north) -- (enhanced_\nRep);
\foreach \x in {0, ..., 2}{
 \pgfmathtruncatemacro{\colorPerc}{round(\x*30 + 15)}
\pgfmathsetlengthmacro{\sOff}{(\x - 1) * \shadowOff}
\ifthenelse{\x = 2}{
    \def\shadowColor{black}
}{
    \pgfmathtruncatemacro{\colorPerc}{round(\x*10 + 20)}
    \def\shadowColor{black!\colorPerc!white}
}
\draw [arrow,color=\shadowColor] ($(noisy_input_\nRep.north) + (-\sOff,-\sOff)$)
        -- ($(ssf_\nRep.south) + (-\sOff, 0)$);
\node[unfoldedSplit, above=of noisy_input_\nRep, xshift=-\sOff, yshift=-\sOff, color=\shadowColor] (noisy_input_split_\nRep_\x) {};
\draw[arrow, color=\shadowColor]
        (noisy_input_split_\nRep_\x.center)
        -| ($(tst_\nRep.south) + (-\sOff, 0)$);
}

}

\end{tikzpicture}
    \vspace\flowVspaceUnfolded
  \end{minipage}
  }
 \hfill
\subfloat[MISO-AR.\label{fig:weak_ssf_miso_unfolded}]{
  \hspace*{0pt}\begin{minipage}{0.29\linewidth}%
    \centering

\newcommand\unboundedBoxStartX{9mm}
\newcommand\unboundedBoxEndX{3.7cm}
\newcommand\unfoldedOff{1.3cm} 
\begin{tikzpicture}[node distance=0.5*\nodeDistance and 1*\nodeDistance, background rectangle/.style={fill=none}] 
\path[use as bounding box, draw, dashed, bboxcolorUnfolded] (\unboundedBoxStartX,\unboundedBoxStartY) rectangle (\unboundedBoxEndX,\unboundedBoxEndY);

\pgfmathsetlengthmacro{\condShift}{
    \pfHeight + 0.5 * \nodeDistance
}
\pgfmathsetlengthmacro{\ssfOff}{
    \dnnWidth - 1.25 * \nodeDistance
}
\pgfmathsetlengthmacro{\ssfShift}{
    \pfWidth - \dnnWidth
}


\foreach \nRep in {1, ..., \numRep}{
\pgfmathsetlengthmacro{\xtstshift}{\nRep * \unfoldedOff}
 \node[draw,
    rectangle, xshift=\xtstshift, minimum width=\dnnBoxWidth, minimum height=\dnnBoxHeight, align=center,
      inner sep=0pt,
      outer sep=0pt] (tst_\nRep) {};
\node at (tst_\nRep.center) {\unfoldedBigFont TST};

\pgfmathsetlengthmacro{\xtstshift}{\nRep * \unfoldedOff}
 \node[draw,
    rectangle, above=of tst_\nRep, xshift=\unfoldedSSFshift, minimum width=\dnnBoxWidth, minimum height=\dnnBoxHeight, align=center,
      inner sep=0pt,
      outer sep=0pt] (ssf_\nRep) {};
\node at (ssf_\nRep.center) {\unfoldedBigFont SSF};

\node [unfoldedSplit, above=of ssf_\nRep] (split_ssf_\nRep) {};
\ifnum\nRep=1
    \node [below=of ssf_\nRep, yshift=\unfoldedNoisyOff] (noisy_input_\nRep) {\unfoldedFlowMathFont$\mathbf{Y}_{\delayIdx{t}}$};
\else
    \node [below=of ssf_\nRep, yshift=\unfoldedNoisyOff] (noisy_input_\nRep) {\unfoldedFlowMathFont$\mathbf{Y}_t$};
\fi 


\node [unfoldedSplit, above=of ssf_\nRep] (split_ssf_\nRep) {};
\ifnum\nRep=1
    \node [left=of ssf_\nRep, xshift=\unfoldedSSFlabelShift]{\unfoldedFlowMathFont$\hat{\theta}_{\delayIdx{t}}$};
\else
    \node [left=of ssf_\nRep, xshift=\unfoldedSSFlabelShift] {\unfoldedFlowMathFont$\hat{\theta}_t$\,};
\fi 

\node [unfoldedSplit, above=of ssf_\nRep] (split_ssf_\nRep) {};
\ifnum\nRep=1
    \node [above=of split_ssf_\nRep] (enhanced_\nRep) {\unfoldedFlowMathFont$\hat{S}_{\delayIdx{t}}$};
\else
    \node [above=of split_ssf_\nRep] (enhanced_\nRep) {\unfoldedFlowMathFont$\hat{S}_t$};
\fi 

\draw [arrow] (tst_\nRep.north) |- (ssf_\nRep.west);
\draw [arrow] (ssf_\nRep.north) -- (enhanced_\nRep);
\foreach \x in {0, ..., 2}{
 \pgfmathtruncatemacro{\colorPerc}{round(\x*30 + 15)}
\pgfmathsetlengthmacro{\sOff}{(\x - 1) * \shadowOff}
\ifthenelse{\x = 2}{
    \def\shadowColor{black}
}{
    \pgfmathtruncatemacro{\colorPerc}{round(\x*10 + 20)}
    \def\shadowColor{black!\colorPerc!white}
}
\draw [arrow,color=\shadowColor] ($(noisy_input_\nRep.north) + (-\sOff,-\sOff)$)
        -- ($(ssf_\nRep.south) + (-\sOff, 0)$);
\node[unfoldedSplit, above=of noisy_input_\nRep, xshift=-\sOff, yshift=-\sOff, color=\shadowColor] (noisy_input_split_\nRep_\x) {};

}

}

\pgfmathtruncatemacro{\lastRep}{\numRep -1}
\foreach \nRep in {1, ..., \lastRep}{
\pgfmathtruncatemacro{\nextRep}{\nRep + 1}

\draw[arrow]
    (split_ssf_\nRep.center)
    -| ($ (ssf_\nRep.east)!0.4!(tst_\nextRep.west)$)
    |- (tst_\nextRep.west);

\foreach \x in {0, ..., 2}{
 \pgfmathtruncatemacro{\colorPerc}{round(\x*30 + 15)}
\pgfmathsetlengthmacro{\sOff}{(\x - 1) * \shadowOff}
\ifthenelse{\x = 2}{
    \def\shadowColor{black}
}{
    \pgfmathtruncatemacro{\colorPerc}{round(\x*10 + 20)}
    \def\shadowColor{black!\colorPerc!white}
}

    \edef\nextTST{tst_\nextRep}
    \draw[arrow, color=\shadowColor]
        (noisy_input_split_\nRep_\x.center)
        -| ($(tst_\nextRep.south) + (-\sOff, 0)$);
}
}

\foreach \x in {0, ..., 2}{
 \pgfmathtruncatemacro{\colorPerc}{round(\x*30 + 15)}
\pgfmathsetlengthmacro{\sOff}{(\x - 1) * \shadowOff}
\ifthenelse{\x = 2}{
    \def\shadowColor{black}
}{
    \pgfmathtruncatemacro{\colorPerc}{round(\x*10 + 20)}
    \def\shadowColor{black!\colorPerc!white}
}

    \draw[arrow, dash pattern=on 2pt off 1pt, color=\shadowColor]
        ($(noisy_input_split_1_\x.center) + (-1.25 * \dnnBoxWidth, 0mm)$)
        -| ($(tst_1.south) + (-\sOff, 0)$);
}

\foreach \x in {0, ..., 2}{
 \pgfmathtruncatemacro{\colorPerc}{round(\x*30 + 15)}
\pgfmathsetlengthmacro{\sOff}{(\x - 1) * \shadowOff}
\draw[dash pattern=on 2pt off 1pt]
        (split_ssf_\numRep.center)-- ($(split_ssf_\numRep.center) + (0.5 * \dnnBoxWidth, 0mm)$);

\ifthenelse{\x = 2}{
    \def\shadowColor{black}
}{
    \pgfmathtruncatemacro{\colorPerc}{round(\x*10 + 20)}
    \def\shadowColor{black!\colorPerc!white}
}

    \draw[dash pattern=on 2pt off 1pt, color=\shadowColor]
        (noisy_input_split_\numRep_\x.center)-- ($(noisy_input_split_\numRep_\x.center) + (0.5 * \dnnBoxWidth, 0mm)$);
}

\end{tikzpicture}
    \vspace\flowVspaceUnfolded
  \end{minipage}
  }
 \hfill
\subfloat[MIMO-AR.\label{fig:weak_ssf_mimo_unfolded}]{
  \hspace*{-2pt}\begin{minipage}{0.30\linewidth}%
    \centering

\newcommand\unboundedBoxStartX{9mm}
\newcommand\unboundedBoxEndX{3.7cm}
\newcommand\unfoldedOff{1.4cm}  
\begin{tikzpicture}[node distance=0.5*\nodeDistance and 1*\nodeDistance, background rectangle/.style={fill=none}] 
\path[use as bounding box, draw, dashed, bboxcolorUnfolded] (\unboundedBoxStartX,\unboundedBoxStartY) rectangle (\unboundedBoxEndX,\unboundedBoxEndY);

\pgfmathsetlengthmacro{\condShift}{
    \pfHeight + 0.5 * \nodeDistance
}
\pgfmathsetlengthmacro{\ssfOff}{
    \dnnWidth - 1.25 * \nodeDistance
}
\pgfmathsetlengthmacro{\ssfShift}{
    \pfWidth - \dnnWidth
}


\foreach \nRep in {1, ..., \numRep}{
\pgfmathsetlengthmacro{\xtstshift}{\nRep * \unfoldedOff}
 \node[draw,
    rectangle, xshift=\xtstshift, minimum width=\dnnBoxWidth, minimum height=\dnnBoxHeight, align=center,
      inner sep=0pt,
      outer sep=0pt] (tst_\nRep) {};
\node at (tst_\nRep.center) {\unfoldedBigFont TST};

\pgfmathsetlengthmacro{\xtstshift}{\nRep * \unfoldedOff}
 \node[draw,
    rectangle, above=of tst_\nRep, xshift=\unfoldedSSFshift, minimum width=\dnnBoxWidth, minimum height=\dnnBoxHeight, align=center,
      inner sep=0pt,
      outer sep=0pt] (ssf_\nRep) {};
\node at (ssf_\nRep.center) {\unfoldedBigFont SSF};

\ifnum\nRep=1
    \node [below=of ssf_\nRep, yshift=\unfoldedNoisyOff] (noisy_input_\nRep) {\unfoldedFlowMathFont$\mathbf{Y}_{\delayIdx{t}}$};
\else
    \node [below=of ssf_\nRep, yshift=\unfoldedNoisyOff] (noisy_input_\nRep) {\unfoldedFlowMathFont$\mathbf{Y}_t$};
\fi 


\node [unfoldedSplit, above=of ssf_\nRep, color=white] (split_ssf_\nRep) {};

\ifnum\nRep=1
    \node [left=of ssf_\nRep, xshift=\unfoldedSSFlabelShift]{\unfoldedFlowMathFont$\hat{\theta}_{\delayIdx{t}}$};
\else
    \node [left=of ssf_\nRep, xshift=\unfoldedSSFlabelShift] {\unfoldedFlowMathFont$\hat{\theta}_t\,$};
\fi 

\ifnum\nRep=1
    \node [above=of split_ssf_\nRep] (enhanced_\nRep) {\unfoldedFlowMathFont$\hat{S}_{\delayIdx{t}}$};
\else
    \node [above=of split_ssf_\nRep] (enhanced_\nRep) {\unfoldedFlowMathFont$\hat{S}_t$};
\fi 

\draw [arrow] (tst_\nRep.north) |- (ssf_\nRep.west);
\draw [arrow] ($(ssf_\nRep.north) + (-\shadowOff, 0)$) -- ($(enhanced_\nRep.south) + (-\shadowOff, 0)$);
\foreach \x in {0, ..., 2}{
 \pgfmathtruncatemacro{\colorPerc}{round(\x*30 + 15)}
\pgfmathsetlengthmacro{\sOff}{(\x - 1) * \shadowOff}
\ifthenelse{\x = 2}{
    \def\shadowColor{black}
}{
    \pgfmathtruncatemacro{\colorPerc}{round(\x*10 + 20)}
    \def\shadowColor{black!\colorPerc!white}
}
\draw [arrow,color=\shadowColor] ($(noisy_input_\nRep.north) + (-\sOff,-\sOff)$)
        -- ($(ssf_\nRep.south) + (-\sOff, 0)$);

}

}

\pgfmathtruncatemacro{\lastRep}{\numRep -1}
\foreach \nRep in {1, ..., \lastRep}{
\pgfmathtruncatemacro{\nextRep}{\nRep + 1}


\foreach \x in {0, ..., 2}{
 \pgfmathtruncatemacro{\colorPerc}{round(\x*30 + 15)}
\pgfmathsetlengthmacro{\sOff}{(\x - 1) * \shadowOff}
\ifthenelse{\x = 2}{
    \def\shadowColor{black}
}{
    \pgfmathtruncatemacro{\colorPerc}{round(\x*10 + 20)}
    \def\shadowColor{black!\colorPerc!white}
}
    \node (tmp) at ($(ssf_\nRep.east)!0.5!(tst_\nextRep.west)+ (-\sOff, 0)$) {};
    \draw [color=\shadowColor] ($(ssf_\nRep.north) + (-\sOff, 0)$) -- ($(split_ssf_\nRep.center) + (-\sOff, -\sOff)$);
    \draw[arrow, color=\shadowColor]
    ($(split_ssf_\nRep.center)+ (-\sOff, -\sOff)$)
    -| ($ (tmp) $)
    -- (noisy_input_split_\nRep_\x.center -| tmp) -| ($(tst_\nextRep.south) + (-\sOff, 0)$);
}

\node [unfoldedSplit, above=of ssf_\nRep, xshift=-\shadowOff, yshift=-\shadowOff] (split_ssf_\nRep_tmp) {};
}

\foreach \x in {0, ..., 2}{
 \pgfmathtruncatemacro{\colorPerc}{round(\x*30 + 15)}
\pgfmathsetlengthmacro{\sOff}{(\x - 1) * \shadowOff}
\ifthenelse{\x = 2}{
    \def\shadowColor{black}
}{
    \pgfmathtruncatemacro{\colorPerc}{round(\x*10 + 20)}
    \def\shadowColor{black!\colorPerc!white}
}

    \draw[arrow, dash pattern=on 2pt off 1pt, color=\shadowColor]
        ($(noisy_input_split_1_\x.center) + (-1.25 * \dnnBoxWidth, 0mm)$)
        -| ($(tst_1.south) + (-\sOff, 0)$);
    
}

\foreach \x in {0, ..., 2}{
 \pgfmathtruncatemacro{\colorPerc}{round(\x*30 + 15)}
\pgfmathsetlengthmacro{\sOff}{(\x - 1) * \shadowOff}

\ifthenelse{\x = 2}{
    \def\shadowColor{black}
}{
    \pgfmathtruncatemacro{\colorPerc}{round(\x*10 + 20)}
    \def\shadowColor{black!\colorPerc!white}
}
    \draw[dash pattern=on 2pt off 1pt, color=\shadowColor]
        ($(split_ssf_\numRep.center) + (-\sOff, -\sOff)$) -- ($(split_ssf_\numRep.center) + (0.5 * \dnnBoxWidth-\sOff, -\sOff)$);
    \draw [dash pattern=on 2pt off 1pt, color=\shadowColor] ($(ssf_\numRep.north) + (-\sOff, 0)$) -- ($(split_ssf_\numRep.center) + (-\sOff, -\sOff)$);
\node [unfoldedSplit, above=of ssf_\numRep, xshift=-\shadowOff, yshift=-\shadowOff] (split_ssf_\numRep_tmp) {};
}
\end{tikzpicture}
    \vspace\flowVspaceUnfolded
  \end{minipage}
  }
\caption{Sequential processing of weakly guided speaker extraction frameworks unfolded over time. \Acl{tst} (\acs{tst}) estimates the target's direction $\theta_t$ to steer a \acl{ssf} (\acs{ssf}) for enhancement. Our proposed \acl{ar} (\acs{ar}) methods in (b) and (c) utilize the processed speech from the previous frame $t\!-\!1$ during tracking in the current frame $t$.}
\label{fig:unfolded_flowcharts}
\end{figure}

\section{Dataset}\label{sec:dataset}
\subsection{Acoustic Dataset Parametrization}\label{sec:dataset_parametrization}
To facilitate development and evaluation under controlled acoustic conditions, we generate a synthetic dataset of noisy and reverberant recordings containing two moving speakers.  
For this, we use utterances from the \mbox{LibriSpeech} corpus \cite{panayotov15librispeech} and pair them according to \mbox{Libri2Mix} \cite{cosentino20librimix}.
For spatialization, we simulate \acp{rir} for shoe-box shaped rooms via gpuRIR~\cite{diaz18gpu_rir}, a GPU accelerated implementation of the image method \cite{allen79image_method}. %
We parameterize each acoustic scenario according to the randomized setup of Tesch and Gerkmann \cite{tesch24ssf_journal}, using reverberation times between 0.2\,s and 0.5\,s and a circular three-microphone array with 10\,cm diameter.
In contrast to \cite{tesch24ssf_journal}, we enlarge the room dimensions to 4–8\,m and place the array within the central 20\,\% of the room to induce more movement around the array.
Speakers are initially separated by at least 15° in azimuth and move along trajectories in the horizontal plane at a constant height during each recording.
The temporal discretization of the trajectories is aligned with the \ac{stft} parametrization, for which we use a $\sqrt{\text{Hann}}$ window of length 32\,ms and 16\,ms hop-size at 16\,kHz.
While our focus is speaker extraction, we add spatially diffuse, spectrally white, stationary 
Gaussian noise~\cite{habets07isotropic_noise_generation} at 20–30\,dB \ac{snr} to improve robustness to mild additive interference. 

\begin{figure*}[t!]
\begin{tikzpicture}
    
\newcommand\doaWidth{1.5cm}
\newcommand\doaHeight{8mm}
\newcommand\doaDist{2mm}
\newcommand\lrDist{10mm}
\newcommand\tbDist{8mm}
\newcommand\tXoff{5.5mm}
\newcommand\tYoff{2.5mm}
\newcommand\plotStartX{0mm}
\newcommand\plotStartY{0mm}
\newcommand\legendTickTop{-0.25mm}
\newcommand\legendTickBottom{-0.5mm}
\pgfmathsetlengthmacro{\legendTickMiddle}{0.5 * \legendTickTop + 0.5 * \legendTickBottom}
\newcommand\legendXBoxOff{1mm} 
\newcommand\legendXBoxOffBack{4mm}
\newcommand\legendYBoxOff{2.9cm}
\pgfmathsetlengthmacro{\legendXBoxWidth}{0.995*\textwidth}
\newcommand\legendYBoxHeight{4.5mm} 
\newcommand\legendSkip{4mm}


\foreach \x / \w / \h / \frameIdx / \i [
    remember=\dw as \lastWidth (initially 0), 
    evaluate=\w as \dw using \w+\lastWidth,
]in {
    0/7.19/4.00/0000/2798,
    1/7.19/4.00/0030/494,
    2/7.19/4.00/0050/849,
    3/7.19/4.00/0070/2409,
    4/7.19/4.00/0130/2798,
    5/7.19/4.00/0150/494,
    6/7.19/4.00/0170/849,
    7/7.19/4.00/0190/2409
    }{

    \def\fileName{snapshot_\frameIdx}
    
    \newcommand\relScale{5.4}
    \pgfmathsetlengthmacro{\dOff}{\lastWidth / 5.375 * \doaWidth} 
    \pgfmathsetlengthmacro{\W}{\w / \relScale * \doaWidth} 
    \pgfmathsetlengthmacro{\H}{\h / \relScale * \doaWidth} 
    \node[anchor=center, rotate=-90] at (\plotStartX + \lrDist + \dOff + \x * \doaDist + 0.5 * \W, \plotStartY + 1*\doaHeight + \tbDist + 0.5 * \H) {%
        \pgfimage[height=\W, width=\H]{images/social_force_model/\fileName.pdf} 
    };
    \pgfmathsetmacro{\widthNum}{int(\w)}
    \pgfmathsetlengthmacro{\widthOff}{\doaWidth / \relScale}
    \foreach \wn in {0,...,\widthNum}{
        \pgfmathtruncatemacro\tmp{int(\wn/2) * 2}
        \ifnum\tmp=\wn
            \pgfmathsetmacro{\tickLen}{\majorTick}  
            \node[anchor=north] at (\plotStartX  + \lrDist + \dOff + \x * \doaDist + \wn * \widthOff, \plotStartY + \doaHeight + \tbDist) {\tickFont \wn};
        \else
            \pgfmathsetmacro{\tickLen}{\minorTick}
        \fi
        \draw[line width=\tickWidth] (\plotStartX  + \lrDist + \dOff + \x * \doaDist + \wn * \widthOff, \plotStartY + \doaHeight + \tbDist) --  (\plotStartX  + \lrDist + \dOff + \x * \doaDist + \wn * \widthOff, \plotStartY - \tickLen + \doaHeight + \tbDist);
    }
    \pgfmathsetmacro{\heightNum}{int(\h)}
    \pgfmathsetlengthmacro{\heightOff}{\doaWidth / \relScale}
    \foreach \hn in {0,...,\heightNum}{
        \pgfmathtruncatemacro\tmp{int(\hn/2) * 2}
        \ifnum\tmp=\hn
            \pgfmathsetmacro{\tickLen}{\majorTick}
            \ifthenelse{\x = 0}{
                \node[anchor=east] at (\plotStartX  + \lrDist + \dOff + \x * \doaDist, \plotStartY + \doaHeight + \tbDist + \hn * \heightOff) {\tickFont \hn};
            }{}
        \else
            \pgfmathsetmacro{\tickLen}{\minorTick}
        \fi
        \draw[line width=\tickWidth] (\plotStartX  + \lrDist + \dOff + \x * \doaDist - \tickLen, \plotStartY + \doaHeight + \tbDist + \hn * \heightOff) --  (\plotStartX  + \lrDist + \dOff + \x * \doaDist, \plotStartY + \doaHeight + \tbDist + \hn * \heightOff);
    }

    \pgfmathtruncatemacro{\frameInt}{\frameIdx}
    \pgfmathsetmacro{\timeval}{\frameInt /20}
    \node[anchor=center, xshift=\tXoff, yshift=\tYoff] at (\plotStartX + \lrDist + \dOff + \x * \doaDist + 0.5 * \W, \plotStartY + 1*\doaHeight + \tbDist + 0.5 * \H) {\scriptsize \timeval\,s};
    
}

\node[anchor=north] at (0.5 * \linewidth + 2mm, \plotStartY- 3*\majorTick + \doaHeight + \tbDist) {\footnotesize room width [m]};
\node[anchor=south, rotate=90] at (\plotStartX  + \lrDist - 2mm, \plotStartY + \doaHeight+ 0.5 * \doaHeight + \tbDist + 1mm) {\footnotesize length [m]};

\newcommand\labelColor{tab_orange}
\newcommand\legendDeltaX{4mm}
\draw[draw, rounded corners=\legendRectEdge, line width=\tickWidth]
            (\plotStartX + \legendXBoxOffBack, \plotStartY + \legendYBoxOff) rectangle (\plotStartX + \legendXBoxWidth + \legendXBoxOffBack, \plotStartY + \legendYBoxOff + \legendYBoxHeight);
            
 \pgfmathsetlengthmacro{\micLegendDist}{
 2mm+\legendDeltaX
 }
 \node[anchor=west] at (\plotStartX + \legendXBoxOff + \micLegendDist, \plotStartY + \legendYBoxOff+ 0.5 * \legendYBoxHeight) {\labelFont mic.\,array pos. $\mathbf{r}^{\scriptscriptstyle(\mathrm{A})}$};
 \node[
    regular polygon,
    regular polygon sides=3,
    draw,
    fill=black,
    minimum size=5pt,
    inner sep=0pt,
    anchor=center,
    xshift=-0.5mm,
    yshift=-0.25mm
] at (\plotStartX + \legendXBoxOff + \micLegendDist, \plotStartY + \legendYBoxOff+ 0.5 * \legendYBoxHeight) {};

 \pgfmathsetlengthmacro{\oracleDist}{
 2.6cm+2*\legendDeltaX
 }
 \newcommand\yRoomOff{0mm}
 \draw[color=\labelColor, line width=1.5pt,opacity=0.5,decorate,
  decoration={snake, amplitude=3pt, segment length=10pt}] (\plotStartX + \legendXBoxOff + \oracleDist, \plotStartY + \legendYBoxOff+ 0.5 * \legendYBoxHeight + \yRoomOff) -- coordinate (snakeend) (\plotStartX + \legendXBoxOff + \oracleDist+ \legendSkip, \plotStartY + \legendYBoxOff+ 0.5 * \legendYBoxHeight + \yRoomOff);
  \fill[\labelColor] (\plotStartX + \legendXBoxOff + \oracleDist+ \legendSkip, \plotStartY + \legendYBoxOff+ 0.5 * \legendYBoxHeight + \yRoomOff) circle (2pt);
 \node[anchor=west, xshift=0.5mm] at (\plotStartX + \legendXBoxOff + \oracleDist+ \legendSkip, \plotStartY + \legendYBoxOff+ 0.5 * \legendYBoxHeight){\labelFont trajectory to speaker pos. $\mathbf{r}_i$\,({\protect\tikz \protect\fill[\labelColor] (0mm,0mm) ++ (0mm,0.75mm) circle[radius=1.75pt];})};

 \pgfmathsetlengthmacro{\drivingDist}{
 6.9cm+3*\legendDeltaX
 }
 \fill[\labelColor,opacity=0.5] (\plotStartX + \legendXBoxOff + \drivingDist, \plotStartY + \legendYBoxOff+ 0.5 * \legendYBoxHeight) circle (4pt);
 \draw (\plotStartX + \legendXBoxOff + \drivingDist, \plotStartY + \legendYBoxOff+ 0.5 * \legendYBoxHeight) 
 node[star,star points=5,draw,fill=\labelColor,minimum size=4pt,star point ratio=2.25, anchor=center, inner sep=0pt, rounded corners=0pt, line width=0pt]{};
 \node[xshift=1mm, anchor=west] at (\plotStartX + \legendXBoxOff + \drivingDist, \plotStartY + \legendYBoxOff+ 0.5 * \legendYBoxHeight) {
 	\labelFont driving goal $\mathbf{r}_i^{\scriptscriptstyle(\mathrm{D})}$
};

 \pgfmathsetlengthmacro{\micStartDist}{
 9.25cm+4*\legendDeltaX
 }
\draw[
  -{Triangle[length=1.125mm,width=1.125mm]},
  color=\labelColor,
  line width=1.0pt,
  yshift=-1.1mm
]
(\plotStartX + \legendXBoxOff + \micStartDist-1.75mm,
 \plotStartY + \legendYBoxOff + 0.5 * \legendYBoxHeight)
-- ++(1.125mm,2.5mm);
\draw[
  -{Triangle[length=1.125mm,width=1.125mm]},
  color=red,
  line width=1.0pt,
  yshift=-1.1mm,
  xshift=0.8mm
]
(\plotStartX + \legendXBoxOff + \micStartDist,
 \plotStartY + \legendYBoxOff + 0.5 * \legendYBoxHeight)
-- ++(1.125mm,2.5mm);
 \node[] at (\plotStartX + \legendXBoxOff + \micStartDist, \plotStartY + \legendYBoxOff+ 0.5 * \legendYBoxHeight) {
 	\labelFont $/$
 };
 \node[xshift=1.75mm, anchor=west] at (\plotStartX + \legendXBoxOff + \micStartDist, \plotStartY + \legendYBoxOff+ 0.5 * \legendYBoxHeight) {
 	\labelFont driving force $\mathbf{f}_i^{\scriptscriptstyle(\mathrm{D})}$\hspace*{-2pt}$/$\,repulsive force $\mathbf{f}_i^{\scriptscriptstyle(\mathrm{R})}$
 };

\def\boundaryHatchWidth{6pt}
 \pgfmathsetlengthmacro{\boundaryDist}{
 14.25cm+5*\legendDeltaX
 }
 \fill[
  red!10,
  xshift=-1mm,
  anchor=center
]
  (\plotStartX + \legendXBoxOff + \boundaryDist - 0.5 * \boundaryHatchWidth,
   \plotStartY + \legendYBoxOff + 0.5 * \legendYBoxHeight - 0.5 * \boundaryHatchWidth)
  rectangle ++(\boundaryHatchWidth,\boundaryHatchWidth);
 
 \fill[
  boundaryhatch,
  draw=red,
  xshift=-1mm,
  anchor=center
]
  (\plotStartX + \legendXBoxOff + \boundaryDist - 0.5 * \boundaryHatchWidth,
   \plotStartY + \legendYBoxOff + 0.5 * \legendYBoxHeight - 0.5 * \boundaryHatchWidth)
  rectangle ++(\boundaryHatchWidth,\boundaryHatchWidth);
 
 \node[anchor=west] at (\plotStartX + \legendXBoxOff + \boundaryDist, \plotStartY + \legendYBoxOff+ 0.5 * \legendYBoxHeight) {
 	\labelFont boundary region
 };

\end{tikzpicture}
\vspace*{-20pt}
\caption{
Social force motion model adapted from \cite{helbing95social_force_model} to simulate planar two speaker ({\protect\tikz[baseline=-1.0ex]  \protect\draw[tab_blue, line width=1pt] (0,0mm) -- (1mm,0mm);}/{\protect\tikz[baseline=-0.4ex] \protect\draw[tab_orange, line width=1pt] (0,0mm) -- (1mm,0mm);}) trajectories in an enclosed room. An underlying Newtonian formulation enforces smooth motion patterns while satisfying boundary constraints. Dataset generation code and further visualizations are available online\textsuperscript{\ref{code_page}}.
}
\label{fig:social_force_model}
\end{figure*}

\subsection{Social Force Motion Model}\label{sec:social_force_model}
While \mbox{gpuRIR}~\cite{diaz18gpu_rir} enables efficient simulation of moving speakers, realistic motion models are essential to ensure generalization of data-driven tracking and enhancement to real-world recordings. 
A common approach samples start and end points within the simulation boundaries and connects them via linear trajectories at constant velocity~\cite{ochiai23moving_speaker_attention_mvdr, tammen2024linear_movement_array_agnostic_beamformer}, or optionally use sinusoidally modulated trajectories~\cite{diaz21srp_phat}.
However, finite path lengths couple speaker velocity to room size and recording duration.
Circular trajectories~\cite{kienegger25wg_ssf,kienegger25sg_ssf,rusrus23cirular_movement_doa_estimation} avoid this issue, but enforce an unrealistic fixed array distance.
To overcome these limitations, we propose adopting the \textit{social force model} of Helbing et al.~\cite{helbing95social_force_model}, originally introduced in the context of environmentally aware pedestrian dynamics, 
to simulate speaker movement in enclosed acoustic scenarios.
Via a Newtonian formulation, smooth trajectories of arbitrary length and velocity profiles can be generated that satisfy environmental constraints.
In particular, Newton's second law of motion [\citen{goldstein02classical_mechanics}, Eq.~1.3] is employed to couple the positions $\mathbf{r}_i$ of all $i\in\mathcal{I}$ speakers to the unit mass driving forces $\mathbf{f}_i^{\scriptscriptstyle(\mathrm{D})}$  
and repulsive forces $\mathbf{f}_i^{\scriptscriptstyle(\mathrm{R})}$ through the differential equation
\begin{equation}\label{eq:social_force_model}
    \dot{\mathbf{v}}_i = \mathbf{f}_i^{\scriptscriptstyle(\mathrm{D})} + \mathbf{f}_i^{\scriptscriptstyle(\mathrm{R})} \, , \ \mathbf{v}_i = \dot{\mathbf{r}}_i \, .
\end{equation} 
The driving force $\mathbf{f}_i^{\scriptscriptstyle(\mathrm{D})}$ represents the $i$-th speaker's desire to move towards a fictitious goal $\mathbf{r}_i^{\scriptscriptstyle(\mathrm{D})}$ at a desired velocity $\left\lVert\mathbf{v}_i^{\scriptscriptstyle(\mathrm{D})}\right\rVert$, with relaxation time $\tau$ influencing the acceleration behavior
\begin{equation}\label{eq:driving_force}
    \mathbf{f}_i^{\scriptscriptstyle(\mathrm{D})} = \frac{1}{\tau} \left(\mathbf{v}_i^{\scriptscriptstyle(\mathrm{D})} - \mathbf{v}_i\right) \, , \  \mathbf{v}_i^{\scriptscriptstyle(\mathrm{D})} = \frac{\mathbf{r}^{\scriptscriptstyle(\mathrm{D})}_i - \mathbf{r}_i}{\left\lVert \mathbf{r}^{\scriptscriptstyle(\mathrm{D})}_i - \mathbf{r}_i \right\rVert} \left\lVert\mathbf{v}_i^{\scriptscriptstyle(\mathrm{D})}\right\rVert \, .
\end{equation}
We sample the driving velocity $\left\lVert\mathbf{v}_i^{\scriptscriptstyle(\mathrm{D})}\right\rVert$ from a Gaussian distribution with mean 1.34\,m$/$s and standard deviation 0.26\,m$/$s (clamped at zero) \cite{helbing95social_force_model}, which corresponds to typical walking speeds
\cite{ murtagh21outdoor_walking_speed}.
The fictitious goal $\mathbf{r}^{\scriptscriptstyle(\mathrm{D})}_i$ is randomly initialized and resampled when the speaker comes within 0.5\,m, with relaxation time $\tau\!=\,$1\,s enforcing smooth directional changes.
While the driving force $\mathbf{f}_i^{\scriptscriptstyle(\mathrm{D})}$ guides each speaker along an intended path, the repulsive force $\mathbf{f}_i^{\scriptscriptstyle(\mathrm{R})}$
  in \cref{eq:social_force_model} incorporates environmental constraints and thereby shapes trajectories to ensure physical feasibility. 
  We decompose $\mathbf{f}_i^{\scriptscriptstyle(\mathrm{R})}$ into boundary forces from walls $\mathbf{f}_i^{\scriptscriptstyle(\mathrm{W})}$, the microphone array $\mathbf{f}_i^{\scriptscriptstyle(\mathrm{A})}$, and inter-speaker forces $\mathbf{f}_i^{\scriptscriptstyle(\mathrm{S})}$	
  that preserve comfortable distances,
\begin{equation}\label{eq:repulsive_decomposition}
    \mathbf{f}_i^{\scriptscriptstyle(\mathrm{R})} = \mathbf{f}_i^{\scriptscriptstyle(\mathrm{W})} + \mathbf{f}_i^{\scriptscriptstyle(\mathrm{A})} + \mathbf{f}_i^{\scriptscriptstyle(\mathrm{S})} \, .
\end{equation} 
We model the wall forces $\mathbf{f}_i^{\scriptscriptstyle(\mathrm{W})}$ as the sum of gradients of per-wall repulsive, exponential potentials $U_{iw}^{\scriptscriptstyle(\mathrm{W})}$ \cite{helbing95social_force_model}
\begin{equation}\label{eq:boundary_force}
    \mathbf{f}_i^{\scriptscriptstyle(\mathrm{W})} = - \displaystyle \sum_{w=1}^4\nabla_{\hspace{-2pt}\mathbf{r}_i} U_{iw}^{\scriptscriptstyle(\mathrm{W})} \, , \ U_{iw}^{\scriptscriptstyle(\mathrm{W})} = A^{\scriptscriptstyle(\mathrm{W})}_ie^{-\left\lVert \mathbf{d}^{\scriptscriptstyle(\mathrm{W})}_{iw} \right\rVert \left/ B^{\scriptscriptstyle(\mathrm{W})}\right.} \, ,
\end{equation}
where distance $\mathbf{d}^{\scriptscriptstyle(\mathrm{W})}_{iw} = \mathbf{r}_i - \mathbf{r}^{\scriptscriptstyle(\mathrm{W})}_{iw}$ and $\mathbf{r}^{\scriptscriptstyle(\mathrm{W})}_{iw}$ denotes the point on wall $w$ closest to the speaker’s position $\mathbf{r}_i$.
To parametrize $A^{\scriptscriptstyle(\mathrm{W})}_i$ for maintaining a minimum distance $\scalebox{1.2}{$\varepsilon$}^{\scriptscriptstyle(\mathrm{W})}$ to the walls, we consider the limiting case of a head-on approach at  speed $\lVert\mathbf{v}_i\rVert = \left\lVert\mathbf{v}_i^{\scriptscriptstyle(\mathrm{D})}\right\rVert$.
Specifically, we equate the speaker's kinetic energy [\citen{goldstein02classical_mechanics}, Eq.~1.3] to the work of the wall force for deceleration.
This is approximated by an infinite deceleration path to $\scalebox{1.2}{$\varepsilon$}^{\scriptscriptstyle(\mathrm{W})}$, justified by the small exponential scale $B^{\scriptscriptstyle(\mathrm{W})}$ of 0.2\,m~\cite{helbing95social_force_model}.
Since the integration of the wall force component toward the $w$-th wall cancels the gradient, work amounts to the $w$-th potential $U_{iw}^{\scriptscriptstyle(\mathrm{W})}$ in~\cref{eq:boundary_force} at $\scalebox{1.2}{$\varepsilon$}^{\scriptscriptstyle(\mathrm{W})}$.
Minding sign-convention, the resulting equality can be solved for $A^{\scriptscriptstyle(\mathrm{W})}_i$, leading to
\begin{equation}\label{eq:boundary_force_parametrization}
    A^{\scriptscriptstyle(\mathrm{W})}_i = \frac{1}{2} {\left\lVert\mathbf{v}_i^{\scriptscriptstyle(\mathrm{D})}\right\rVert}^2 e^{\scalebox{1.0}{$\varepsilon$}^{\scriptscriptstyle(\mathrm{W})}\!\left/ B^{\scriptscriptstyle(\mathrm{W})}\right.} \, ,
\end{equation}
which we parametrize according to a minimum distance $\scalebox{1.2}{$\varepsilon$}^{\scriptscriptstyle(\mathrm{W})}$ of 0.5\,m.
Interaction forces from the microphone array $\mathbf{f}_i^{\scriptscriptstyle(\mathrm{A})}$ and other speakers $\mathbf{f}_i^{\scriptscriptstyle(\mathrm{S})}$ are modeled by repulsive potentials with elliptical contours, inducing realistic evasion maneuvers for point-like obstacles. 
For the microphone array, this gives
\begin{equation}\label{eq:array_force}
    \mathbf{f}_i^{\scriptscriptstyle(\mathrm{A})} = - \nabla_{\hspace{-2pt}\mathbf{r}_i} U_i^{\scriptscriptstyle(\mathrm{A})} \, , \  U_i^{\scriptscriptstyle(\mathrm{A})} = A^{\scriptscriptstyle(\mathrm{A})}e^{-2b_{i}^{\scriptscriptstyle(\mathrm{A})} \!\left/ B^{\scriptscriptstyle(\mathrm{A})}\right.} \, ,
\end{equation}
and semi-minor axis $b_{i}^{\scriptscriptstyle(\mathrm{A})}$ of the equipotential lines defined as
\begin{equation}\label{eq:array_semi_minor}
    2b_{i}^{\scriptscriptstyle(\mathrm{A})} = \sqrt{
    \left( \left\lVert \mathbf{d}_{i}^{\scriptscriptstyle(\mathrm{A})} \right\rVert + \left\lVert \mathbf{d}_{i}^{\scriptscriptstyle(\mathrm{A})} + \Delta t \, \mathbf{v}_{i}\right\rVert \right)^2 - \left( \Delta t \,  \lVert \mathbf{v}_{i}\rVert \right)^2
    } \, ,
\end{equation}
with distance $\mathbf{d}^{\scriptscriptstyle(\mathrm{A})}_{i} = \mathbf{r}_i - \mathbf{r}^{\scriptscriptstyle(\mathrm{A})}$ and array center $\mathbf{r}^{\scriptscriptstyle(\mathrm{A})}$.
Thus, the non-central focal point of the ellipse is shifted toward the $i$-th speaker proportional with factor $\Delta t$ of 2\,s to the velocity $\mathbf{v}_{i}$, causing earlier interaction for fast, head-on approaches. 
To maintain far-field conditions, we parametrize $A^{\scriptscriptstyle(\mathrm{A})}_i$ to ensure a minimum distance $\scalebox{1.2}{$\varepsilon$}^{\scriptscriptstyle(\mathrm{A})}$ of 0.5\,m from speaker to array. 
However, since the semi-minor in \cref{eq:array_semi_minor} depends on both position and velocity, the previous energy-based method becomes intractable.  
Instead, we adopt a quasi-static approximation ($\mathbf{v}_i = 0$), which guarantees to overestimate the deceleration force.  
Following the same derivation as for \cref{eq:boundary_force_parametrization} yields
\begin{equation}
    A^{\scriptscriptstyle(\mathrm{A})}_i = \frac{1}{2} {\left\lVert\mathbf{v}_i^{\scriptscriptstyle(\mathrm{D})}\right\rVert}^2 e^{2\scalebox{1.0}{$\varepsilon$}^{\scriptscriptstyle(\mathrm{A})}\!\left/ B^{\scriptscriptstyle(\mathrm{A})}\right.} \, .
\end{equation}
While the array is a stationary obstacle, the interfering speakers are moving. 
Accordingly, we employ the modified definition of repulsive potentials in \cite{johansson07social_force_video_tracking}, 
which uses the speaker velocity difference $\mathbf{v}_{ij} = \mathbf{v}_{i}\!-\!\mathbf{v}_{j}$ to orient semi-minor $b_{ij}^{\scriptscriptstyle(\mathrm{S})}$
\begin{equation}\label{eq:social_semi_minor}
    2b_{ij}^{\scriptscriptstyle(\mathrm{S})} = \sqrt{
    \left( \lVert \mathbf{d}_{ij} \rVert + \lVert \mathbf{d}_{ij} + \Delta t \, \mathbf{v}_{ij}\rVert \right)^2 - \left( \Delta t \,  \lVert \mathbf{v}_{ij}\rVert \right)^2
    }
\end{equation}
and results after accumulation in the inter-speaker force 
\begin{equation}\label{eq:social_force}
    \mathbf{f}_i^{\scriptscriptstyle(\mathrm{S})} = -\hspace{-2mm}\sum_{j\in\mathcal{I}\backslash \{i\}} \hspace{-2mm}\nabla_{\hspace{-2pt}\mathbf{r}_i} U_{ij}^{\scriptscriptstyle(\mathrm{S})} \, , \ U_{ij}^{\scriptscriptstyle(\mathrm{S})} = A^{\scriptscriptstyle(\mathrm{S})} e^{-2b_{ij}^{\scriptscriptstyle(\mathrm{S})}\!\left/ B^{\scriptscriptstyle(\mathrm{S})}\right.} \, .
\end{equation}
For parametrization, we adopt the originally proposed values of $A^{\scriptscriptstyle(\mathrm{S})} = \ $2.1\,m$^2/$s$^2$ and $B^{\scriptscriptstyle(\mathrm{S})} = \ $0.3\,m in \cite{helbing95social_force_model}.
During simulation, we solve the resulting nonlinear, coupled differential equations for the \newdiffComm{speaker's}{speakers'} positions $\mathbf{r}_i$ in \cref{eq:social_force_model} using Euler's method. 
\Cref{fig:social_force_model} illustrates how the interplay between driving and repulsive forces determines the \newdiffComm{speaker's}{speakers'} movement patterns. 
Further trajectories are shown in \cref{fig:synthetic_trajectories}, with additional visualizations and dataset generation code available online\footnote{\label{code_page}\codePage}.

\begin{figure}[t!]
\hspace{-1mm}\input{images/synthetic_trajectories/synthetic_trajectories_large.tikz}
\vspace*{-17.5pt}
\caption{
Two-speaker ({\protect\tikz[baseline=-1.0ex]  \protect\draw[tab_blue, line width=1pt] (0,0mm) -- (1mm,0mm);}/{\protect\tikz[baseline=-0.4ex] \protect\draw[tab_orange, line width=1pt] (0,0mm) -- (1mm,0mm);}) \ac{doa} estimates using concatenative~(\cref{fig:weak_ssf}) and our autoregressive (\mbox{MISO-AR}:~\cref{fig:weak_ssf_miso}, \mbox{MIMO-AR}:~\cref{fig:weak_ssf_mimo}) trackers.
Magnified insets show how autoregression helps suppressing the spatial cues of the interferer and maintain robust tracking for close and crossing speakers.
}
\label{fig:synthetic_trajectories}
\vspace*{-0pt}
\end{figure}

\section{Experimental Setup}\label{sec:experimental_setup}
\subsection{Model and Algorithm Parametrization}\label{sec:model_parametrization}
\textit{\textbf{Spatially Selective Filter}}
We employ \mbox{SpatialNet}~\cite{quan24spatialnet} as a deep, non-linear multichannel speech enhancement architecture.
\mbox{SpatialNet} demonstrates exceptional spatial filtering capabilities by utilizing repeated narrow- and wideband processing modules.
Specifically, we employ its frame-wise causal version using Mamba blocks for narrowband processing \cite{gu2024mamba, quan24online_spatialnet} and the steering mechanism from [\citen{wu25trajectories_universal_sound_separation}, Fig.~5]. 
In total, this amounts to \diffComm{a computational cost of 18.8\,GMACs$/$s and 1.74\,M parameters, with the \ac{mimo} extension using the microphone array and \ac{stft} configuration of \cref{sec:dataset_parametrization} adding fewer than 500 parameters and about 800\,kMACs$/$s per kHz bandwidth.}{1.74\,M parameters at a computational cost of 18.8\,GMACs$/$s to extract a single target speaker, with the \ac{mimo} extension using the microphone array and \ac{stft} configuration of \cref{sec:dataset_parametrization} adding fewer than 500 parameters and 6.2\,MMACs$/$s.
To put the complexity into context, we profile the streaming capability of \mbox{SpatialNet} using an \texttt{NVIDIA RTX 4080} GPU on a commercially available laptop. 
With the \ac{stft} parametrization from \cref{sec:dataset_parametrization} (32\,ms window size and a 16\,ms hop), \mbox{SpatialNet} satisfies the real-time processing constraint by achieving \newdiffComm{a}{an} \ac{rtf} of 0.29~\cite{long26diffusion_buffer_journal}, and a processing latency of 48\,ms.}

\textit{\textbf{Target Speaker Tracking}} 
In the concatenative, weakly guided case, we use the \mbox{Wrapped\,KF} and \diffComm{\mbox{Boostrap\,PF}}{\mbox{Bootstrap\,PF}} from \cref{sec:weak_tse} for tracking, with generic algorithmic implementations found in [\citen{traa13wrapped_kalman_filter}, Alg.~1] and  [\citen{lehmann06pf_resampling_Neff}, Alg.~1]\diffComm{}{,} respectively.
Our proposed modifications in \cref{sec:self_tse} can be incorporated by changing the order of prediction and update steps with modified likelihood definitions, as demonstrated for the \diffComm{\mbox{Boostrap\,PF}}{\mbox{Bootstrap\,PF}} in \cref{alg:bootstrap_filter}.
The complexity of the \mbox{Wrapped\,KF} \diffComm{filter}{} is mainly governed by the \ac{ipd} computation and \ac{ls} operation in the linear-phase \ac{doa} estimators in \cref{eq:doa_estimator,eq:miso_doa_estimator}. 
Since the spatial aliasing frequency is \diffComm{already}{} at 2\,kHz for the circular three-microphone array (\cref{sec:dataset_parametrization}), the computational load is only approximately 300\,kMACs$/$s to track a single speaker.
The \mbox{Bootstrap\,PF} is also dominated by the likelihood evaluation, which has to be done $N=$ 50 particle times, yielding about 2.5\,MMACs$/$s for both Watson and Gaussian likelihoods in \cref{eq:watson_likelihood} and \cref{eq:pf_miso_likelihood}\newdiffComm{}{,} respectively.
\diffComm{}{Including the trackers to guide \mbox{SpatialNet} increases the \ac{rtf} from 0.29 to 0.31$-$0.37.
Note that during frame-wise inference, both concatenative and \ac{ar} methods require the same processing overhead.
To extract multiple speakers, tracking and enhancement need to be executed repeatedly, linearly scaling computational load with the number of speakers.}

\subsection{Training and Optimization Details}\label{sec:optimization_details}
To ensure a robust interplay between tracking (TST) and enhancement (SSF) while minimizing \ac{nn} training overhead, we adopt a \diffComm{multi-stage optimization strategy, which has lead to \newdiffComm{significant}{substantial} performance improvements in prior work~\cite{kienegger25wg_ssf}.}{two-stage optimization strategy consisting of \textit{pretraining} and \textit{fine-tuning}}.
\diffComm{}{Both stages utilize our synthetic dataset with moving speakers from \cref{sec:dataset} for optimization.}

\textit{\textbf{Pretraining}} In the pretraining stage, we train the deep \ac{ssf} \mbox{SpatialNet} in a strongly guided setup (oracle \ac{doa}), adopting the joint time- and frequency domain loss $\mathcal{L}^{\tinySuper{(\mathrm{MISO})}}$ from~\cite{tesch24ssf_journal}
\begin{equation}\label{eq:tesch_L1_loss}
    \mathcal{L}^{\tinySuper{(\mathrm{MISO})}}(s, \hat{s}) = \alpha^{\tinySuper{(\ell_1)}} \lVert s-\hat{s} {\rVert}_1 + \big\lVert |S|-|\hat{S}| {\big\rVert}_1 \, .
\end{equation}
The $\ell_1$ norms are computed over temporal waveform and \ac{stft} time-frequency bins respectively, with $\alpha^{\tinySuper{(\ell_1)}}=10$ balancing both domains~\cite{tesch24ssf_journal}.
For the \ac{mimo} extension in \cref{eq:mimo_ssf_sequential}, $\mathcal{L}^{\tinySuper{(\mathrm{MISO})}}$ is averaged across all microphone channels, yielding
\begin{equation}\label{eq:mimo_loss}
    \mathcal{L}^{\tinySuper{(\mathrm{MIMO})}}(\mathbf{s}, \hat{\mathbf{s}}) = \frac{1}{M}\sum_{m=1}^M \mathcal{L}^{\tinySuper{(\mathrm{MISO})}}(s^m, \hat{s}^m) \, .
\end{equation}
Pretraining runs for 50 epochs using the Adam optimizer with an initial learning rate of 10$^{-3}$.
\newdiffComm{Exponential decay with a factor}{An exponential decay factor} of 0.955 \newdiffComm{decimates}{reduces} the learning rate \newdiffComm{during}{to 10$^{-4}$ over} pretraining.

\begin{figure}[t!]
\vspace{-5pt}
\centering
\captionsetup[subfloat]{
    justification=justified,
    singlelinecheck=false,
    margin=3pt
}
\subfloat[KF parameter sweep of likelihood variance $\sigma^2_\Phi$~\cref{eq:wrappedkf_measurement_model} and state-transition variance $\sigma^2_\nu$~\cref{eq:cv_motion_model} during the fine-tuning stage.\label{fig:kalman_parameter_finetuning}]{
  \begin{minipage}[b]{0.53\linewidth} %
    \vspace{0pt}
    \centering
\hspace*{-2mm}\begin{tikzpicture}

\newcommand\figSize{2.0cm}
\pgfmathsetlengthmacro{\figWidth}{\figSize * 1.5}
\pgfmathsetlengthmacro{\figHeight}{\figSize * 0.9}
\newcommand\plotStartX{0mm}
\newcommand\plotStartY{0mm}

\node[anchor=center] at (\plotStartX + 0.5 * \figWidth, \plotStartY + 0.5 * \figHeight) {%
    \pgfimage[height=\figHeight, width=\figWidth]{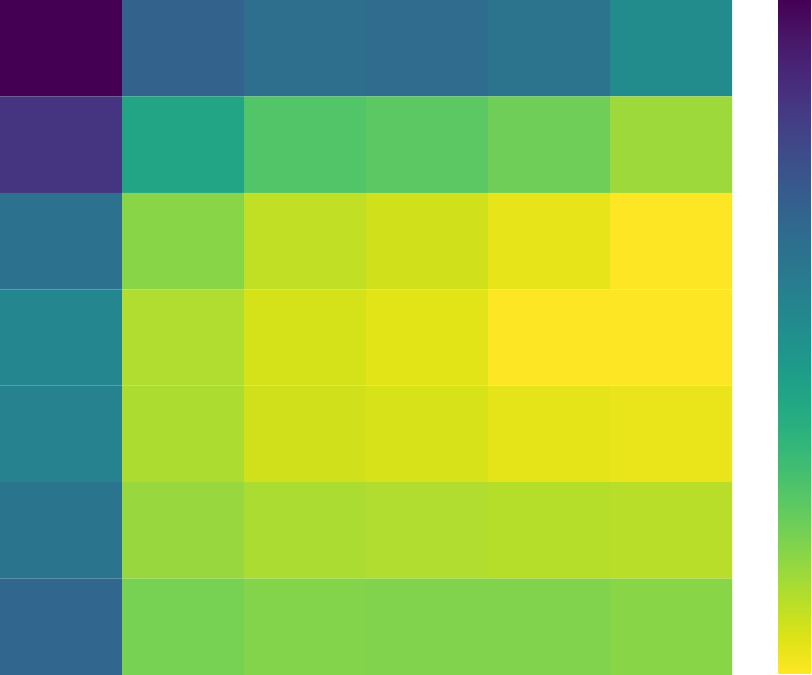} 
};

\pgfmathsetlengthmacro{\overlayXstart}{
    \plotStartX + 2.25mm
}
\pgfmathsetlengthmacro{\overlayYstart}{
    \plotStartY + \figHeight - 1.2mm
}
\pgfmathsetlengthmacro{\overlayXdelta}{4.5mm}   
\pgfmathsetlengthmacro{\overlayYdelta}{2.575mm}   

\node[font=\tiny, text=white] at ({\overlayXstart + 0*\overlayXdelta},{\overlayYstart - 0*\overlayYdelta}) {34.57};
\node[font=\tiny, text=white] at ({\overlayXstart + 1*\overlayXdelta},{\overlayYstart - 0*\overlayYdelta}) {25.76};
\node[font=\tiny, text=white] at ({\overlayXstart + 2*\overlayXdelta},{\overlayYstart - 0*\overlayYdelta}) {24.33};
\node[font=\tiny, text=white] at ({\overlayXstart + 3*\overlayXdelta},{\overlayYstart - 0*\overlayYdelta}) {24.70};
\node[font=\tiny, text=white] at ({\overlayXstart + 4*\overlayXdelta},{\overlayYstart - 0*\overlayYdelta}) {23.72};
\node[font=\tiny, text=white] at ({\overlayXstart + 5*\overlayXdelta},{\overlayYstart - 0*\overlayYdelta}) {20.86};

\node[font=\tiny, text=white] at ({\overlayXstart + 0*\overlayXdelta},{\overlayYstart - 1*\overlayYdelta}) {30.21};
\node[font=\tiny] at ({\overlayXstart + 1*\overlayXdelta},{\overlayYstart - 1*\overlayYdelta}) {17.95};
\node[font=\tiny] at ({\overlayXstart + 2*\overlayXdelta},{\overlayYstart - 1*\overlayYdelta}) {13.92};
\node[font=\tiny] at ({\overlayXstart + 3*\overlayXdelta},{\overlayYstart - 1*\overlayYdelta}) {13.32};
\node[font=\tiny] at ({\overlayXstart + 4*\overlayXdelta},{\overlayYstart - 1*\overlayYdelta}) {12.46};
\node[font=\tiny] at ({\overlayXstart + 5*\overlayXdelta},{\overlayYstart - 1*\overlayYdelta}) {10.36};

\node[font=\tiny, text=white] at ({\overlayXstart + 0*\overlayXdelta},{\overlayYstart - 2*\overlayYdelta}) {23.96};
\node[font=\tiny] at ({\overlayXstart + 1*\overlayXdelta},{\overlayYstart - 2*\overlayYdelta}) {11.29};
\node[font=\tiny] at ({\overlayXstart + 2*\overlayXdelta},{\overlayYstart - 2*\overlayYdelta}) {8.96};
\node[font=\tiny] at ({\overlayXstart + 3*\overlayXdelta},{\overlayYstart - 2*\overlayYdelta}) {8.27};
\node[font=\tiny] at ({\overlayXstart + 4*\overlayXdelta},{\overlayYstart - 2*\overlayYdelta}) {7.25};
\node[font=\tiny] at ({\overlayXstart + 5*\overlayXdelta},{\overlayYstart - 2*\overlayYdelta}) {\textbf{6.22}};

\node[font=\tiny, text=white] at ({\overlayXstart + 0*\overlayXdelta},{\overlayYstart - 3*\overlayYdelta}) {\textbf{21.53}};
\node[font=\tiny] at ({\overlayXstart + 1*\overlayXdelta},{\overlayYstart - 3*\overlayYdelta}) {\textbf{9.54}};
\node[font=\tiny] at ({\overlayXstart + 2*\overlayXdelta},{\overlayYstart - 3*\overlayYdelta}) {\textbf{8.02}};
\node[font=\tiny] at ({\overlayXstart + 3*\overlayXdelta},{\overlayYstart - 3*\overlayYdelta}) {\textbf{7.54}};
\node[font=\tiny] at ({\overlayXstart + 4*\overlayXdelta},{\overlayYstart - 3*\overlayYdelta}) {\textbf{6.32}};
\node[font=\tiny] at ({\overlayXstart + 5*\overlayXdelta},{\overlayYstart - 3*\overlayYdelta}) {6.27};

\node[font=\tiny, text=white] at ({\overlayXstart + 0*\overlayXdelta},{\overlayYstart - 4*\overlayYdelta}) {22.10};
\node[font=\tiny] at ({\overlayXstart + 1*\overlayXdelta},{\overlayYstart - 4*\overlayYdelta}) {9.66};
\node[font=\tiny] at ({\overlayXstart + 2*\overlayXdelta},{\overlayYstart - 4*\overlayYdelta}) {8.31};
\node[font=\tiny] at ({\overlayXstart + 3*\overlayXdelta},{\overlayYstart - 4*\overlayYdelta}) {7.96};
\node[font=\tiny] at ({\overlayXstart + 4*\overlayXdelta},{\overlayYstart - 4*\overlayYdelta}) {7.33};
\node[font=\tiny] at ({\overlayXstart + 5*\overlayXdelta},{\overlayYstart - 4*\overlayYdelta}) {7.19};

\node[font=\tiny, text=white] at ({\overlayXstart + 0*\overlayXdelta},{\overlayYstart - 5*\overlayYdelta}) {23.70};
\node[font=\tiny] at ({\overlayXstart + 1*\overlayXdelta},{\overlayYstart - 5*\overlayYdelta}) {10.65};
\node[font=\tiny] at ({\overlayXstart + 2*\overlayXdelta},{\overlayYstart - 5*\overlayYdelta}) {9.86};
\node[font=\tiny] at ({\overlayXstart + 3*\overlayXdelta},{\overlayYstart - 5*\overlayYdelta}) {9.60};
\node[font=\tiny] at ({\overlayXstart + 4*\overlayXdelta},{\overlayYstart - 5*\overlayYdelta}) {9.41};
\node[font=\tiny] at ({\overlayXstart + 5*\overlayXdelta},{\overlayYstart - 5*\overlayYdelta}) {9.28};

\node[font=\tiny, text=white] at ({\overlayXstart + 0*\overlayXdelta},{\overlayYstart - 6*\overlayYdelta}) {25.23};
\node[font=\tiny] at ({\overlayXstart + 1*\overlayXdelta},{\overlayYstart - 6*\overlayYdelta}) {12.03};
\node[font=\tiny] at ({\overlayXstart + 2*\overlayXdelta},{\overlayYstart - 6*\overlayYdelta}) {11.53};
\node[font=\tiny] at ({\overlayXstart + 3*\overlayXdelta},{\overlayYstart - 6*\overlayYdelta}) {11.54};
\node[font=\tiny] at ({\overlayXstart + 4*\overlayXdelta},{\overlayYstart - 6*\overlayYdelta}) {11.63};
\node[font=\tiny] at ({\overlayXstart + 5*\overlayXdelta},{\overlayYstart - 6*\overlayYdelta}) {11.25};

    \pgfmathsetlengthmacro{\plotYoff}{
     -1/14 * \figHeight
    }
    \pgfmathsetlengthmacro{\labelDist}{
        3/7 * \figHeight
    }
    \foreach \y / \yLabel in {
        0/\myExp{3},1/\myExp{2},2/\myExp{1}
    } {
        \ifnum\y<2
            \foreach \logY in {0,...,9} {
                \pgfmathsetlengthmacro{\logLabelDist}{
                    \labelDist * ln(\logY + 1) / ln(10)
                }
                \ifnum\logY=0
                    \def\tickColor{black}
                \else\ifnum\logY=1
                    \def\tickColor{black}
                \else\ifnum\logY=4
                    \def\tickColor{black}
                \else
                    \def\tickColor{lightgray}
                \fi\fi\fi

                \ifnum\logY=0
                    \def\tickLen{\majorTick}
                \else
                    \def\tickLen{\minorTick}
                \fi
                \draw[line width=\tickWidth, color=\tickColor] (\plotStartX,\plotStartY+ \figHeight - \y * \labelDist  - \logLabelDist + \plotYoff) -- (\plotStartX - \tickLen  ,\plotStartY+ \figHeight - \y * \labelDist  - \logLabelDist + \plotYoff);
                
            }
            \node[anchor=east, align=right, xshift=1mm] at (\plotStartX, \plotStartY + \figHeight - \y * \labelDist + \plotYoff) {\tickFont \yLabel};
        \else
            \draw[line width=\tickWidth] (\plotStartX,\plotStartY + \figHeight - \y * \labelDist  + \plotYoff) -- (\plotStartX - \majorTick,\plotStartY+ \figHeight - \y * \labelDist  + \plotYoff);
            \node[anchor=east, align=right, xshift=1.9mm] at (\plotStartX, \plotStartY + \figHeight - \y * \labelDist + \plotYoff) {\tickFont \yLabel};
        \fi
    }
    \node[anchor=south, rotate=90, yshift=5.5mm] at (\plotStartX, \plotStartY + 0.5 * \figHeight) {\labelFont var. ratio $\sigma_{\Phi}^2/\sigma_{\nu}^2$};

    \pgfmathsetlengthmacro{\effFigWidth}{
         \figWidth - 3mm
    }
    \pgfmathsetlengthmacro{\labelDist}{
        \effFigWidth / 6
    }
    \foreach \x / \xLabel in {
        0/1,1/4,2/8,3/12,4/16,5/20
    } {
        \node[anchor=north] at (\plotStartX + \x * \labelDist + 0.5 * \labelDist,\plotStartY) {\tickFont \xLabel};
        \draw[line width=\tickWidth] (\plotStartX + \x * \labelDist + 0.5 * \labelDist,\plotStartY) -- (\x * \labelDist + 0.5 * \labelDist,\plotStartY- \majorTick);
    }
    \node[anchor=north, yshift=-2.5mm] at (\plotStartX + 0.5 * \effFigWidth,\plotStartY) {\labelFont fine-tuning epoch};

    \pgfmathsetlengthmacro{\effFigHeight}{
        \figHeight - 4mm
    }
    \pgfmathsetlengthmacro{\labelDist}{
        \effFigHeight / 4
    }
    \foreach \y / \yLabel in {
        0/10,1/15,2/20,3/25,4/30
    } {
        \pgfmathtruncatemacro\tmp{int(\y/2) * 2}
            \ifnum\tmp=\y
            \node[anchor=west] at (\plotStartX+ \figWidth,\plotStartY + \y * \labelDist + 0.5 * \labelDist) {\tickFont \yLabel};
            \def\tickLen{\majorTick}
            \else
            \def\tickLen{\minorTick}
            \fi
        \draw[line width=\tickWidth] (\plotStartX+ \figWidth,\plotStartY + \y * \labelDist + 0.5 * \labelDist) -- (\plotStartX+\tickLen+ \figWidth,\plotStartY + \y * \labelDist + 0.5 * \labelDist);
        
    }
    \node[rotate=90, anchor=north, yshift=-3mm] at (\plotStartX + \figWidth, \plotStartY + 0.5 * \figHeight) {\labelFont MAE\,[°]\,$\leftarrow$};

\end{tikzpicture} %
\vspace{-15pt}
  \end{minipage}
}
\hfill
\subfloat[Influence of IPD-loss in \cref{eq:mimo_ar_loss} on enhancement (PESQ) and tracking (MAE) performance.\label{fig:ipd_alpha}]{
  \begin{minipage}[b]{0.41\linewidth} %
    \vspace{0pt}
    \centering
    \hspace*{-2.5mm}\begin{tikzpicture}

\newcommand\figSize{2.0cm}
\pgfmathsetlengthmacro{\figWidth}{\figSize * 1.2}
\pgfmathsetlengthmacro{\figHeight}{\figSize * 0.9}
\newcommand\plotStartX{0mm}
\newcommand\plotStartY{0mm}
\node[anchor=center] at (\plotStartX + 0.5 * \figWidth, \plotStartY + 0.5 * \figHeight) {%
    \pgfimage[height=\figHeight, width=\figWidth]{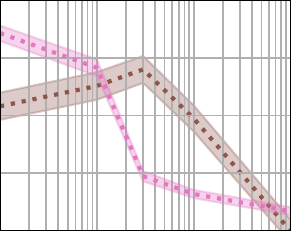} 
};

    \pgfmathsetlengthmacro{\labelDist}{
        \figWidth / 3
    }
    \foreach \x / \xLabel in {
        0/\myExp{4},1/\myExp{3},2/\myExp{2},3/\myExp{1}
    } {
        \pgfmathsetmacro{\exponent}{
        int(\x - 4)
        }
        \node[anchor=north] at (\plotStartX + \x * \labelDist, \plotStartY+0.25mm) {\tickFont \xLabel};
        \ifnum\x<3
            \foreach \logX in {0,...,9} {
                \pgfmathsetlengthmacro{\logLabelDist}{
                    \labelDist * ln(\logX + 1) / ln(10)
                }
                \ifnum\logX=0
                    \def\tickLen{\majorTick}
                \else
                    \def\tickLen{\minorTick}
                \fi
                \draw[line width=\tickWidth] (\plotStartX + \x * \labelDist + \logLabelDist,\plotStartY) -- (\plotStartX + \x * \labelDist + \logLabelDist ,\plotStartY - \tickLen);
                
            }
        \else
            \draw[line width=\tickWidth] (\plotStartX + \x * \labelDist,\plotStartY) -- (\plotStartX + \x * \labelDist ,\plotStartY - \majorTick);
        \fi
    }
    \node[anchor=north, yshift=-3mm] at (\plotStartX + 0.5 * \figWidth, \plotStartY) {\labelFont $\alpha^{\tinySuper{(\mathrm{IPD})}}$};

    \pgfmathsetlengthmacro{\labelDist}{
        \figHeight / 4
    }
    \foreach \y / \yLabel in {
        0/1.85,1/1.9,2/1.95,3/2.0,4/2.05
    } {
        \node[anchor=east] at (\plotStartX,\plotStartY + \y * \labelDist) {\tickFont \yLabel};
        \draw[line width=\tickWidth] (\plotStartX,\plotStartY + \y * \labelDist) -- (\plotStartX- \majorTick,\plotStartY + \y * \labelDist);
    }
    \node[rotate=90, anchor=south, yshift=4.5mm] at (\plotStartX, \plotStartY + 0.5 * \figHeight) {\labelFont \textcolor{tab_brown}{PESQ}\,$\rightarrow$};

    \pgfmathsetlengthmacro{\labelDist}{
        \figHeight / 4
    }
    \foreach \y / \yLabel in {
        0/4,1/6,2/8,3/10,4/12
    } {
        \node[anchor=west] at (\plotStartX+ \figWidth,\plotStartY + \y * \labelDist) {\tickFont \yLabel};
        \draw[line width=\tickWidth] (\plotStartX+ \figWidth,\plotStartY + \y * \labelDist) -- (\plotStartX+\majorTick+ \figWidth,\plotStartY + \y * \labelDist);
        
    }
    \node[rotate=90, anchor=north, yshift=-3mm] at (\plotStartX + \figWidth, \plotStartY + 0.5 * \figHeight) {\labelFont \textcolor{tab_pink}{MAE}\,[°]\,$\leftarrow$};

\end{tikzpicture}
    \vspace{-16pt} %
  \end{minipage}
}

\caption{Closed-loop (AR) parameter optimization on the validation set using the Wrapped\,KF for TST and \mbox{SpatialNet}-MIMO as SSF (MIMO-AR, \cref{fig:weak_ssf_mimo}).}
\label{fig:hyperparameter_tuning}
\vspace{0pt}
\end{figure}

\textbf{\textit{Fine-tuning}} 
After pretraining, we fine-tune \mbox{SpatialNet} with the \ac{doa} estimates of the weakly guided Bayesian \ac{tst} algorithms.
Fine-tuning continues with the reduced learning rate of 10$^{-4}$ and lasts for 20 additional epochs.
To avoid the inherent non-parallelizability of the \ac{ar} \ac{tse} pipelines, we adapt the pseudo-\ac{ar} training strategy used in~\cite{shen25arise,wang17recurrent_deep_stacking} for our setup. 
Specifically, \diffComm{non-\ac{ar} (open-loop) tracking results, which are based on noisy measurements at training start and later incorporate enhanced speech, are stored during each epoch and used for \ac{ssf} guidance in the following, thereby emulating \ac{ar} (closed-loop) inference.
Although the \ac{ar} tracking algorithms are inherently dependent on the \ac{ssf} performance, which evolves during fine-tuning, fixing their parameters after the first epoch and then performing a final parameter sweep proved sufficient, see \cref{fig:kalman_parameter_finetuning}.}{in each fine-tuning epoch, we load cached \ac{doa} estimates from the previous epoch, which are initialized from noisy measurements at training start and later incorporate enhanced speech.
We use these trajectories to train the \ac{ssf} fully parallelized and then leverage the resulting processed speech signals to compute open-loop (non-AR) \ac{doa} estimates, which we store for the next epoch.
Since the \ac{ssf} is optimized with degraded tracking results based on processed speech signals, the pseudo-AR training strategy emulates closed-loop error propagation.
While the \ac{ar} tracking algorithms are inherently dependent on the \ac{ssf} performance, we found that their optimal parameterization only varies slightly throughout fine-tuning, as shown in 
\cref{fig:kalman_parameter_finetuning}.
As a result, we limit the number of \ac{tst} optimization steps to a parameter sweep after the first and the final fine-tuning epoch, both conducted on the validation subset.
Note that the final parameter sweep solely affects inference.}
To preserve spatial cues in the estimates of \mbox{SpatialNet}-\ac{mimo}, we incorporate the \ac{ipd} loss $\mathcal{L}^{\tinySuper{(\mathrm{IPD})}}$~\cite{battula25MIMO_localization}
\begin{equation}\label{eq:mimo_ar_loss}
    \mathcal{L}^{\tinySuper{(\mathrm{MIMO\!-\!AR})}}(\mathbf{s}, \hat{\mathbf{s}}) = \alpha^{\tinySuper{(\mathrm{IPD})}} \mathcal{L}^{\tinySuper{(\mathrm{IPD})}}(\mathbf{s}, \hat{\mathbf{s}})+ \mathcal{L}^{\tinySuper{(\mathrm{MIMO})}}(\mathbf{s}, \hat{\mathbf{s}}) \, ,
\end{equation}
with $\alpha^{\tinySuper{(\mathrm{IPD})}}$ balancing both optimization objectives.
Fine-tuning \mbox{SpatialNet}-MIMO for different values of  $\alpha^{\tinySuper{(\mathrm{IPD})}}$, as shown in \cref{fig:ipd_alpha}, proves how stronger spatial cue preservation consistently improves tracking in terms of \ac{mae}. 
However, increasing $\alpha^{\tinySuper{(\mathrm{IPD})}}$ de-emphasizes the signal reconstruction loss $\mathcal{L}^{\tinySuper{(\mathrm{MIMO})}}$ in \cref{eq:mimo_ar_loss}, resulting in a \diffComm{\textit{tradeoff}}{\textit{trade-off}} between more precise guidance and \ac{ssf} performance\diffComm{}{,} with a distinct optimum for closed-loop enhancement (PESQ).

\section{Evaluation}\label{sec:evaluation}
During evaluation, we utilize our synthetic dataset from \cref{sec:dataset} together with real-world recordings to provide a detailed analysis in a controlled acoustic scenario as well as test generalization capabilities to unseen acoustic conditions.

\subsection{Spatially Guided Extraction of Moving Speakers}\label{sec:synthetic_eval}
\Cref{tab:results} \diffComm{summarizes}{lists} the results of all presented \acl{tse} (\acs{tse}) pipelines using \mbox{SpatialNet} as \acl{ssf} (\acs{ssf}).
With the synthetic dataset \newdiffComm{availing}{providing} ground truth speaker trajectories and speech signals, we employ intrusive metrics during evaluation.
Specifically, we report utterance-wise \ac{mae} and \ac{acc} with a 10° threshold \cite{wang2023fnssl_full_band_narrow_band_tracking,xiao25tf_mamba_ssl} to assess tracking performance as well as \diffComm{\mbox{PESQ}~\cite{rix01pesq} and \mbox{ESTOI}~\cite{jensen16estoi} as measures for perceptual speech quality and intelligibility}{\mbox{SI-SDR}~\cite{roux19sisdr_half_baked}, \mbox{PESQ}~\cite{rix01pesq} and \mbox{ESTOI}~\cite{jensen16estoi} as measures for speech signal distortion, perceptual quality, and intelligibility}, respectively.

Under strong guidance (oracle \ac{doa}), the \ac{mimo} extension of \mbox{SpatialNet}~\diffComm{(2)}{(\picLegend{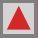})} shows only negligible degradation in \diffComm{}{interference suppression and} intelligibility while matching the perceptual quality of the initial \ac{miso} implementation~\diffComm{(1)}{(\picLegend{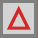})}, resulting in comparable starting conditions across all weakly guided methods following pretraining.
After subsequent fine-tuning with the Bayesian trackers from \cref{sec:weak_tse}, 
the enhancement performance in the concatenative \ac{tse} pipeline (Concat, \cref{fig:weak_ssf}) drops significantly due to imprecise guidance \diffComm{}{across all metrics}, with the more accurate \mbox{Bootstrap\,PF}~\diffComm{(6)}{(\picLegend{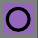})} outperforming the \mbox{Wrapped\,KF}~\diffComm{(3)}{(\picLegend{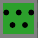})}.
With a \ac{mae} above 30°, 
the latter performs particularly poorly, reflecting the limited modeling capacity of the \diffComm{\ac{kf}}{\ac{kf}'s linear-Gaussian state-space on top of its \newdiffComm{bandwidth-constraint}{bandwidth constraint} due to spatial aliasing}.
\diffComm{}{By enforcing a linear-Gaussian state-space, the likelihood cannot account for the multi-modality due to overlapping speakers.
Combined with the reduced bandwidth due to spatial aliasing, the \ac{doa} estimates $\Phi_t$ in \cref{eq:doa_estimator} intrinsically bias the \mbox{Wrapped\,KF} towards the dominant speaker in the lower frequency band.}

\diffComm{}{To quantify the impact of imprecise steering on the \ac{ssf} enhancement performance, we train \mbox{SpatialNet} from scratch solely using the starting \ac{doa} as directional conditioning, i.e. $\hat{\theta}_t = \theta_0$,} 
\newdiffComm{to act as a lower performance bound for all tracking based methods}{and thereby provide a baseline without explicit tracking}~(\picLegend{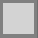}). \newdiffComm{}{For a fair comparison, we train \mbox{SpatialNet} for 70 epochs, which matches the combined pretraining and fine-tuning optimization budget of the tracking-based methods and use the corresponding dataset and learning rate schedule.}
\diffComm{}{A comparison with (\picLegend{images/tst_accuracy/wrapped_kf.pdf}) demonstrates how the \ac{doa} inaccuracies of the \mbox{Wrapped\,KF} are indeed so severe that there are \newdiffComm{no significant}{only marginal} enhancement improvements towards no explicit tracking~(\picLegend{images/tst_accuracy/unguided.pdf}).}
\diffComm{By}{However, by} autoregressively incorporating the processed speech into the filtering formulations (MISO-AR, \cref{fig:weak_ssf_miso}), \diffComm{spurious modes of interfering speakers can be suppressed in the underlying statistical models, increasing robustness for both Bayesian trackers (5,\,7)}{the spatial cues of the interfering speakers can be effectively suppressed, increasing the accuracy of both Bayesian trackers (\picLegend{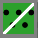},\,\picLegend{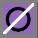})}. 
This becomes especially evident for closely spaced or crossing speakers, as shown in the example trajectories in \cref{fig:synthetic_trajectories} and on our project page\footnote{\label{project_page}\projectPage}.
Incorporating the multichannel estimates of \mbox{SpatialNet} (MIMO-AR, \cref{fig:weak_ssf_mimo}) can further amplify this effect, achieving superior tracking and enhancement for the \mbox{Wrapped\,KF}~\diffComm{(5)}{(\picLegend{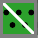})}.

\def\tableW{5pt}
\begin{table}[t!]
 \caption{
 Concatenative and autoregressive (AR) extraction pipelines
}
\label{tab:results}
\resizebox{\linewidth}{!}{
\renewcommand{\arraystretch}{0.9}
\footnotesize
\begin{tabular}{
c@{\hspace{\tableW}}l@{\hspace{1pt}}c@{\hspace{3pt}}c@{\hspace{\tableW}}c@{\hspace{-5pt}}c@{\hspace{0pt}}c@{\hspace{-5pt}}c@{\hspace{-5pt}}c
} %
 \toprule[1.5pt]
 & \multicolumn{3}{c}{\hspace{-1mm}\textbf{Extraction Method}} & \multicolumn{2}{c}{\hspace{-1mm}\textbf{Tracking Results}} & \multicolumn{3}{c}{\hspace{-1mm}\textbf{Enhancement Results}} \\[1pt]
 \cmidrule(l{ -0.25mm }r){2-4} \cmidrule(l{ -0.25mm }r){5-6} \cmidrule(l{ -0.25mm }r){7-9} 
\textbf{ID} & Tracking & MIMO & AR & ACC\,[\%]\,$\uparrow$ & \acs{mae}\,[°]\,$\downarrow$ & \acs{sisdr}\,[dB]\,$\uparrow$ & \acs{pesq}\,$\uparrow$ & \acs{estoi}\,[\%]\,$\uparrow$ \\ \midrule
$-$ & $-$ & $-$ & $-$ & $-$ & $-$ & \hspace*{-1mm}-6.67\myPM.07 & 1.10\myPM.06 & 41.8\myPM.3 \\[-1pt] \cmidrule(lr){1-9}
\picTable{images/tst_accuracy/strong.pdf} & Oracle & \textcolor{red}{\ding{55}} & $-$ & $-$ & $-$ & \textbf{9.29\myPM.07} & \textbf{2.14\myPM.01} & \textbf{81.6\myPM.2} \\
\picTable{images/tst_accuracy/strong_mimo.pdf} & Oracle & \textcolor{green}{\ding{51}} & $-$ & $-$ & $-$ & 9.22\myPM.07 & \textbf{2.14\myPM.01} & 81.4\myPM.2 \\[-1pt] \cmidrule(lr){1-9}
\picTable{images/tst_accuracy/unguided.pdf} & Unguided\,$^\dagger$ & \textcolor{red}{\ding{55}} & $-$ & 19.7\myPM.4 & 62.72\myPM.72 & 7.74\myPM.07 & 1.89\myPM.01 & 77.3\myPM.2 \\[-1pt] \cmidrule(lr){1-9}
\picTable{images/tst_accuracy/wrapped_kf.pdf} & Wrapped\,\acs{kf} & \textcolor{red}{\ding{55}} & \textcolor{red}{\ding{55}} & 33.2\myPM.3 & 32.67\myPM.36 & \hspace{\sigSymbolOff}7.75\myPM.06$^\sigSymbol$ & \hspace{\sigSymbolOff}1.89\myPM.01$^\sigSymbol$ & \hspace{\sigSymbolOff}77.6\myPM.2$^\sigSymbol$ \\
\picTable{images/tst_accuracy/wrapped_kf_miso_ar.pdf} & Wrapped\,\acs{kf} & \textcolor{red}{\ding{55}} & \textcolor{green}{\ding{51}} & \hspace{\sigSymbolOffTwo}47.1\myPM.3$^{\sigSymbolTwo}$ & \hspace{\sigSymbolOffTwo}17.94\myPM.27$^{\sigSymbolTwo}$ & \hspace{\sigSymbolOffTwo}7.99\myPM.07$^{\sigSymbolTwo}$ & \hspace{\sigSymbolOffTwo}1.94\myPM.01$^{\sigSymbolTwo}$ & \hspace{\sigSymbolOffTwo}78.3\myPM.2$^{\sigSymbolTwo}$ \\
\picTable{images/tst_accuracy/wrapped_kf_mimo_ar.pdf} & Wrapped\,\acs{kf} & \textcolor{green}{\ding{51}} & \textcolor{green}{\ding{51}} & \hspace{\sigSymbolOffTwo}\textbf{86.4\myPM.3}$^{\sigSymbolTwo}$ & \hspace{\sigSymbolOffTwo}\phantom{0}\textbf{6.65\myPM.18}$^{\sigSymbolTwo}$ & \hspace{\sigSymbolOffTwo}\textbf{8.53\myPM.07}$^{\sigSymbolTwo}$ & \hspace{\sigSymbolOffTwo}\textbf{1.98\myPM.01}$^{\sigSymbolTwo}$ & \hspace{\sigSymbolOffTwo}\textbf{79.8\myPM.2}$^{\sigSymbolTwo}$ \\[-1pt] \cmidrule(lr){1-9}
\picTable{images/tst_accuracy/bootstrap_pf.pdf} & Bootstrap\,\acs{pf} & \textcolor{red}{\ding{55}} & \textcolor{red}{\ding{55}} & 56.2\myPM.4 & 21.65\myPM.37 & \hspace{\sigSymbolOff}7.93\myPM.06$^\sigSymbol$ & \hspace{\sigSymbolOff}1.93\myPM.01$^\sigSymbol$ & \hspace{\sigSymbolOff}78.2\myPM.2$^\sigSymbol$ \\
\picTable{images/tst_accuracy/bootstrap_pf_miso_ar.pdf} & Bootstrap\,\acs{pf} & \textcolor{red}{\ding{55}} & \textcolor{green}{\ding{51}} & \hspace{\sigSymbolOffTwo}\textbf{87.6\myPM.3}$^{\sigSymbolTwo}$ & \hspace{\sigSymbolOffTwo}\phantom{0}\textbf{6.47\myPM.19}$^{\sigSymbolTwo}$ & \hspace{\sigSymbolOffTwo}\textbf{8.71\myPM.07}$^{\sigSymbolTwo}$ & \hspace{\sigSymbolOffTwo}\textbf{2.04\myPM.01}$^{\sigSymbolTwo}$ & \hspace{\sigSymbolOffTwo}\textbf{80.4\myPM.2}$^{\sigSymbolTwo}$ \\
\picTable{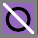} & Bootstrap\,\acs{pf} & \textcolor{green}{\ding{51}} & \textcolor{green}{\ding{51}} & \hspace{\sigSymbolOffTwo}86.6\myPM.4$^{\sigSymbolTwo}$ & \hspace{\sigSymbolOffTwo}\phantom{0}8.07\myPM.25$^{\sigSymbolTwo}$ & \hspace{\sigSymbolOffTwo}8.59\myPM.06$^{\sigSymbolTwo}$ & \hspace{\sigSymbolOffTwo}1.98\myPM.01$^{\sigSymbolTwo}$ & \hspace{\sigSymbolOffTwo}79.9\myPM.2$^{\sigSymbolTwo}$ \\
 \bottomrule[1.5pt]
\end{tabular}
}

\vspace{1mm}
\footnotesize
{%
\spaceskip=0.9\fontdimen2\font
  plus 0.5\fontdimen3\font
  minus 1.5\fontdimen4{2mm} Reported values are sample means with 95\% confidence intervals%
\newdiffComm{}{, computed utterance-wise across the 3000 test mixtures using Student's $t$-distribution. Statistical significance is assessed via paired $t$-tests at a 5\% significance level.}%
} \\

\vspace{-8pt}
\hspace{1.5mm} \newdiffComm{}{$^\sigSymbol$Significantly worse compared to oracle tracking~(\picLegend{images/tst_accuracy/strong.pdf}).} \\

\vspace{-8pt}
\hspace{0.125mm} \newdiffComm{}{$^{\sigSymbolTwo}$Significantly better than with the corresponding non-AR tracker~(\picLegend{images/tst_accuracy/wrapped_kf.pdf},\,\picLegend{images/tst_accuracy/bootstrap_pf.pdf}).} \\

\vspace{-8pt}
\hspace{1.5mm} $^\dagger$Utilizes the initialization direction $\theta_0$ as \ac{doa} estimate $\hat{\theta}_t = \theta_0$.

\end{table}

Nevertheless, \diffComm{the \ac{ssf} guided by our proposed MISO-AR formulation of the \mbox{Bootstrap\,PF}~(7) in \cref{alg:bootstrap_filter} achieves the best performance overall, 
emphasizing the potential of \textit{accurate guidance} without enforcing spatial cue preservation.}
{as spatial cue preservation degrades enhancement (see \cref{sec:optimization_details} and \cref{fig:ipd_alpha}), the \ac{mimo}-\ac{ssf} only benefits tracking algorithms that cannot incorporate the processed speech sufficiently. 
While this is the case for the \mbox{Wrapped\,KF}, which utilizes only a narrow bandwidth of the target speech, the improved modeling capacity of the \mbox{Bootstrap\,PF} achieves accurate tracking without spatial cue preservation.
Thus, \textit{without \ac{ssf} modification}, the MISO-AR configuration with the \mbox{Bootstrap\,PF}~(\picLegend{images/tst_accuracy/bootstrap_pf_miso_ar.pdf}) yields the best overall enhancement\newdiffComm{.}{, significantly outperforming its non-AR counterpart~(\picLegend{images/tst_accuracy/bootstrap_pf.pdf}).}
}

\begin{figure}[t!]
\vspace{-3pt}
\hspace{-3mm}\begin{tikzpicture}

\newcommand\figSize{1.6cm}
\pgfmathsetlengthmacro{\figWidth}{\figSize * 3.625 / 2.5}  
\pgfmathsetlengthmacro{\figDist}{4.25mm} 
\pgfmathsetlengthmacro{\figHeight}{\figSize * 0.9}
\newcommand\plotStartX{0mm}
\newcommand\plotStartY{0mm}
\colorlet{trainingColor}{red!70}
\colorlet{testColor}{blue!70}
\newcommand\rowSplitSkip{-0.75mm}
\node[
    anchor=south east,
    fill=white,
    font=\scriptsize,
    xshift=-4.5mm,  
    yshift=-9.5mm 
] at (\plotStartX,\plotStartY) {
\setlength{\tabcolsep}{2pt} 
\renewcommand{\arraystretch}{1.2} 
\begin{tabular}{c l c}
\textbf{ID} & \textbf{Tracking} & \textbf{MACs} \\
\hline
\picTable{images/tst_accuracy/wrapped_kf.pdf},\,\picTable{images/tst_accuracy/wrapped_kf_miso_ar.pdf} & Wrapped\,KF & 0.3\,[M/s] \\[\rowSplitSkip]
\picTable{images/tst_accuracy/wrapped_kf_mimo_ar.pdf} & +MIMO-SSF & 1.8\,[M/s] \\
\picTable{images/tst_accuracy/bootstrap_pf.pdf},\,\picTable{images/tst_accuracy/bootstrap_pf_miso_ar.pdf} & Bootstrap\,PF & 2.5\,[M/s] \\[\rowSplitSkip]
\picTable{images/tst_accuracy/bootstrap_pf_mimo_ar.pdf} & +MIMO-SSF & 8.7\,[M/s] \\
\picTable{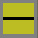},\,\picTable{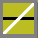} & SELDnet & 70\,[M/s] \\[\rowSplitSkip]
\picTable{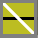} & +MIMO-SSF & 76\,[M/s] \\
\picTable{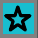},\,\picTable{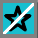} & CNN/LSTM & \hspace{-1mm}0.8,\,1.1\,[G/s]\hspace{-2.5mm} \\[\rowSplitSkip]
\picTable{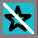} & +MIMO-SSF & 0.8\,[G/s] \\
\end{tabular}
};

\foreach \plotIdx/\metricLabel in {%
    0/mae,%
    1/acc_10%
}{
    \node[anchor=center] at (\plotStartX + 0.5 * \figWidth + \plotIdx * \figWidth + \plotIdx * \figDist, \plotStartY + 0.5 * \figHeight) {%
        \pgfimage[height=\figHeight, width=\figWidth]{images/tst_accuracy/tst_\metricLabel.pdf} 
    };

    \foreach \xRel / \xLabel in {
        0.19/Concat,0.51/MISO-AR,0.83/MIMO-AR
    } {
        \node[anchor=east, rotate=30, yshift=-1mm, xshift=0.5mm] at (\plotStartX + \xRel * \figWidth + \plotIdx * \figWidth + \plotIdx * \figDist,\plotStartY) {\tickFont \xLabel};
        \draw[line width=\tickWidth] (\plotStartX + \xRel * \figWidth + \plotIdx * \figWidth + \plotIdx * \figDist,\plotStartY) -- (\plotStartX + \xRel * \figWidth+ \plotIdx * \figWidth + \plotIdx * \figDist,\plotStartY- \majorTick);
    }

    \pgfmathsetlengthmacro{\labelDist}{
        \figHeight / 7
    }
    \foreach \y in {%
        0,...,7%
    } {%
        \pgfmathtruncatemacro{\tmp}{int(\y / 2) * 2}
        \ifnum\plotIdx=0
            \pgfmathsetmacro{\yLabel}{int(\y * 5)}
            \ifnum\tmp=\y
                \def\tickLen{\majorTick}
                \node[anchor=east] at (\plotStartX + \plotIdx * \figWidth + \plotIdx * \figDist,\plotStartY + \y * \labelDist) {\tickFont \yLabel};
            \else
                \def\tickLen{\minorTick}
            \fi
        \else
            \pgfmathsetmacro{\yLabel}{int(\y * 10 + 30)}
            \ifnum\tmp=\y
                \def\tickLen{\minorTick}
            \else
                \def\tickLen{\majorTick}
                \node[anchor=east] at (\plotStartX + \plotIdx * \figWidth + \plotIdx * \figDist,\plotStartY + \y * \labelDist) {\tickFont \yLabel};
            \fi
        \fi
        
        \draw[line width=\tickWidth] (\plotStartX + \plotIdx * \figWidth + \plotIdx * \figDist,\plotStartY + \y * \labelDist) -- (\plotStartX +\plotIdx * \figWidth + \plotIdx * \figDist- \tickLen,\plotStartY + \y * \labelDist);
    }
    \ifnum\plotIdx=0
    \node[anchor=south, yshift=0mm, xshift=1.5mm] at (\plotStartX + \plotIdx * \figWidth + \plotIdx * \figDist,\plotStartY + \figHeight) {\labelFont MAE\,[°]\,$\downarrow$};
    \else 
    \node[anchor=south, yshift=0mm, xshift=1.5mm] at (\plotStartX + \plotIdx * \figWidth + \plotIdx * \figDist,\plotStartY + \figHeight) {\labelFont ACC\,[\%]\,$\uparrow$};
    \fi
}

\end{tikzpicture}
\vspace*{-20pt}
\caption{
Performance (MAE, ACC) and computational cost (MACs) of Bayesian filters (\mbox{Wrapped\,KF}, \mbox{Bootstrap\,PF}) relative to \acp{dnn} (\mbox{SELDnet}, \mbox{CNN/LSTM}) for tracking a single speaker.
MACs account for the additional cost of the modified spatial filter (MIMO-SSF) for MIMO-AR tracking.
}
\label{fig:tst_acc}
\vspace*{-5pt}
\end{figure}

\subsection{Comparison with Deep Neural Tracking Methods}\label{sec:dnn_tracking}
To contextualize the performance of the Bayesian filters, we compare against data-driven methods for \acl{tst} (\acs{tst}).
As a strong reference, we use the \mbox{CNN/LSTM} architecture from our prior work in \cite{kienegger25wg_ssf}, which we adapted from \cite{bohlender21ssl_temporal_context}.
Additionally, we include \mbox{SELDnet}~\cite{adavanne19seldnet} as a low-complexity baseline, following \cite{kienegger25deep_joint_ar_ssf,xiao25tf_mamba_ssl}.
To adapt it for our setup, we incorporate the modifications for causality according to~\cite{yasuda24causal_seldnet_6dof}, condition the GRU layers of \mbox{SELDnet} with the initial \ac{doa}~\cite{tesch24ssf_journal,kienegger25wg_ssf} and solely use the \mbox{GCC-PHAT} input features \cite{adavanne19seldnet} due to the array's compact size. 
\Cref{fig:tst_acc} presents the \diffComm{tracking}{} performance and computational complexity of all Bayesian and data-driven tracking methods.
\diffComm{}{For a fair comparison, we account for the additional computational cost of the \ac{mimo}-\ac{ssf} modifications, which is less for the \mbox{Wrapped\,KF} due to its limited bandwidth of 2\,kHz.}
In their original formulation (Concat), the Bayesian filters \mbox{Wrapped\,KF}~(\picLegend{images/tst_accuracy/wrapped_kf.pdf}) and \mbox{Bootstrap\,PF}~(\picLegend{images/tst_accuracy/bootstrap_pf.pdf}) are greatly outperformed by the neural trackers~(\picLegend{images/tst_accuracy/seldnet.pdf},\,\picLegend{images/tst_accuracy/cnn_lstm.pdf}).
However, autoregressively incorporating \mbox{SpatialNet} as \ac{ssf} into our proposed reformulation of the \mbox{Bootstrap\,PF} (MISO-AR), (\picLegend{images/tst_accuracy/bootstrap_pf_miso_ar.pdf})  as well as for both Bayesian filters with the \ac{ssf}-\ac{mimo} extension (MIMO-AR), (\picLegend{images/tst_accuracy/wrapped_kf_mimo_ar.pdf},\,\picLegend{images/tst_accuracy/bootstrap_pf_mimo_ar.pdf}) achieves \diffComm{competetive}{competitive} performance to the data-driven methods \diffComm{}{without \ac{ar}}~(\picLegend{images/tst_accuracy/seldnet.pdf},\,\picLegend{images/tst_accuracy/cnn_lstm.pdf}). 
Most notably, our model-based \mbox{Bootstrap\,PF} in MISO-AR configuration~(\picLegend{images/tst_accuracy/bootstrap_pf_miso_ar.pdf})
consistently outperforms \diffComm{}{non-\ac{ar}} \mbox{SELDnet}~(\picLegend{images/tst_accuracy/seldnet.pdf})\newdiffComm{ at less than a \textit{tenth} of the computational cost.}{, while requiring no supervised \ac{doa} training, using an interpretable state-space formulation, and less than a tenth of the tracking complexity.}

\diffComm{}{As an upper bound, we also include \ac{ar} formulations of the data-driven trackers\newdiffComm{, for which we follow our previous work in~\cite{kienegger25deep_joint_ar_ssf} and modify the pseudo-\ac{ar} fine-tuning to train both tracking and enhancement \acp{dnn} with a dual-optimization strategy.}{.
In contrast to our Bayesian filters, this requires adjusting the pseudo-\ac{ar} fine-tuning stage in \cref{sec:optimization_details} to jointly train tracking and enhancement \acp{dnn}, for which we employ the dual-optimization strategy from our previous work~\cite{kienegger25deep_joint_ar_ssf}.}
Regarding the \mbox{MISO-AR} configuration, we incorporate the enhanced speech as weighting factor into the GCC-PHAT in \mbox{SELDnet}, (see, e.g.,~\cite{li23gcc-speaker}) and employ input concatenation for \mbox{CNN/LSTM}. 
While for \mbox{SELDnet} both \ac{ar} formulations yield similar improvements (\picLegend{images/tst_accuracy/seldnet_miso_ar.pdf},\,\picLegend{images/tst_accuracy/seldnet_mimo_ar.pdf}),  \mbox{CNN/LSTM} in \mbox{MISO-AR} configuration~(\picLegend{images/tst_accuracy/cnn_lstm_miso_ar.pdf})  achieves the best overall tracking. 
In addition to a higher complexity, access to both unprocessed and processed speech allows the \ac{dnn} to \textit{compensate} for processing degradation of the \ac{ssf}.}

\subsection{Influence of Speaker Motion Patterns on Enhancement}\label{sec:locata_eval}
While prior work demonstrated the necessity of training a deep \ac{ssf} with moving speakers for robust speech enhancement under dynamic conditions~\cite{kienegger25wg_ssf}, the role of the motion patterns remains unclear.
For further analysis, we cross-evaluate \mbox{SpatialNet} as strongly guided \ac{ssf} trained on different speaker trajectories. 
Specifically, we retain the acoustic setup of \cref{sec:dataset_parametrization} while varying speaker motion between stationary, circular~\cite{kienegger25wg_ssf,kienegger25sg_ssf}, and our social force model (\cref{sec:social_force_model}).
As a benchmark, we use real-world trajectories from Task~4 of the \mbox{LOCATA} Challenge~\cite{evers20locata_challenge} with a stationary array and two moving speakers.
\Cref{fig:locata_trajectories} presents the performance results in terms of perceptual quality~(\acs{pesq}) and intelligibility~(\acs{estoi}).
Due to the same span of room dimensions, the distribution of speaker-array distances varies throughout datasets, yielding different input \acp{snr} \cite{roux19sisdr_half_baked}, ranging from \mbox{-5.7\,dB} (circular) to \mbox{-7.6\,dB} (\mbox{LOCATA}).
As expected, \mbox{SpatialNet} trained on stationary speakers performs poorly across all motion types.
However, due to constant speaker-array distances, also the circular dataset results in \newdiffComm{significant}{substantial} enhancement degradation when evaluated on other movement patterns. 
Only \mbox{SpatialNet} trained on our proposed social force model remains robust over all datasets while achieving a 0.3~\acs{pesq} gain on the \mbox{LOCATA} trajectories, 
underlining the importance of \textit{motion diversity}.

\begin{figure}[t!]
\begin{tikzpicture}

\newcommand\figSize{1.5cm}
\pgfmathsetlengthmacro{\figWidth}{\figSize * 1.4}
\pgfmathsetlengthmacro{\figDist}{\figWidth * 1.175} 
\pgfmathsetlengthmacro{\figHeight}{\figSize * 0.9}
\newcommand\plotStartX{0mm}
\newcommand\plotStartY{0mm}
\colorlet{trainingColor}{blue!8}
\colorlet{testColor}{red!8} 

\node[anchor=north, yshift=-2mm, xshift=-5mm] (testnode) at (\plotStartX + \figWidth+ 0.5 * \figDist,\plotStartY) {\phantom{\labelFont test dataset}};
\def\testNodeWidthL{2.75cm}
\def\testNodeWidthR{2.75cm}
\def\testNodeHeight{1.5mm}
\fill[
    rounded corners=2pt,
    fill=testColor
]
($(testnode.north west)+(-\testNodeWidthL,\testNodeHeight)$)
rectangle
($(testnode.south east)+(\testNodeWidthR,-\testNodeHeight)$);
\node[] at (testnode) {\labelFont test dataset};

\foreach \plotIdx/\metricName in {%
    0/pesq,%
    1/estoi%
}{
    \node[anchor=center] at (\plotStartX + 0.5 * \figWidth + \plotIdx * \figWidth + \plotIdx * \figDist, \plotStartY + 0.5 * \figHeight) {%
        \pgfimage[height=\figHeight, width=\figWidth]{images/locata_trajectories/\metricName_heatmap.pdf} 
    };

    \node[anchor=south east, yshift=-1mm] (trainnode) at (\plotStartX + \plotIdx * \figWidth + \plotIdx * \figDist,\plotStartY + \figHeight) {\phantom{\labelFont training}};
    \def\trainNodeWidthL{3mm}
    \def\trainNodeWidthR{-0.5mm}
    \def\trainNodeHeightTop{-1mm}
    \def\trainNodeHeightBottom{1.25cm}
    \fill[
        rounded corners=2pt,
        fill=trainingColor
    ]
    ($(trainnode.north west)+(-\trainNodeWidthL,\trainNodeHeightTop)$)
    rectangle
    ($(trainnode.south east)+(\trainNodeWidthR,-\trainNodeHeightBottom)$);
    \node[] at (trainnode) {\labelFont training};
    
    \pgfmathsetlengthmacro{\effFigWidth}{
         \figWidth * 0.92
    }
    \pgfmathsetlengthmacro{\labelDist}{
        \effFigWidth / 4
    }
    \foreach \x / \xLabel in {
        0/stationary,1/circular,2/proposed,3/LOCATA
    } {
        \node[anchor=north east, rotate=30] at (\plotStartX + \x * \labelDist + 0.5 * \labelDist+ \plotIdx * \figWidth + \plotIdx * \figDist,\plotStartY + 0.8mm) {\tickFont \xLabel};
        \draw[line width=\tickWidth] (\plotStartX + \x * \labelDist + 0.5 * \labelDist+ \plotIdx * \figWidth + \plotIdx * \figDist,\plotStartY) -- (\x * \labelDist + 0.5 * \labelDist+ \plotIdx * \figWidth + \plotIdx * \figDist,\plotStartY- \majorTick);
    }

    \pgfmathsetlengthmacro{\labelDist}{
        \figHeight / 4
    }
    \foreach \y/\yLabel in {%
        0/proposed,1/circular,2/stationary,3/unprocessed%
    } {%
        \node[anchor=east] at (\plotStartX + \plotIdx * \figWidth + \plotIdx * \figDist,\plotStartY + \y * \labelDist+ 0.5 * \labelDist) {\tickFont \yLabel};
        \draw[line width=\tickWidth] (\plotStartX + \plotIdx * \figWidth + \plotIdx * \figDist,\plotStartY + \y * \labelDist+ 0.5 * \labelDist) -- (\plotStartX +\plotIdx * \figWidth + \plotIdx * \figDist- \majorTick,\plotStartY + \y * \labelDist+ 0.5 * \labelDist);
    }

}

\pgfmathsetlengthmacro{\overlayXstart}{
    \plotStartX + 2.4mm
}
\pgfmathsetlengthmacro{\overlayYstart}{
    \plotStartY + \figHeight - 1.6mm
}
\pgfmathsetlengthmacro{\overlayXdelta}{4.8mm}   
\pgfmathsetlengthmacro{\overlayYdelta}{3.4mm}   

\node[font=\tiny, text=white] at ({\overlayXstart + 0*\overlayXdelta},{\overlayYstart - 0*\overlayYdelta}) {1.09};
\node[font=\tiny, text=white] at ({\overlayXstart + 1*\overlayXdelta},{\overlayYstart - 0*\overlayYdelta}) {1.10};
\node[font=\tiny, text=white] at ({\overlayXstart + 2*\overlayXdelta},{\overlayYstart - 0*\overlayYdelta}) {1.10};
\node[font=\tiny, text=white] at ({\overlayXstart + 3*\overlayXdelta},{\overlayYstart - 0*\overlayYdelta}) {1.10};

\node[font=\tiny] at ({\overlayXstart + 0*\overlayXdelta},{\overlayYstart - 1*\overlayYdelta}) {\textbf{2.16}};
\node[font=\tiny] at ({\overlayXstart + 1*\overlayXdelta},{\overlayYstart - 1*\overlayYdelta}) {1.63};
\node[font=\tiny, text=white] at ({\overlayXstart + 2*\overlayXdelta},{\overlayYstart - 1*\overlayYdelta}) {1.50};
\node[font=\tiny] at ({\overlayXstart + 3*\overlayXdelta},{\overlayYstart - 1*\overlayYdelta}) {1.70};

\node[font=\tiny] at ({\overlayXstart + 0*\overlayXdelta},{\overlayYstart - 2*\overlayYdelta}) {2.09};
\node[font=\tiny] at ({\overlayXstart + 1*\overlayXdelta},{\overlayYstart - 2*\overlayYdelta}) {\textbf{2.27}};
\node[font=\tiny] at ({\overlayXstart + 2*\overlayXdelta},{\overlayYstart - 2*\overlayYdelta}) {1.68};
\node[font=\tiny] at ({\overlayXstart + 3*\overlayXdelta},{\overlayYstart - 2*\overlayYdelta}) {1.67};

\node[font=\tiny] at ({\overlayXstart + 0*\overlayXdelta},{\overlayYstart - 3*\overlayYdelta}) {2.05};
\node[font=\tiny] at ({\overlayXstart + 1*\overlayXdelta},{\overlayYstart - 3*\overlayYdelta}) {2.08};
\node[font=\tiny] at ({\overlayXstart + 2*\overlayXdelta},{\overlayYstart - 3*\overlayYdelta}) {\textbf{2.14}};
\node[font=\tiny] at ({\overlayXstart + 3*\overlayXdelta},{\overlayYstart - 3*\overlayYdelta}) {\textbf{2.02}};

\pgfmathsetlengthmacro{\overlayXstart}{
    \plotStartX + 2.4mm + \figWidth + \figDist
}

\node[font=\tiny, text=white] at ({\overlayXstart + 0*\overlayXdelta},{\overlayYstart - 0*\overlayYdelta}) {40.2};
\node[font=\tiny, text=white] at ({\overlayXstart + 1*\overlayXdelta},{\overlayYstart - 0*\overlayYdelta}) {44.4};
\node[font=\tiny, text=white] at ({\overlayXstart + 2*\overlayXdelta},{\overlayYstart - 0*\overlayYdelta}) {41.8};
\node[font=\tiny, text=white] at ({\overlayXstart + 3*\overlayXdelta},{\overlayYstart - 0*\overlayYdelta}) {40.9};

\node[font=\tiny] at ({\overlayXstart + 0*\overlayXdelta},{\overlayYstart - 1*\overlayYdelta}) {\textbf{82.1}};
\node[font=\tiny] at ({\overlayXstart + 1*\overlayXdelta},{\overlayYstart - 1*\overlayYdelta}) {73.9};
\node[font=\tiny] at ({\overlayXstart + 2*\overlayXdelta},{\overlayYstart - 1*\overlayYdelta}) {69.0};
\node[font=\tiny] at ({\overlayXstart + 3*\overlayXdelta},{\overlayYstart - 1*\overlayYdelta}) {73.2};

\node[font=\tiny] at ({\overlayXstart + 0*\overlayXdelta},{\overlayYstart - 2*\overlayYdelta}) {80.2};
\node[font=\tiny] at ({\overlayXstart + 1*\overlayXdelta},{\overlayYstart - 2*\overlayYdelta}) {\textbf{84.4}};
\node[font=\tiny] at ({\overlayXstart + 2*\overlayXdelta},{\overlayYstart - 2*\overlayYdelta}) {75.3};
\node[font=\tiny] at ({\overlayXstart + 3*\overlayXdelta},{\overlayYstart - 2*\overlayYdelta}) {74.1};

\node[font=\tiny] at ({\overlayXstart + 0*\overlayXdelta},{\overlayYstart - 3*\overlayYdelta}) {79.4};
\node[font=\tiny] at ({\overlayXstart + 1*\overlayXdelta},{\overlayYstart - 3*\overlayYdelta}) {82.1};
\node[font=\tiny] at ({\overlayXstart + 2*\overlayXdelta},{\overlayYstart - 3*\overlayYdelta}) {\textbf{81.6}};
\node[font=\tiny] at ({\overlayXstart + 3*\overlayXdelta},{\overlayYstart - 3*\overlayYdelta}) {\textbf{79.3}};

    \pgfmathsetlengthmacro{\effFigHeight}{
        \figHeight - 5mm
    }
    \pgfmathsetlengthmacro{\labelDist}{
        \effFigHeight / 4
    }
    \pgfmathsetlengthmacro{\labelOff}{
        1.5mm
    }
    \foreach \y / \yLabel in {
        0/1.2,1/1.4,2/1.6,3/1.8,4/2.0,5/2.2
    } {
        \pgfmathtruncatemacro\tmp{int(\y/2) * 2}
            \ifnum\tmp=\y
            \node[anchor=west] at (\plotStartX+ \figWidth,\plotStartY + \y * \labelDist + \labelOff) {\tickFont \yLabel};
            \def\tickLen{\majorTick}
            \else
            \def\tickLen{\minorTick}
            \fi
        \draw[line width=\tickWidth] (\plotStartX+ \figWidth,\plotStartY + \y * \labelDist + \labelOff) -- (\plotStartX+\tickLen+ \figWidth,\plotStartY + \y * \labelDist  + \labelOff);
        
    }
    \node[rotate=90, anchor=north, yshift=-4mm] at (\plotStartX + \figWidth, \plotStartY + 0.5 * \figHeight) {\labelFont PESQ\,$\rightarrow$};

    \pgfmathsetlengthmacro{\effFigHeight}{
        \figHeight - 5mm
    }
    \pgfmathsetlengthmacro{\labelDist}{
        \effFigHeight / 3
    }
    \pgfmathsetlengthmacro{\labelOff}{
        0.5mm
    }
    \foreach \y / \yLabel in {
        0/40,1/50,2/60,3/70,4/80
    } {
        \pgfmathtruncatemacro\tmp{int(\y/2) * 2}
            \ifnum\tmp=\y
            \node[anchor=west] at (\plotStartX+ 2 * \figWidth + \figDist,\plotStartY + \y * \labelDist + \labelOff) {\tickFont \yLabel};
            \def\tickLen{\majorTick}
            \else
            \def\tickLen{\minorTick}
            \fi
        \draw[line width=\tickWidth] (\plotStartX+ 2 * \figWidth + \figDist,\plotStartY + \y * \labelDist + \labelOff) -- (\plotStartX+\tickLen+ 2 * \figWidth + \figDist,\plotStartY + \y * \labelDist  + \labelOff);
        
    }
    \node[rotate=90, anchor=north, yshift=-3.5mm] at (\plotStartX + 2 * \figWidth + \figDist, \plotStartY + 0.5 * \figHeight) {\labelFont ESTOI\,\footnotesize{[\%]}\,$\rightarrow$};
    
\end{tikzpicture}
\vspace*{-21pt}
\caption{
Generalization from training to inference under mismatched speaker trajectories using \mbox{SpatialNet} as \ac{ssf} with strong guidance (oracle DoA, \picLegend{images/tst_accuracy/strong.pdf}).
}
\label{fig:locata_trajectories}
\vspace*{-5pt}
\end{figure}

\begin{figure}[t!]
\centering
\captionsetup[subfloat]{
    justification=justified,
    singlelinecheck=false,
    margin=3pt
}
\subfloat[Open-loop sensitivity\,to\,interfering\,speaker leakage ($\alpha^{\tinySuper{(\mathrm{\interferingSpeaker})}}\,\uparrow$).\label{fig:plug_in_approximation}]{
  \begin{minipage}[b]{0.3\linewidth} %
    \vspace{0pt}
    \centering
    \hspace*{-1mm}\begin{tikzpicture}

\newcommand\figSize{1.8cm}
\pgfmathsetlengthmacro{\figWidth}{\figSize} 
\pgfmathsetlengthmacro{\figHeight}{\figSize}
\newcommand\plotStartX{0mm}
\newcommand\plotStartY{0mm}
\node[anchor=center] at (\plotStartX + 0.5 * \figWidth, \plotStartY + 0.5 * \figHeight) {%
    \pgfimage[height=\figHeight, width=\figWidth]{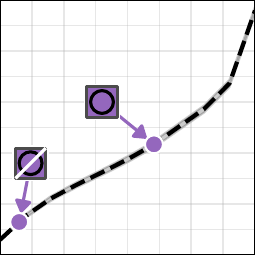} 
};

    \pgfmathsetlengthmacro{\labelDist}{
        \figWidth / 8
    }
    \foreach \x / \xLabel in {
    	0/0.0,1/0.125,2/0.25,3/0.375,4/0.5,5/0.625,6/0.75,7/0.875,8/1.0
	}{
	\pgfmathtruncatemacro\tmp{int(\x/2) * 2}
	\ifnum\x=\tmp
		 \pgfmathsetmacro{\tickLen}{\majorTick}
		 \node[anchor=north] at (\plotStartX + \x * \labelDist, \plotStartY+0.25mm) {\tickFont \xLabel};
	\else
		 \pgfmathsetmacro{\tickLen}{\minorTick}
	\fi
	\draw[line width=\tickWidth] (\plotStartX + \x * \labelDist,\plotStartY) -- (\plotStartX + \x * \labelDist,\plotStartY - \tickLen);
	
	}
%
    \node[anchor=north, yshift=-3mm] at (\plotStartX + 0.5 * \figWidth, \plotStartY) {\labelFont $\alpha^{\tinySuper{(\mathrm{\interferingSpeaker})}}$ in \cref{eq:plug_in_approximation}};

    \pgfmathsetlengthmacro{\labelDist}{
        \figHeight / 10
    }
    \foreach \y in {
        0,...,10
    } {
    \pgfmathtruncatemacro\tmp{int(\y/2) * 2}
	\ifnum\y=\tmp
        \pgfmathsetmacro{\yLabel}{int(\y * 5)}
        \node[anchor=east] at (\plotStartX,\plotStartY + \y * \labelDist) {\tickFont \yLabel};
        \pgfmathsetmacro{\tickLen}{\majorTick}
    \else
        \pgfmathsetmacro{\tickLen}{\minorTick}
    \fi
    \draw[line width=\tickWidth] (\plotStartX,\plotStartY + \y * \labelDist) -- (\plotStartX- \tickLen,\plotStartY + \y * \labelDist);
    }
    \node[rotate=90, anchor=south, yshift=3.25mm] at (\plotStartX, \plotStartY + 0.5 * \figHeight) {\labelFont MAE\,[°]\,$\leftarrow$};

\end{tikzpicture}
    \vspace{-16pt} %
  \end{minipage}
}
\hfill
\subfloat[Closed-loop~(AR) sensitivity to perturbations of duration $ t^{\tinySuper{(\mathrm{\interferingSpeaker})}}$.\label{fig:error_sensitivity}]{
  \begin{minipage}[b]{0.3\linewidth} %
    \vspace{0pt}
    \centering
    \hspace*{-1mm}\begin{tikzpicture}

\newcommand\figSize{1.8cm}
\pgfmathsetlengthmacro{\figWidth}{\figSize}
\pgfmathsetlengthmacro{\figHeight}{\figSize}
\newcommand\plotStartX{0mm}
\newcommand\plotStartY{0mm}
\node[anchor=center] at (\plotStartX + 0.5 * \figWidth, \plotStartY + 0.5 * \figHeight) {%
    \pgfimage[height=\figHeight, width=\figWidth]{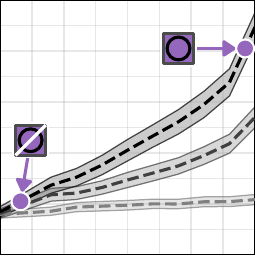} 
};

    \pgfmathsetlengthmacro{\labelDist}{
        \figWidth / 8
    }
    \foreach \x / \xLabel in {
    	0/0.0,1/0.125,2/0.25,3/0.375,4/0.5,5/0.625,6/0.75,7/0.875,8/1.0
	}{
	\pgfmathtruncatemacro\tmp{int(\x/2) * 2}
	\ifnum\x=\tmp
		 \pgfmathsetmacro{\tickLen}{\majorTick}
		 \node[anchor=north] at (\plotStartX + \x * \labelDist, \plotStartY+0.25mm) {\tickFont \xLabel};
	\else
		 \pgfmathsetmacro{\tickLen}{\minorTick}
	\fi
	\draw[line width=\tickWidth] (\plotStartX + \x * \labelDist,\plotStartY) -- (\plotStartX + \x * \labelDist,\plotStartY - \tickLen);
	
	}
%
    \node[anchor=north, yshift=-3mm] at (\plotStartX + 0.5 * \figWidth, \plotStartY) {\labelFont $\alpha^{\tinySuper{(\mathrm{\interferingSpeaker})}}$ in \cref{eq:plug_in_approximation}};

    \pgfmathsetlengthmacro{\labelDist}{
        \figHeight / 10
    }
    \foreach \y in {
        0,...,10
    } {
    \pgfmathtruncatemacro\tmp{int(\y/2) * 2}
	\ifnum\y=\tmp
        \pgfmathsetmacro{\yLabel}{int(\y * 5)}
        \node[anchor=east] at (\plotStartX,\plotStartY + \y * \labelDist) {\tickFont \yLabel};
        \pgfmathsetmacro{\tickLen}{\majorTick}
    \else
        \pgfmathsetmacro{\tickLen}{\minorTick}
    \fi
    \draw[line width=\tickWidth] (\plotStartX,\plotStartY + \y * \labelDist) -- (\plotStartX- \tickLen,\plotStartY + \y * \labelDist);
    }
    \node[rotate=90, anchor=south, yshift=3.25mm] at (\plotStartX, \plotStartY + 0.5 * \figHeight) {\labelFont MAE\,[°]\,$\leftarrow$};

 \node[anchor=north, yshift=5mm, xshift=-1.5mm] at (\plotStartX + 0.5 * \figWidth, \plotStartY + 0.5 * \figHeight) {\tickFont \contour{white}{$t^{\tinySuper{(\mathrm{\interferingSpeaker})}}$=\,1.0\,s}};

 \definecolor{labelgray}{gray}{0.3}
  \node[anchor=north, yshift=-1.5mm, xshift=6.4mm] at (\plotStartX + 0.5 * \figWidth, \plotStartY + 0.5 * \figHeight) {\tickFont \contour{white}{\textcolor{labelgray}{0.5\,s}}};

 \definecolor{labelgray}{gray}{0.5}
 \node[anchor=north, yshift=-5mm, xshift=4mm] at (\plotStartX + 0.5 * \figWidth, \plotStartY + 0.5 * \figHeight) {\tickFont \contour{white}{\textcolor{labelgray}{$t^{\tinySuper{(\mathrm{\interferingSpeaker})}}$=\,0.1\,s}}};

\end{tikzpicture}
    \vspace{-16pt} %
  \end{minipage}
}
\hfill
\subfloat[Error response relative to unperturbed \mbox{trajectory} ($\Delta$MAE).\label{fig:error_propagation}]{
  \begin{minipage}[b]{0.30\linewidth} %
    \vspace{0pt}
    \centering
    \hspace*{-1mm}\begin{tikzpicture}

\newcommand\figSize{1.8cm}
\pgfmathsetlengthmacro{\figWidth}{\figSize}
\pgfmathsetlengthmacro{\figHeight}{\figSize}
\newcommand\plotStartX{0mm}
\newcommand\plotStartY{0mm}
\node[anchor=center] at (\plotStartX + 0.5 * \figWidth, \plotStartY + 0.5 * \figHeight) {%
    \pgfimage[height=\figHeight, width=\figWidth]{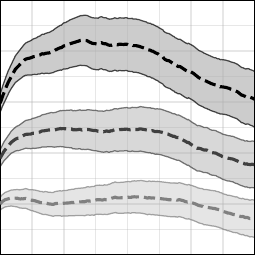} 
};

    \pgfmathsetlengthmacro{\labelDist}{
        \figWidth / 8
    }
    \foreach \x in {
    	0,...,8
	}{
	\pgfmathsetmacro{\xLabel}{int(\x / 2)}
	\pgfmathtruncatemacro\tmp{int(\x/2) * 2}
	\ifnum\x=\tmp
		 \pgfmathsetmacro{\tickLen}{\majorTick}
		 \node[anchor=north] at (\plotStartX + \x * \labelDist, \plotStartY+0.25mm) {\tickFont \xLabel};
	\else
		 \pgfmathsetmacro{\tickLen}{\minorTick}
	\fi
	\draw[line width=\tickWidth] (\plotStartX + \x * \labelDist,\plotStartY) -- (\plotStartX + \x * \labelDist,\plotStartY - \tickLen);
	
	}
%
    \node[anchor=north, yshift=-3mm, xshift=-2mm] at (\plotStartX + 0.5 * \figWidth, \plotStartY) {\labelFont time after perturb. [s]};

    \pgfmathsetlengthmacro{\labelDist}{
        \figHeight / 10
    }
    \foreach \y in {
        0,...,10
    } {
    \pgfmathtruncatemacro\tmp{int(\y/2) * 2}
	\ifnum\y=\tmp
        \pgfmathsetmacro{\yLabel}{int(\y * 2.5)}
        \node[anchor=east] at (\plotStartX,\plotStartY + \y * \labelDist) {\tickFont \yLabel};
        \pgfmathsetmacro{\tickLen}{\majorTick}
    \else
        \pgfmathsetmacro{\tickLen}{\minorTick}
    \fi
    \draw[line width=\tickWidth] (\plotStartX,\plotStartY + \y * \labelDist) -- (\plotStartX- \tickLen,\plotStartY + \y * \labelDist);
    }
    \node[rotate=90, anchor=south, yshift=3.25mm] at (\plotStartX, \plotStartY + 0.5 * \figHeight) {\labelFont \textcolor{black}{$\Delta$MAE}\,[°]\,$\leftarrow$};

 \node[anchor=north, yshift=10mm, xshift=4mm] at (\plotStartX + 0.5 * \figWidth, \plotStartY + 0.5 * \figHeight) {\tickFont \contour{white}{$\alpha^{\tinySuper{(\mathrm{\interferingSpeaker})}}$=\,1.0}};

 \definecolor{labelgray}{gray}{0.3}
 \node[anchor=north, yshift=4mm] at (\plotStartX + 0.5 * \figWidth, \plotStartY + 0.5 * \figHeight) {\tickFont \contour{white}{\textcolor{labelgray}{$\alpha^{\tinySuper{(\mathrm{\interferingSpeaker})}}$=\,0.8}}};

 \definecolor{labelgray}{gray}{0.5}
 \node[anchor=north, yshift=-5mm] at (\plotStartX + 0.5 * \figWidth, \plotStartY + 0.5 * \figHeight) {\tickFont \contour{white}{\textcolor{labelgray}{$\alpha^{\tinySuper{(\mathrm{\interferingSpeaker})}}$=\,0.4}}};

\end{tikzpicture}
    \vspace{-16pt} %
  \end{minipage}
}

\caption{Sensitivity and robustness of our proposed autoregressive (AR) tracking formulation with the MISO-AR \mbox{Bootstrap\,PF} (\cref{fig:weak_ssf_miso} and \cref{alg:bootstrap_filter}).
\newdiffComm{}{We report sample means~(\meanpictogram) with shaded 95\% confidence intervals~(\confidencepictogram).}}
\label{fig:ar_eval}
\vspace{0pt}
\end{figure}

\begin{diffEnv}

\subsection{Sensitivity and Robustness of Autoregressive Tracking}\label{sec:ar_tracking_analysis}
To evaluate the sensitivity and robustness of our \ac{ar} tracking algorithms to enhancement errors, we use the \mbox{Bootstrap\,PF} in MISO-AR configuration (\cref{fig:weak_ssf_miso} and \cref{alg:bootstrap_filter}).

\textbf{\textit{Open-Loop}} 
During inference, we replace the latent target speech signal $S_{tk}$ with the enhanced speech $\hat{S}_{tk}$ as plug-in approximation in \cref{eq:miso_posterior_prediction,eq:pf_miso_likelihood}, which neglects signal degradation from the \ac{ssf}.
To quantify the impact of the approximation error on tracking performance, we replace $\hat{S}_{tk}$ with an interpolation between target and interfering speech signals $S_{tk}$ and $S^{\tinySuper{(\mathrm{\interferingSpeaker})}}_{tk}$, respectively, to simulate controlled processing artifacts in form of interfering speaker leakage via
\begin{equation}\label{eq:plug_in_approximation}
    \hat{S}_{tk} = (1 - \alpha^{\tinySuper{(\mathrm{\interferingSpeaker})}})S_{tk} + \alpha^{\tinySuper{(\mathrm{\interferingSpeaker})}}S^{\tinySuper{(\mathrm{\interferingSpeaker})}}_{tk} \, .
\end{equation} 
The open-loop tracking results shown in \cref{fig:plug_in_approximation} reveal a strong correlation between estimation accuracy (MAE,\,ACC) and interference suppression, demonstrating the capability of our tracking formulation to exploit the provided target speech signal for improved performance.
Nonetheless, retaining superior accuracy compared to the original \mbox{Bootstrap\,PF} formulation~(\picLegend{images/tst_accuracy/bootstrap_pf.pdf}) even at dominating interference~($\alpha^{\tinySuper{(\mathrm{\interferingSpeaker})}}$\,$>$\,0.5), \newdiffComm{proves}{shows} \textit{robustness} to \newdiffComm{significant}{substantial} \textit{speaker leakage} \newdiffComm{}{under the evaluated conditions}.
The remaining gap between no signal degradation~($\alpha^{\tinySuper{(\mathrm{\interferingSpeaker})}}$\,$=$\,0) and the closed-loop (\ac{ar}) performance with \mbox{SpatialNet}~(\picLegend{images/tst_accuracy/bootstrap_pf_miso_ar.pdf})\newdiffComm{,}{} 
indicates that current research on increasingly effective enhancement architectures, see, e.g., ~\cite{wang26parallel_spectral_spatial_architecture}, will further improve our \ac{ar} tracking methods.

\textbf{\textit{Closed-Loop}}  
To analyze the closed-loop~(\ac{ar}) sensitivity \newdiffComm{and stability}{} to enhancement errors, we insert controlled perturbations in finite intervals and analyze the \ac{ar} tracking response.
In particular, our experiment design introduces a 2\,s long warm-up phase to allow algorithm initialization, after which we replace the enhanced speech signal $\hat{S}_{tk}$ with controlled interference leakage via~\cref{eq:plug_in_approximation} in continuous bursts of length $t^{\tinySuper{(\mathrm{\interferingSpeaker})}}$.
We use the subsequent 4\,s recovery phase as basis for analysis and restrict the test subset such that target and interfering speakers are both active during the evaluation period\newdiffComm{}{, leaving 415$/$3000 two-speaker mixtures}.
\Cref{fig:error_sensitivity} illustrates the resulting angular error averaged over the recovery phase (\ac{mae}) as an indicator of tracking sensitivity to varying levels of interference. 
Most notably, the tracker is only negligibly affected by speaker confusions ($\alpha^{\tinySuper{(\mathrm{\interferingSpeaker})}}$\,$=$\,1) in short perturbation intervals $ t^{\tinySuper{(\mathrm{\interferingSpeaker})}}$\,$=$\,100\,ms, which can effectively \textit{counter error propagation} through the \ac{ssf}. 
However, with increasing duration, tracking degrades, until matching the original formulation of the \mbox{Bootstrap\,PF}~(\picLegend{images/tst_accuracy/bootstrap_pf.pdf}) at $t^{\tinySuper{(\mathrm{\interferingSpeaker})}}$\,$=$\,1\,s. 
To assess \newdiffComm{stability}{finite-duration robustness}, we compute the average angular error difference between perturbed and unperturbed trajectories in each time-step~($\Delta$\,MAE) for $t^{\tinySuper{(\mathrm{\interferingSpeaker})}}=$\,500\,ms.
\Cref{fig:error_propagation} illustrates how \newdiffComm{the resulting angular error remains \textit{bounded} across varying levels of speaker leakage $\alpha^{\tinySuper{(\mathrm{\interferingSpeaker})}}$ and exhibits a \textit{decreasing trend} toward the end of the recovery phase.}{increasing interference ($\alpha^{\tinySuper{(\mathrm{\interferingSpeaker})}}\,\uparrow$) yields a larger angular error overshoot. Nevertheless, all levels of perturbation exhibit a \textit{decreasing trend} in angular error toward the end of the recovery phase.
}

\end{diffEnv}

\begin{figure}[t!]
\centering
\captionsetup[subfloat]{
    justification=justified,
    singlelinecheck=false,
    margin=3pt
}
\subfloat[Dependency between enhancement performance and initial ($t=0$) \ac{doa} accuracy.\label{fig:doa_perturbation}]{
  \begin{minipage}[b]{0.575\linewidth} %
    \vspace{0pt}
    \centering
\hspace*{-2mm}\begin{tikzpicture}

\newcommand\figSize{1.8cm}
\pgfmathsetlengthmacro{\figWidth}{\figSize * 2.25 * 1.1}
\pgfmathsetlengthmacro{\figDist}{8.5mm} 
\pgfmathsetlengthmacro{\figHeight}{\figSize * 1.0}
\newcommand\plotStartX{0mm}
\newcommand\plotStartY{0mm}
\colorlet{trainingColor}{red!70}
\colorlet{testColor}{blue!70}
\newcommand\rowSplitSkip{-0.75mm}

\foreach \plotIdx/\metricLabel in {%
    0/mae%
}{
    \node[anchor=center] at (\plotStartX + 0.5 * \figWidth + \plotIdx * \figWidth + \plotIdx * \figDist, \plotStartY + 0.5 * \figHeight) {%
        \pgfimage[height=\figHeight, width=\figWidth]{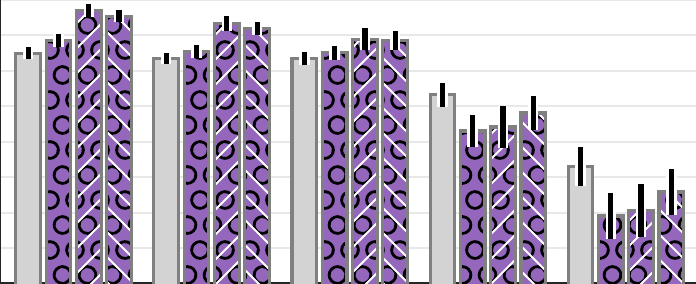} 
    };

    \foreach \xRel / \xLabel in {
        0.1/{[0, 3)},0.3/{[3, 6)},0.5/{[6, 9)},0.7/{[9, 12)},0.9/{[12, 15]}
    } {
        \node[anchor=north] at (\plotStartX + \xRel * \figWidth + \plotIdx * \figWidth + \plotIdx * \figDist,\plotStartY) {\tickFont \xLabel};
        \draw[line width=\tickWidth] (\plotStartX + \xRel * \figWidth + \plotIdx * \figWidth + \plotIdx * \figDist,\plotStartY) -- (\plotStartX + \xRel * \figWidth+ \plotIdx * \figWidth + \plotIdx * \figDist,\plotStartY- \majorTick);
    }

    \pgfmathsetlengthmacro{\labelDist}{
        \figHeight / 8
    }
    \foreach \y in {%
        0,...,8%
    } {%
        \pgfmathtruncatemacro{\tmp}{int(\y / 2) * 2}
        \ifnum\plotIdx=0
            \pgfmathsetmacro{\yLabel}{int(\y * 1+ 8) }
            \ifnum\tmp=\y
                \def\tickLen{\majorTick}
                \node[anchor=east] at (\plotStartX + \plotIdx * \figWidth + \plotIdx * \figDist,\plotStartY + \y * \labelDist) {\tickFont \yLabel};
            \else
                \def\tickLen{\minorTick}
            \fi
        \else
            \pgfmathsetmacro{\yLabel}{int(\y * 10 + 30)}
            \ifnum\tmp=\y
                \def\tickLen{\minorTick}
            \else
                \def\tickLen{\majorTick}
                \node[anchor=east] at (\plotStartX + \plotIdx * \figWidth + \plotIdx * \figDist,\plotStartY + \y * \labelDist) {\tickFont \yLabel};
            \fi
        \fi
        
        \draw[line width=\tickWidth] (\plotStartX + \plotIdx * \figWidth + \plotIdx * \figDist,\plotStartY + \y * \labelDist) -- (\plotStartX +\plotIdx * \figWidth + \plotIdx * \figDist- \tickLen,\plotStartY + \y * \labelDist);
    }
    \ifnum\plotIdx=0
    \node[anchor=south, rotate=90, yshift=3.0mm] at (\plotStartX + \plotIdx * \figWidth + \plotIdx * \figDist,\plotStartY + 0.5 * \figHeight) {\labelFont $\Delta$SI-SDR\,[dB]\,$\rightarrow$};
    \else 
    \node[anchor=south, rotate=90, yshift=3.5mm, xshift=0mm] at (\plotStartX + \plotIdx * \figWidth + \plotIdx * \figDist,\plotStartY + 0.5 * \figHeight) {\labelFont ACC\,[\%]\,$\rightarrow$};
    \fi
}

\node[anchor=north, yshift=-3mm] at (\plotStartX + 0.5 * \figWidth, \plotStartY) {\labelFont initial DoA perturbation $\Delta \theta_0$\,[°]};

\end{tikzpicture} %
\vspace{-15pt}
  \end{minipage}
}
\hfill
\subfloat[Conversational speech enhancement performance.\label{fig:sparse_performance}]{
  \begin{minipage}[b]{0.35\linewidth} %
    \vspace{0pt}
    \centering
    \hspace*{-2.5mm}\begin{tikzpicture}

\newcommand\figSize{1.8cm}
\pgfmathsetlengthmacro{\figWidth}{\figSize * 2.25 * 2 / 3} 
\pgfmathsetlengthmacro{\figDist}{8.5mm} 
\pgfmathsetlengthmacro{\figHeight}{\figSize * 1.0}
\newcommand\plotStartX{0mm}
\newcommand\plotStartY{0mm}
\colorlet{trainingColor}{red!70}
\colorlet{testColor}{blue!70}
\newcommand\rowSplitSkip{-0.75mm}

\foreach \plotIdx/\metricLabel in {%
    0/mae%
}{
    \node[anchor=center] at (\plotStartX + 0.5 * \figWidth + \plotIdx * \figWidth + \plotIdx * \figDist, \plotStartY + 0.5 * \figHeight) {%
        \pgfimage[height=\figHeight, width=\figWidth]{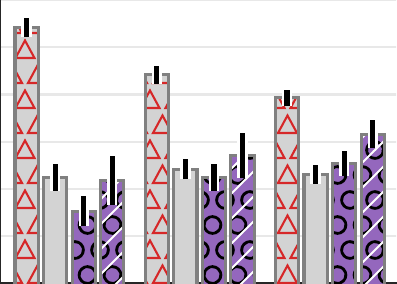} 
    };

    \foreach \xRel / \xLabel in {
        0.167/{0$-$20},0.5/{40$-$60},0.833/{80$-$100}
    } {
        \node[anchor=north] at (\plotStartX + \xRel * \figWidth + \plotIdx * \figWidth + \plotIdx * \figDist,\plotStartY) {\tickFont \xLabel};
        \draw[line width=\tickWidth] (\plotStartX + \xRel * \figWidth + \plotIdx * \figWidth + \plotIdx * \figDist,\plotStartY) -- (\plotStartX + \xRel * \figWidth+ \plotIdx * \figWidth + \plotIdx * \figDist,\plotStartY- \majorTick);
    }

    \pgfmathsetlengthmacro{\labelDist}{
        \figHeight / 6
    }
    \foreach \y in {%
        0,...,6%
    } {%
        \pgfmathtruncatemacro{\tmp}{int(\y / 2) * 2}
        \ifnum\plotIdx=0
            \pgfmathsetmacro{\yLabel}{int(\y * 1+ 12) }
            \ifnum\tmp=\y
                \def\tickLen{\majorTick}
                \node[anchor=east] at (\plotStartX + \plotIdx * \figWidth + \plotIdx * \figDist,\plotStartY + \y * \labelDist) {\tickFont \yLabel};
            \else
                \def\tickLen{\minorTick}
            \fi
        \else
            \pgfmathsetmacro{\yLabel}{int(\y * 10 + 30)}
            \ifnum\tmp=\y
                \def\tickLen{\minorTick}
            \else
                \def\tickLen{\majorTick}
                \node[anchor=east] at (\plotStartX + \plotIdx * \figWidth + \plotIdx * \figDist,\plotStartY + \y * \labelDist) {\tickFont \yLabel};
            \fi
        \fi
        
        \draw[line width=\tickWidth] (\plotStartX + \plotIdx * \figWidth + \plotIdx * \figDist,\plotStartY + \y * \labelDist) -- (\plotStartX +\plotIdx * \figWidth + \plotIdx * \figDist- \tickLen,\plotStartY + \y * \labelDist);
    }
    \ifnum\plotIdx=0
    \node[anchor=south, rotate=90, yshift=3.0mm] at (\plotStartX + \plotIdx * \figWidth + \plotIdx * \figDist,\plotStartY + 0.5 * \figHeight) {\labelFont $\Delta$SI-SDR\,[dB]\,$\rightarrow$};
    \else 
    \node[anchor=south, rotate=90, yshift=3.5mm, xshift=0mm] at (\plotStartX + \plotIdx * \figWidth + \plotIdx * \figDist,\plotStartY + 0.5 * \figHeight) {\labelFont ACC\,[\%]\,$\rightarrow$};
    \fi
}

\node[anchor=north, yshift=-3mm] at (\plotStartX + 0.5 * \figWidth, \plotStartY) {\labelFont overlap ratio [\%]};

\end{tikzpicture}
    \vspace{-15pt} %
  \end{minipage}
}

\caption{Robustness to parameter and scenario mismatch using oracle~(\picLegend{images/tst_accuracy/strong.pdf}), unguided~(\picLegend{images/tst_accuracy/unguided.pdf}) and tracking-based methods with the \mbox{Bootstrap\,PF}~(\picLegend{images/tst_accuracy/bootstrap_pf.pdf},\,\picLegend{images/tst_accuracy/bootstrap_pf_miso_ar.pdf},\,\picLegend{images/tst_accuracy/bootstrap_pf_mimo_ar.pdf}). $\Delta$\ac{sisdr} denotes signal distortion improvement w.r.t. the noisy recording.}
\label{fig:robustness_eval}
\vspace{0pt}
\end{figure}

\begin{diffEnv}
\subsection{\newdiffComm{Sensitvity}{Sensitivity} to DoA Initialization Accuracy}\label{sec:doa_init}

While the weakly guided extraction methods solely require the target speaker's starting \ac{doa} $\theta_0$, uncertainty in $\theta_0$ is not explicitly accounted for in this work.
However, even with oracle initialization on our synthetic dataset, 
the actual \ac{doa} at utterance start may deviate from $\theta_0$ as the LibriSpeech recordings are not trimmed to the speech onset.
An energy-based \ac{vad} reveals deviations above 5° for about 7\% of the test samples. 
To analyze the impact of \ac{doa} inaccuracies in a controlled scenario, we artificially align all utterances at $t=0$ and provide a perturbed cue $\tilde{\theta}_0 = \theta_0 + \Delta\theta_0$ with $\Delta\theta_0$ between -15° and 15° for algorithm initialization.
\Cref{fig:doa_perturbation} compares the resulting enhancement performance using \mbox{SpatialNet} as \ac{ssf} without guidance (\picLegend{images/tst_accuracy/unguided.pdf}) and with the \mbox{Bootstrap\,PF} (\picLegend{images/tst_accuracy/bootstrap_pf.pdf},\,\picLegend{images/tst_accuracy/bootstrap_pf_miso_ar.pdf},\,\picLegend{images/tst_accuracy/bootstrap_pf_mimo_ar.pdf}) for tracking at varying levels of \ac{doa} perturbation.
All tracking-based methods yield a slight increase in speech distortions (lower $\Delta$\acs{sisdr}) for perturbations up to 9°.
Since, without tracking, the \ac{ssf}~(\picLegend{images/tst_accuracy/unguided.pdf}) can only implicitly utilize spatial information, it is unaffected by minor \ac{doa} initialization inaccuracies.
Nevertheless, our \ac{ar}-guided \acp{ssf} maintain superior enhancement even under performance degradation~(\picLegend{images/tst_accuracy/bootstrap_pf_miso_ar.pdf},\,\picLegend{images/tst_accuracy/bootstrap_pf_mimo_ar.pdf}).
Only for major perturbations above 9° the enhancement worsens \newdiffComm{significantly}{substantially} for all methods, with the unguided \ac{ssf} being affected the least~(\picLegend{images/tst_accuracy/unguided.pdf}).
Note that the PF also allows explicit modeling of initial \ac{doa} uncertainty for improved robustness using a prior over $\theta_0$~(\cref{alg:bootstrap_filter}, line 1), which we leave for future work.
\end{diffEnv}

\vspace{-12pt}

\begin{diffEnv}
\subsection{Conversational Speech Enhancement}\label{sec:conversational_enhancement} 
Fully overlapping speech mixtures, the center of this work, are representative of \textit{cocktail-party}-style recording environments, where groups of speakers are entangled in independent conversations.
On the other hand, \textit{meeting scenarios} only involve a single conversation, \newdiffComm{significantly}{considerably} reducing the overlap between participants~\cite{cetin06overlaps_meetings}.
As a result, the challenge for \acl{tse}~(\acs{tse}) algorithms shifts from speaker disentanglement to maintaining speaker identity during the absence of target speech~\cite{delcroix22speakerbeam_false_alarms}.
Without explicitly modeling target inactivity in our framework, the \ac{ssf} needs to be steered away from any interferer during periods of silence.
While our \ac{mimo}-\ac{ar} trackers rely on spatial cue preservation, which becomes ambiguous during non-speech periods, the \ac{miso}-\ac{ar} \newdiffComm{}{likelihood} formulation \newdiffComm{of}{in} the \mbox{Bootstrap\,PF}\newdiffComm{}{~\cref{eq:pf_miso_likelihood}} \newdiffComm{implicitly models this aspect.
Without target speech, the likelihood in \cref{eq:pf_miso_likelihood} becomes invariant to \ac{doa} $\theta_t$, yielding uniform particle weight updates independent of the spatial cues of the interfering speaker.}{becomes \ac{doa}-invariant and therefore insensitive to the interfering speaker's directional features.
As a result, particle trajectories diffuse under the state-transition model, 
retaining support across increasingly diverse motion hypotheses, which enables re-attachment to a moving target after periods of silence.
However, under the white-noise acceleration model in \cref{eq:cv_motion_model}, particle trajectories become increasingly erratic over time, limiting the realism of the motion hypotheses after long periods of target inactivity.}

For further analysis, we replace Libri2Mix with SparseLibri2Mix~\cite{cosentino20librimix} in ~\cref{sec:dataset}\newdiffComm{,}{} to generate a new test subset with speech mixtures from 0$-$100\% overlap in 20\% steps.
\Cref{fig:examples_sparse_librimix} \newdiffComm{illustrates}{shows} the enhancement and tracking performance of the MISO-AR \mbox{Bootstrap\,PF}~({\protect\tikz[baseline=-0.6ex]  \protect\draw[color=tab_orange, line width=1.0pt, decorate,dash pattern=on 3pt off 1pt, decoration={snake, amplitude=3pt, segment length=10pt}] (0, 0) -- (4mm, 0);}) on samples without overlap. 
The increasing spread of the \newdiffComm{particles}{particle trajectories}~(\particlePictogram) indicates the \newdiffComm{growing uncertainty in the \ac{doa} estimates due to the absence of target speech}{diffusing motion hypotheses during target inactivity}.
However, \newdiffComm{}{in the shown examples,} as long as the \newdiffComm{trajectories}{resulting \ac{doa} estimates} remain sufficiently disjoint, the \ac{ssf} successfully attenuates the interferer (\cref{fig:examples_sparse_librimix}a$-$e).
Most notably, \newdiffComm{}{the examples qualitatively demonstrate how} the \mbox{Bootstrap\,PF} \newdiffComm{is able to}{can} \textit{re-attach} to the target after \newdiffComm{periods of silence}{speech pauses} \newdiffComm{as long as}{if} particles are within vicinity. 
Nevertheless, closely aligned trajectories and similar speaker characteristics can lead to speaker leakage and confusion, see~\cref{fig:examples_sparse_librimix}f.

\Cref{fig:sparse_performance} illustrates the resulting enhancement performance across all levels of overlap. 
With decreasing overlap, the tracking-based algorithms~(\picLegend{images/tst_accuracy/bootstrap_pf.pdf},\,\picLegend{images/tst_accuracy/bootstrap_pf_miso_ar.pdf}) cause increasing distortion (lower $\Delta$\ac{sisdr}) and eventually provide no gain over the unguided \ac{ssf}~(\picLegend{images/tst_accuracy/unguided.pdf}).
Only oracle steering~(\picLegend{images/tst_accuracy/strong.pdf}) lets the \ac{ssf} exploit the improved speaker separability at low overlap, revealing the \textit{potential} of accurate guidance for meeting scenarios.
\end{diffEnv}

\begin{figure*}[t!]
\vspace*{-5pt}
\begin{tikzpicture}[spy using outlines={magnification=1.5, draw=black!65,size=0.5cm, connect spies,every spy on node/.style={
      draw=black!65,
      line width=0.8pt
    },
    every spy in node/.style={
      draw=black!65,
      line width=0.8pt
    },
    every spy connection path/.style={
      draw=black!65,
      line width=0.8pt
    }}]
     
\newcommand\doaWidth{2.325cm} 
\newcommand\doaHeight{10mm} 
\newcommand\doaDist{2mm} 
\pgfmathsetlengthmacro{\yRoomOff}{2* (\doaHeight + \doaDist)}
\newcommand\doaTickVoff{-0.2mm}
\newcommand\doaTickHoff{-0.15mm}
\newcommand\doaLabelDist{3mm}
\newcommand\lrDist{10mm}
\newcommand\tbDist{8mm}
\newcommand\plotStartX{0mm}
\newcommand\plotStartY{0mm}
\newcommand\timeDist{3.1mm} 
\newcommand\legendTick{4mm}
\newcommand\legendTickTop{-0.25mm}
\newcommand\legendTickBottom{-0.5mm}
\pgfmathsetlengthmacro{\legendTickMiddle}{0.5 * \legendTickTop + 0.5 * \legendTickBottom}
\newcommand\legendOff{1mm}
\newcommand\legendYOff{-2mm} 
\newcommand\centerOff{-1.1cm}
\newcommand\legendWidth{4cm}
\newcommand\legendSkip{4mm}
\newcommand\legendOffProp{2.7cm}
\newcommand\labelWidth{1pt}
\newcommand\descriptX{2.7mm}
\newcommand\descriptY{2.2mm}
\newcommand{\tinymid}{\footnotesize}

\foreach \x / \w / \h / \l / \i / \g / \acc [
    remember=\dl as \lastLen (initially 0), 
    evaluate=\l as \dl using \l+\lastLen,
]in {
    0/4.77/7.43/14.15/2079/mixed/99.5,
    1/5.50/4.48/14.28/2165/mixed/77.1,
    2/4.77/7.43/14.70/2158/mixed/98.2,
    3/5.50/4.48/13.50/2109/same/87.1, 
    4/5.50/4.48/13.48/2093/same/58.4,
    5/4.77/7.43/14.25/2059/same/39.2%
    }{
    
    \pgfmathsetmacro{\filenum}{int(\i)}
    \def\exampleName{example_\filenum}
    \pgfmathsetlengthmacro{\dWidth}{\l / 12 * \doaWidth} 
        \pgfmathsetlengthmacro{\dOff}{\lastLen / 12 * \doaWidth} 
        \pgfmathsetmacro{\dOffVal}{int(abs(\dOff))}
        \pgfmathsetmacro{\dMaxVal}{12 * \doaWidth} 

    \foreach \row in {0, 1}{
        \ifnum\row=0
            \def\smallLegendLabel{\tinymid{MIMO-AR}}
            \def\trajFile{images/examples_sparse_librimix/spatialnet-rbpf-ar_\l_spectrogram_\filenum.pdf}
        \fi

        \ifnum\row=1
            \def\smallLegendLabel{\tinymid{MISO-AR}}
            \def\trajFile{images/examples_sparse_librimix/spatialnet-rbpf-ar_\l_trajectory_\filenum.pdf}
        \fi
    
        \pgfmathsetlengthmacro{\yDoaOff}{\row * (\doaHeight + 2mm)}
        \node[anchor=mid] at (\plotStartX + \lrDist + \dOff + \x * \doaDist + 0.5 * \dWidth, \plotStartY + 0.5 \doaHeight + \yDoaOff) {%
            \pgfimage[height=\doaHeight, width=\dWidth]{\trajFile} 
        };

        \ifnum\x=1
        \pgfmathsetlengthmacro{\spyOnXoff}{\plotStartX + \lrDist + \dOff + \x * \doaDist + 0.5 * \dWidth -3.5mm}
        \pgfmathsetlengthmacro{\spyOnYoff}{\plotStartY + 0.5 \doaHeight + \yDoaOff + 7.75mm}
        \pgfmathsetlengthmacro{\spyAtXoff}{\plotStartX + \lrDist + \dOff + \x * \doaDist + 0.5 * \dWidth -1.5mm}
        \pgfmathsetlengthmacro{\spyAtYoff}{\plotStartY + 0.5 \doaHeight + \yDoaOff +3.25mm}
        \pgfmathsetlengthmacro{\spyRectWidth}{0.9cm}
        \pgfmathsetlengthmacro{\spyRectHeight}{0.4cm}
    \else 
        \ifnum\x=2
            \pgfmathsetlengthmacro{\spyOnXoff}{\plotStartX + \lrDist + \dOff + \x * \doaDist + 0.5 * \dWidth -3.5mm}
            \pgfmathsetlengthmacro{\spyOnYoff}{\plotStartY + 0.5 \doaHeight + \yDoaOff + 0.5mm}
            \pgfmathsetlengthmacro{\spyAtXoff}{\plotStartX + \lrDist + \dOff + \x * \doaDist + 0.5 * \dWidth -1mm}
            \pgfmathsetlengthmacro{\spyAtYoff}{\plotStartY + 0.5 \doaHeight + \yDoaOff +5.25mm}
            \pgfmathsetlengthmacro{\spyRectWidth}{0.9cm}
            \pgfmathsetlengthmacro{\spyRectHeight}{0.35cm}
            \else
         \pgfmathsetlengthmacro{\spyOnXoff}{\plotStartX + \lrDist + \dOff + \x * \doaDist + 0.5 * \dWidth -3mm}
            \pgfmathsetlengthmacro{\spyOnYoff}{\plotStartY + 0.5 \doaHeight + \yDoaOff +6.5mm}
            \pgfmathsetlengthmacro{\spyAtXoff}{\plotStartX + \lrDist + \dOff + \x * \doaDist + 0.5 * \dWidth -2.5mm}
            \pgfmathsetlengthmacro{\spyAtYoff}{\plotStartY + 0.5 \doaHeight + \yDoaOff +2.75mm}
            \pgfmathsetlengthmacro{\spyRectWidth}{1.0cm}
            \pgfmathsetlengthmacro{\spyRectHeight}{0.35cm}
        \fi
        
    \fi

        \def\smallLegendWidth{5mm}
        \def\smallLegendXoff{14.5mm} 
        \def\smallLegendYoff{7.9mm} 
        \def\smallLegendHeight{2mm}
        \ifnum\x=2
        \fi
        

        \ifnum\row=0
        \node[anchor=north] at (\plotStartX + \lrDist + \dOff + \x * \doaDist + 0.925 * \dWidth, \plotStartY + 1.05 *\doaHeight) {\tickFont \contour{white}{(\char\numexpr`a+\x\relax)}};
        \node[anchor=east, xshift=2.5mm] at (\plotStartX + \lrDist + \dOff + \x * \doaDist + 0.925 * \dWidth, \plotStartY + 0.6 *\doaHeight) {\tickFont \contour{white}{\g}};
        \else 
        \node[anchor=east, xshift=3mm, yshift=1.5mm] at (\plotStartX + \lrDist + \dOff + \x * \doaDist + 0.925 * \dWidth, \plotStartY + 2 *\doaHeight + \doaDist) {\tickFont \acc\% Active-ACC};
    \fi
                   
        \pgfmathsetmacro{\timeNum}{int(\l * 2/3)}
        \pgfmathsetmacro{\timeOff}{\doaWidth / 12 * 3/2} 
        \pgfmathsetmacro{\doaOff}{\doaHeight / 4 - 0.025mm}
        \foreach \t in {0, ..., \timeNum}{
            \pgfmathtruncatemacro\tmp{int(\t/2) * 2}
            \ifnum\tmp=\t
                \pgfmathsetmacro{\tickLen}{\majorTick}
                \pgfmathsetmacro{\time}{int(\t * 1.5)}
                \ifnum\row=0
                    \node[anchor=north] at (\plotStartX + \lrDist + \dOff + \x * \doaDist + \t * \timeOff, \plotStartY + \doaTickVoff + \yDoaOff) {\tickFont \time};
                \fi
            \else
                \pgfmathsetmacro{\tickLen}{\minorTick}
            \fi
            \draw[line width=\tickWidth] (\plotStartX + \lrDist + \dOff + \x * \doaDist + \t * \timeOff, \plotStartY + \doaTickVoff + \yDoaOff) --  (\plotStartX + \lrDist + \dOff + \x * \doaDist + \t * \timeOff, \plotStartY - \tickLen + \doaTickVoff + \yDoaOff);
        }
        \foreach \d in {0, ..., 4}{
            \pgfmathtruncatemacro\tmp{int(\d/2) * 2}
            \ifnum\tmp=\d
                \pgfmathsetmacro{\tickLen}{\majorTick}
                 \ifnum\row=1
                \ifthenelse{\x = 0}{
                    \pgfmathsetmacro{\doa}{int(\d * 90 - 180)}
                        \node[anchor=east] at (\plotStartX + \lrDist + \dOff + \x * \doaDist ,\plotStartY + \d * \doaOff + \doaTickHoff + \yDoaOff) {\tickFont \doa};
                }{}
                \else
                \ifthenelse{\x = 0}{
                    \pgfmathsetmacro{\doa}{int(\d * 2)}
                        \node[anchor=east] at (\plotStartX + \lrDist + \dOff + \x * \doaDist ,\plotStartY + \d * \doaOff + \doaTickHoff + \yDoaOff) {\tickFont \doa};
                }{}
                \fi
            \else
                \pgfmathsetmacro{\tickLen}{\minorTick}
            \fi
            \draw[line width=\tickWidth] (\plotStartX + \lrDist + \dOff + \x * \doaDist - \tickLen,\plotStartY + \d * \doaOff + \doaTickHoff + \yDoaOff) --  (\plotStartX + \lrDist + \dOff + \x * \doaDist,\plotStartY + \d * \doaOff + \doaTickHoff + \yDoaOff);
        }
    }

}

\node[anchor=north] at (0.5 * \linewidth + 7mm, \plotStartY- 3*\majorTick + \doaTickVoff) {\footnotesize time [s]};
\node[anchor=south, rotate=90, xshift=-0.25mm] at (\plotStartX  + \lrDist - 2mm, \plotStartY + 0.5 * \doaHeight + 0*\doaDist) {\labelFont freq\,[{\scriptsize kHz}]};
\node[anchor=south, xshift=-1mm, yshift=-0.25mm] at (\plotStartX  + \lrDist, \plotStartY + 2 * \doaHeight + 1* \doaDist) {\labelFont DoA\,[°]};

\end{tikzpicture}
\vspace*{-20pt}
\caption{
Weakly guided target speaker ({\protect\tikz[baseline=-0.6ex]  \protect\draw[color=tab_orange, line width=1.5pt, decorate, draw opacity=0.5,
  decoration={snake, amplitude=3pt, segment length=10pt}] (0, 0) -- (4mm, 0);}) extraction on overlap-free \newdiffComm{two speaker}{two-speaker} mixtures \newdiffComm{}{from SparseLibri2Mix~\cite{cosentino20librimix}} using the \mbox{Bootstrap\,PF} in MISO-AR configuration, see \cref{fig:weak_ssf_miso} and \cref{alg:bootstrap_filter}. \ac{doa} estimates ({\protect\tikz[baseline=-0.6ex]  \protect\draw[color=tab_orange, line width=1.0pt, decorate, dash pattern=on 3pt off 1pt,
  decoration={snake, amplitude=3pt, segment length=10pt}] (0, 0) -- (4mm, 0);}) overlay the particle trajectories (\particlePictogram). \textit{Mixed} and \textit{same} denote different-gender and same-gender speaker pairs, respectively.
  \newdiffComm{}{Active-ACC denotes the tracking accuracy within a 10° threshold, evaluated only on frames with active target speech.}
}
\label{fig:examples_sparse_librimix}
\end{figure*} 

\begin{figure*}[b!]
\begin{tikzpicture}

\newcommand\doaWidth{3.25cm} 
\newcommand\doaHeight{12mm} 
\newcommand\doaDist{4mm} 
\newcommand\yDoaOffConst{2.5mm}
\pgfmathsetlengthmacro{\yRoomOff}{2* (\doaHeight + \doaDist)}
\newcommand\doaTickVoff{-0.2mm}
\newcommand\doaTickHoff{-0.15mm}
\newcommand\doaLabelDist{3mm}
\newcommand\lrDist{10mm}
\newcommand\tbDist{8mm}
\newcommand\plotStartX{0mm}
\newcommand\plotStartY{0mm}
\newcommand\timeDist{3.1mm} 
\newcommand\legendTick{4mm}
\newcommand\legendTickTop{-0.25mm}
\newcommand\legendTickBottom{-0.5mm}
\pgfmathsetlengthmacro{\legendTickMiddle}{0.5 * \legendTickTop + 0.5 * \legendTickBottom}
\newcommand\legendOff{1mm}
\newcommand\legendYOff{-2mm} 
\newcommand\centerOff{-1.1cm}
\newcommand\legendWidth{4cm}
\newcommand\legendSkip{4mm}
\newcommand\legendOffProp{2.7cm}
\newcommand\labelWidth{1pt}
\newcommand\descriptX{2.7mm}
\newcommand\descriptY{2.2mm}
\newcommand\legendXBoxOff{3mm}
\newcommand\legendYBoxOff{3.1cm}
\newcommand\legendXBoxWidth{6.4cm}
\newcommand\legendYBoxHeight{4mm}

\foreach \x / \w / \h / \l / \i [
    remember=\dl as \lastLen (initially 0), 
    evaluate=\l as \dl using \l+\lastLen,
]in {
    0/9.1/6.91/10/2798,
    1/8.36/6.38/10/494,
    2/8.36/6.38/10/494
    }{
    
    \pgfmathsetlengthmacro{\dWidth}{\l / 6 * \doaWidth} 
        \pgfmathsetlengthmacro{\dOff}{\lastLen / 6 * \doaWidth} 
        \pgfmathsetmacro{\dOffVal}{int(abs(\dOff))}
        \pgfmathsetmacro{\dMaxVal}{6 * \doaWidth}

    \foreach \row in {0,...,2}{
        \ifnum\row=0
            \ifnum\x=0
                \def\exampleName{spatialnet-mimo-rbpf-ar+spatialnet-mimo-wrappedkf-ar_low_reverb}
            \else
                \ifnum\x=1
                    \def\exampleName{spatialnet-mimo-rbpf-ar+spatialnet-mimo-wrappedkf-ar_mid_reverb}
                \else
                    \def\exampleName{spatialnet-mimo-rbpf-ar+spatialnet-mimo-wrappedkf-ar_high_reverb}
                \fi
            \fi
        \else
            \ifnum\row=1
                \ifnum\x=0
                    \def\exampleName{spatialnet-rbpf-ar+spatialnet-wrappedkf-ar_low_reverb}
                \else
                    \ifnum\x=1
                        \def\exampleName{spatialnet-rbpf-ar+spatialnet-wrappedkf-ar_mid_reverb}
                    \else
                        \def\exampleName{spatialnet-rbpf-ar+spatialnet-wrappedkf-ar_high_reverb}
                    \fi
                \fi
            \else
                \ifnum\row=2
                    \ifnum\x=0
                        \def\exampleName{spatialnet-rbpf+spatialnet-wrappedkf_low_reverb}
                    \else
                        \ifnum\x=1
                            \def\exampleName{spatialnet-rbpf+spatialnet-wrappedkf_mid_reverb}
                        \else
                            \def\exampleName{spatialnet-rbpf+spatialnet-wrappedkf_high_reverb}
                        \fi
                    \fi
                \fi
            \fi
        \fi
        \pgfmathsetlengthmacro{\yDoaOff}{\row * (\doaHeight + \yDoaOffConst)}
        \node[anchor=mid] at (\plotStartX + \lrDist + \dOff + \x * \doaDist + 0.5 * \dWidth, \plotStartY + 0.5 \doaHeight + \yDoaOff) {%
            \pgfimage[height=\doaHeight, width=\dWidth]{images/recorded_trajectories/\exampleName.pdf} 
        };
        \pgfmathsetmacro{\timeNum}{int(\l)}
        \pgfmathsetmacro{\timeOff}{\doaWidth / 6}
        \pgfmathsetmacro{\doaOff}{\doaHeight / 4 - 0.025mm}
        \foreach \t in {0,...,\timeNum}{
            \pgfmathtruncatemacro\tmp{int(\t/2) * 2}
                \pgfmathsetmacro{\tickLen}{\majorTick}
                \pgfmathsetmacro{\time}{int(\t)}
                \ifnum\row=0
                    \node[anchor=north] at (\plotStartX + \lrDist + \dOff + \x * \doaDist + \t * \timeOff, \plotStartY + \doaTickVoff + \yDoaOff) {\fontsize{\tickSize}{\tickSkip}\selectfont \time};
                \fi
            \draw[line width=\tickWidth] (\plotStartX + \lrDist + \dOff + \x * \doaDist + \t * \timeOff, \plotStartY + \doaTickVoff + \yDoaOff) --  (\plotStartX + \lrDist + \dOff + \x * \doaDist + \t * \timeOff, \plotStartY - \tickLen + \doaTickVoff + \yDoaOff);
        }

        \foreach \d in {0,...,4}{
            \pgfmathtruncatemacro\tmp{int(\d/2) * 2}
                \pgfmathsetmacro{\tickLen}{\majorTick}
                \ifthenelse{\x = 0}{
                    \pgfmathsetmacro{\doa}{int(\d * 90 - 180)}
                        \node[anchor=east] at (\plotStartX + \lrDist + \dOff + \x * \doaDist ,\plotStartY + \d * \doaOff + \doaTickHoff + \yDoaOff) {\fontsize{\tickSize}{\tickSkip}\selectfont \doa};
                }{}
            \draw[line width=\tickWidth] (\plotStartX + \lrDist + \dOff + \x * \doaDist - \tickLen,\plotStartY + \d * \doaOff + \doaTickHoff + \yDoaOff) --  (\plotStartX + \lrDist + \dOff + \x * \doaDist,\plotStartY + \d * \doaOff + \doaTickHoff + \yDoaOff);
        }
    }
    
}

\node[anchor=north, xshift=5mm] at (0.5 * \linewidth, \plotStartY- 3*\majorTick + \doaTickVoff) {\footnotesize time [s]};
\node[anchor=north, xshift=0.5mm, yshift=-1.5mm, rotate=90] at (\plotStartX, \plotStartY + 1.5 * \doaHeight + 1* \doaDist) {\labelFont DoA\,[°]};

\pgfmathsetlengthmacro{\figWidth}{
10 / 6 * \doaWidth
}
\foreach \rIdx/\rTime in {
    0/200,1/350,2/800%
}{
    \node[anchor=south, xshift=1.8mm, yshift=-4.2mm] at (\plotStartX + \figWidth + \rIdx * \figWidth + \rIdx * \doaDist, \plotStartY + 3 * \doaHeight + 2* \yDoaOffConst) {\contour{white}{\labelFont $T_{60\!}=\, $\rTime\,ms}};
}

\end{tikzpicture}
\vspace*{-20pt}
\caption{
Estimated two-speaker ({\protect\tikz[baseline=-1.0ex]  \protect\draw[tab_blue, line width=1pt] (0,0mm) -- (1mm,0mm);}/{\protect\tikz[baseline=-0.4ex] \protect\draw[tab_orange, line width=1pt] (0,0mm) -- (1mm,0mm);}) \ac{doa} trajectories under increasing reverberation (left to right) with (top to bottom) concatenative~(\cref{fig:weak_ssf}) and our autoregressive (\mbox{MISO-AR}:~\cref{fig:weak_ssf_miso}, \mbox{MIMO-AR}:~\cref{fig:weak_ssf_mimo}) tracking methods using the \mbox{Wrapped\,KF}
({\protect\tikz[baseline=-0.6ex]  \protect\draw[color=tab_orange, line width=1pt, decorate, dash pattern=on 1pt off 1pt,
  decoration={snake, amplitude=3pt, segment length=10pt}] (0, 0) -- (4mm, 0);}) and \mbox{Bootstrap\,PF} ({\protect\tikz[baseline=-0.6ex]  \protect\draw[color=tab_orange, line width=1pt, decorate, dash pattern=on 3pt off 1pt,
  decoration={snake, amplitude=3pt, segment length=10pt}] (0, 0) -- (4mm, 0);}). 
  The Watson likelihood from \cref{eq:watson_likelihood} is shown as a background reference (\picLegendLarge{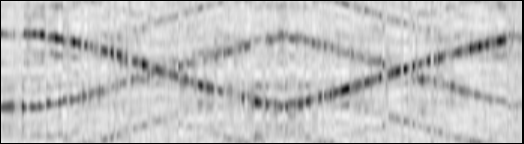}) to illustrate the effect of reverberation.
  Shaded areas (\protect\gridIcon) indicate regional accuracy (Re-ACC) computation, which denotes the fraction of estimates lying on the expected array side after approximating the circular speaker motion by piecewise-linear azimuth segments.
}
\label{fig:recorded_trajectories}
\end{figure*}

\subsection{Generalization and Robustness in Real-World Recordings}\label{sec:lab_recordings_eval}

\textbf{\textit{Recording Setup}}
To assess generalizability and robustness to real-world settings, we include recordings from a variable-acoustics listening room measuring 9.5\,m\,$\times$\,5.1\,m\,$\times$\,2.4\,m \diffComm{with the same}{in three acoustic conditions (reverberation times of 200\,ms, 350\,ms and 800\,ms), using a} centered microphone array as in \cref{sec:dataset_parametrization}.
Each recording features two male non-native English speakers reading Rainbow Passage segments \cite{fairbanks60rainbow_passage} while walking throughout the room.
We adopt the circular trajectories introduced in \cite{kienegger25sg_ssf}, which yield motion patterns suitable for evaluating tracking performance without ground-truth positional data.
Specifically, the speakers start from opposite ends at \diffComm{roughly 1\,m and 2\,m distance}{distances of roughly 1\,m and 2\,m} from the array, traverse to the other end of the room\diffComm{,}{} and back over the duration of their prompt. 
\diffComm{We test three acoustic conditions with reverberation times of 200\,ms, 350\,ms and 800\,ms.}{}
Each configuration includes three 10$-$20\,s recordings, 
\mbox{yielding nine two-speaker mixtures in total}.
Videos of the recording setup are available on our project page\textsuperscript{\ref{project_page}}.

\textbf{\textit{Target Speaker Tracking}} 
\Cref{fig:recorded_trajectories} visualizes the tracking results using the \mbox{Wrapped\,KF}~({\protect\tikz[baseline=-0.6ex]  \protect\draw[color=tab_orange, line width=1pt, decorate, dash pattern=on 1pt off 1pt,
  decoration={snake, amplitude=3pt, segment length=10pt}] (0, 0) -- (4mm, 0);}) and \mbox{Bootstrap\,PF}~({\protect\tikz[baseline=-0.6ex]  \protect\draw[color=tab_orange, line width=1pt, decorate, dash pattern=on 3pt off 1pt,
  decoration={snake, amplitude=3pt, segment length=10pt}] (0, 0) -- (4mm, 0);}) for our 10\,s listening room recordings under varying acoustic conditions. 
  The Watson likelihood from \cref{eq:watson_likelihood}, shown as a background reference~\diffComm{}{(\picLegendLarge{images/recorded_trajectories_large/watson_background})}, \diffComm{clearly}{} exposes the smearing effect of \diffComm{high}{} reverberation on spatial features, which increases tracking difficulty.
  Consistent with the synthetic dataset, the Bayesian filters in their original formulation (top row) yield inaccurate tracking results, especially with increasing reverberation.
   In contrast, the \ac{ar} versions using \ac{ssf} estimates (\mbox{MISO-AR}, center row and \mbox{MIMO-AR}, bottom row) robustly resolve both speaker crossings.
 For a quantitative evaluation, we approximate the \diffComm{speaker's}{speakers'} circular motion patterns with piecewise-linear azimuth trajectories. 
 After segmenting the duration of each recording into four parts, we compute the fraction of estimates on the expected array side (see \protect\gridIcon[1.3]\, in \cref{fig:recorded_trajectories}), termed regional accuracy (\mbox{Re-ACC}), which is \diffComm{particular}{particularly} sensitive to speaker confusions after directional crossings.
 The results in \cref{fig:recorded_metrics} show how both Bayesian filters in MIMO-AR configuration~(\picLegend{images/tst_accuracy/wrapped_kf_mimo_ar.pdf},\,\picLegend{images/tst_accuracy/bootstrap_pf_mimo_ar.pdf}) and the MISO-AR \mbox{Bootstrap\,PF}~(\picLegend{images/tst_accuracy/bootstrap_pf_miso_ar.pdf}) \diffComm{consistently}{} retain robust tracking accuracy, demonstrating \textit{generalizability} to real-world recordings under unseen acoustic conditions. 

\textbf{\textit{Target Speaker Extraction}}
To evaluate perceptual quality without ground truth speech signals, we employ NISQA~\cite{mittag21nisqa}, a data-driven, non-intrusive estimator of the subjective mean opinion score. 
For intelligibility, we leverage transcriptions of a downstream \ac{asr} system and compute the \ac{wer} against the Rainbow Passage reference segments.
Specifically, we utilize the \ac{asr} model \texttt{QuartzNet15x5Base-En}~\cite{kuchaiev9nemo}, which is very sensitive to signal distortions as it is only trained on clean and telephony speech. 
\Cref{fig:recorded_metrics} presents the enhancement results obtained from the listening room recordings using the Bayesian trackers for guidance.
Both perceptual quality (NISQA) and intelligibility (WER) demonstrate how the increased tracking accuracy of the \ac{ar} methods, particularly the MIMO-AR configurations~(\picLegend{images/tst_accuracy/wrapped_kf_mimo_ar.pdf},\,\picLegend{images/tst_accuracy/bootstrap_pf_mimo_ar.pdf}) and the MISO-AR \mbox{Bootstrap\,PF}~(\picLegend{images/tst_accuracy/bootstrap_pf_miso_ar.pdf}), translate into superior enhancement, consistent with the trend on the synthetic data in \cref{tab:results}.
Listening tests, provided on our project page\textsuperscript{\ref{project_page}}, indicate that performance differences are most pronounced at the end of the recordings.
Without reliable guidance, the \mbox{non-AR} approaches (\picLegend{images/tst_accuracy/wrapped_kf.pdf},\,\picLegend{images/tst_accuracy/bootstrap_pf.pdf}) must retain speaker characteristics over time and eventually suffer from signal distortions and speaker leakage.
The accurate tracking provided by our \ac{ar} methods prevents this degradation and yields robust enhancement throughout \diffComm{\textit{long‑form} audio recordings}{both speaker crossings}.
\begin{figure}[t!]
\begin{tikzpicture}
    
\newcommand\figSize{1.35cm}
\pgfmathsetlengthmacro{\figWidth}{\figSize * 1.5}
\pgfmathsetlengthmacro{\figDist}{9mm} 
\pgfmathsetlengthmacro{\figHeight}{\figSize * 1.0}
\newcommand\plotStartX{0mm}
\newcommand\plotStartY{0mm}

\foreach \plotIdx/\metricName in {%
    0/re_acc,%
    1/nisqa,%
    2/wer%
}{
    \ifnum\plotIdx=0
    \pgfmathsetlengthmacro{\newHeight}{1.025*\figHeight}
        \node[anchor=center] at (\plotStartX + 0.5 * \figWidth + \plotIdx * \figWidth + \plotIdx * \figDist, \plotStartY + 0.5 * \newHeight) {%
            \pgfimage[height=\newHeight, width=\figWidth]{images/recorded_metrics/\metricName.pdf} 
        };
    \else
        \node[anchor=center] at (\plotStartX + 0.5 * \figWidth + \plotIdx * \figWidth + \plotIdx * \figDist, \plotStartY + 0.5 * \figHeight) {%
                \pgfimage[height=\figHeight, width=\figWidth]{images/recorded_metrics/\metricName.pdf} 
            };
    \fi
    
     \pgfmathsetlengthmacro{\xTickDist}{
        \figWidth / 3
     }
    \foreach \x / \xLabel in {
        0/Concat,1/MISO-AR,2/MIMO-AR
    } {
        \node[anchor=east, rotate=30, yshift=-1mm, xshift=0.5mm] at (\plotStartX + 0.5 * \xTickDist + \x * \xTickDist + \plotIdx * \figWidth + \plotIdx * \figDist,\plotStartY) {\tickFont \xLabel};
        \draw[line width=\tickWidth] (\plotStartX + 0.5 * \xTickDist + \x * \xTickDist + \plotIdx * \figWidth + \plotIdx * \figDist,\plotStartY) -- (\plotStartX + 0.5 * \xTickDist + \x * \xTickDist+ \plotIdx * \figWidth + \plotIdx * \figDist,\plotStartY- \majorTick);
    }

    \pgfmathsetlengthmacro{\labelDist}{
        \figHeight / 5
    }
    \ifnum\plotIdx=1
        \node[anchor=south, rotate=90, yshift=4.0mm] at (\plotStartX + \plotIdx * \figWidth + \plotIdx * \figDist,\plotStartY + 0.5 * \figHeight) {\labelFont NISQA\,$\rightarrow$};
        \foreach \y/\yLabel in {%
        0/2.5,1/3.0,2/3.5,3/4.0,4/4.5,5/5.0%
    } {%
        \node[anchor=east] at (\plotStartX + \plotIdx * \figWidth + \plotIdx * \figDist,\plotStartY + \y * \labelDist) {\tickFont \yLabel};
        \draw[line width=\tickWidth] (\plotStartX + \plotIdx * \figWidth + \plotIdx * \figDist,\plotStartY + \y * \labelDist) -- (\plotStartX +\plotIdx * \figWidth + \plotIdx * \figDist- \majorTick,\plotStartY + \y * \labelDist);
    }
    \fi
    \ifnum\plotIdx=2
    \pgfmathsetlengthmacro{\labelDist}{
        \figHeight / 8
    }
        \node[anchor=south, rotate=90, yshift=3.5mm] at (\plotStartX + \plotIdx * \figWidth + \plotIdx * \figDist,\plotStartY + 0.5 * \figHeight) {\labelFont WER\,[\%]\,$\leftarrow$};
        \foreach \y in {%
        0,...,8%
    } {%
    \pgfmathsetmacro{\yLabel}{int(\y*5)}
    \pgfmathtruncatemacro{\tmp}{int(\y / 2) * 2}
    \ifnum\tmp=\y
        \node[anchor=east] at (\plotStartX + \plotIdx * \figWidth + \plotIdx * \figDist,\plotStartY + \y * \labelDist) {\tickFont \yLabel};
        \draw[line width=\tickWidth] (\plotStartX + \plotIdx * \figWidth + \plotIdx * \figDist,\plotStartY + \y * \labelDist) -- (\plotStartX +\plotIdx * \figWidth + \plotIdx * \figDist- \majorTick,\plotStartY + \y * \labelDist);
    \else 
        \draw[line width=\tickWidth] (\plotStartX + \plotIdx * \figWidth + \plotIdx * \figDist,\plotStartY + \y * \labelDist) -- (\plotStartX +\plotIdx * \figWidth + \plotIdx * \figDist- \minorTick,\plotStartY + \y * \labelDist);
    \fi
    }
    \fi
    \pgfmathsetlengthmacro{\labelDist}{
        \figHeight / 8
    }
    \ifnum\plotIdx=0
         \node[anchor=south, rotate=90, yshift=3.5mm, xshift=-1.2mm] at (\plotStartX + \plotIdx * \figWidth + \plotIdx * \figDist,\plotStartY + 0.5 * \figHeight) {\labelFont Re-ACC\,[\%]\,$\rightarrow$};
        \foreach \y in {%
        0,...,8%
    } {%
    \pgfmathsetmacro{\yLabel}{int(20 + \y*10)}
    \pgfmathtruncatemacro{\tmp}{int(\y / 2) * 2}
    \ifnum\tmp=\y
        \node[anchor=east] at (\plotStartX + \plotIdx * \figWidth + \plotIdx * \figDist,\plotStartY + \y * \labelDist) {\tickFont \yLabel};
        \draw[line width=\tickWidth] (\plotStartX + \plotIdx * \figWidth + \plotIdx * \figDist,\plotStartY + \y * \labelDist) -- (\plotStartX +\plotIdx * \figWidth + \plotIdx * \figDist- \majorTick,\plotStartY + \y * \labelDist);
    \else 
    \draw[line width=\tickWidth] (\plotStartX + \plotIdx * \figWidth + \plotIdx * \figDist,\plotStartY + \y * \labelDist) -- (\plotStartX +\plotIdx * \figWidth + \plotIdx * \figDist- \minorTick,\plotStartY + \y * \labelDist);
    \fi
    }
    \fi

}

\end{tikzpicture}
\vspace*{-20pt}
\caption{
Tracking (Re-ACC) and enhancement (NISQA, WER) performance of weakly guided TSE methods with \mbox{Wrapped\,KF} (\picLegend{images/tst_accuracy/wrapped_kf.pdf}) and \mbox{Bootstrap\,PF}~(\picLegend{images/tst_accuracy/bootstrap_pf.pdf}).
Unprocessed recordings yield an average of 1.82~NISQA and 81.4\,\%~WER.
}
\label{fig:recorded_metrics}
\end{figure}

\section{Conclusion}
Based on our conference paper~\cite{kienegger25sg_ssf}, we investigated how to improve lightweight Bayesian tracking by autoregressively (AR) incorporating the processed speech signal of a deep \acl{ssf} (\acs{ssf}). 
On top of the multichannel (MIMO) \ac{ssf} extension from~\cite{kienegger25sg_ssf}, we developed novel Bayesian filtering formulations, which integrate the enhanced speech without modifying the \ac{ssf}.
To enable development with realistic motion patterns, we released a synthetic \diffComm{dataset}{data generation framework} based on the social force motion model, which yields superior generalization to \diffComm{real-word}{real-world} trajectories.
\diffComm{A detailed analysis on our synthetic}{Analysis on the resulting} dataset demonstrates significant tracking improvements for our \ac{ar} Bayesian methods with none or negligible additional overhead, achieving competitive accuracy relative to neural methods of much greater complexity. 
Real-world recordings complement these findings, 
with the performance gains of our \textit{autoregressive} methods generalizing to challenging and unseen realistic acoustic conditions.

\bibliographystyle{IEEEtran}
\bibliography{refs, strings}

\vspace{-20pt}

\begin{IEEEbiography}[{\includegraphics[width=1in,height=1.25in,clip,keepaspectratio]{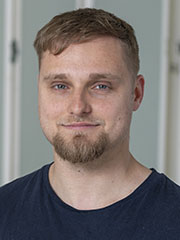}}]{Jakob Kienegger} (Student Member, IEEE) received the B.Sc. degree in  Electrical Engineering from the OWL University of Applied Sciences, Lemgo, Germany, in 2021, and the M.Sc. degree from the University of Paderborn, Paderborn, Germany, in 2024. 
He is currently with the Signal Processing Research Group, University of Hamburg, Hamburg, Germany, under the supervision of Prof. Timo Gerkmann. 
His research interests include statistical signal processing and machine learning applied to sound source localization and multichannel speech enhancement. 
\end{IEEEbiography}

\vspace{-20pt} %

\begin{IEEEbiography}[{\includegraphics[width=1in,height=1.25in,clip,keepaspectratio]{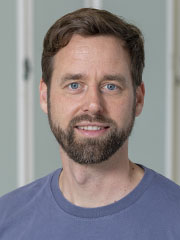}}]{Timo Gerkmann} (Senior Member, IEEE)
is a professor with the University of Hamburg, Germany, where he is the head of the Signal Processing Group. He has previously held positions with Technicolor, University of Oldenburg, Germany, KTH Royal Institute of Technology, Sweden, Ruhr-Universit\"at Bochum, Germany, and Siemens Corporate Research, Princeton, NJ, USA. His research interests include statistical signal processing and machine learning for speech and audio. His work received the VDE ITG award 2022 and 2024.
\end{IEEEbiography}

\vfill

\end{document}